\documentclass[12pt,a4paper]{article}
\usepackage[utf8]{inputenc}
\usepackage[T1]{fontenc}
\usepackage[a4paper,margin=2.5cm]{geometry}
\usepackage{setspace}
\usepackage{hyperref}
\usepackage{titlesec}
\usepackage[authoryear,round]{natbib}
\usepackage{graphicx}
\usepackage{amsmath}
\usepackage{amssymb}
\usepackage{booktabs}
\usepackage{multirow}
\usepackage{tabularx}
\usepackage{todonotes} %
\usepackage{tcolorbox}
\usepackage{fancyvrb}
\usepackage{listings}
\usepackage{placeins}
\usepackage{subcaption}
\tcbuselibrary{listings, skins, breakable}

\lstdefinestyle{systemprompt}{
  basicstyle    = \ttfamily\scriptsize,
  breaklines    = true,
  breakatwhitespace = true,
  frame         = single,
  framerule     = 0.8pt,
  rulecolor     = \color{gray!60},
  backgroundcolor = \color{gray!8},
  xleftmargin   = 3pt,
  xrightmargin  = 3pt,
  aboveskip     = 3pt,
  belowskip     = 3pt,
}

\newtcolorbox{promptbox}[1]{
  colback   = gray!7,
  colframe  = gray!50,
  fonttitle = \bfseries\sffamily,
  title     = #1,              %
  breakable,
  arc       = 4pt,
  listing only,
  listing options = {style=systemprompt, frame=none,
                     aboveskip=0pt, belowskip=0pt},
}

\usepackage{enumitem}
\usepackage{chngcntr}
\usepackage{float}
\newtheorem{hypothesis}{Hypothesis}
\newtheorem{result}{Result}

\newlength{\dashboardpanelheight}
\titleformat{\section}{\large\bfseries}{\thesection.}{0.5em}{}

\title{Rule-Based Pricing Algorithms and Market Outcomes: An Experimental Study\footnote{We thank Leon Musolff, Emilio Calvano, Ginger Jin, and participants at EARIE (Mannheim), seminars in Leicester and Galway, and the AI and the Future of Work Symposium (Berlin) for useful comments. Normann thanks the German Science Foundation (DFG) for financial support (FOR 5392, project NO 429/8  692804). }}

\author{
\begin{tabular}{@{}c@{\qquad\qquad}c@{}}
{Adrian Hillenbrand}\footnote{ZEW -- Leibniz Centre for European Economic Research, Mannheim, and Karlsruhe Institute of Technology (KIT). E-mail: \href{mailto:adrian.hillenbrand@zew.de}{adrian.hillenbrand@zew.de}}
&
{Hans-Theo Normann}\footnote{Düsseldorf Institute for Competition Economics (DICE), Heinrich Heine University Düsseldorf. E-mail: \href{mailto:normann@dice.hhu.de}{normann@hhu.de}}
\\[0.4em]
{Matthias Potarca}\footnote{Karlsruhe Institute of Technology (KIT). E-mail: \href{mailto:matthias.potarca@kit.edu}{matthias.potarca@kit.edu}}&
{Tobias Werner}\footnote{Department of Economics, Maynooth University. E-mail: \href{mailto:tobias.werner@mu.ie}{tobias.werner@mu.ie}}
\end{tabular}%
}

\date{September 22, 2026}

\begin{document}

\maketitle

\begin{abstract}
Rule-based pricing tools are widespread in digital commerce, yet we know little about how their design shapes market outcomes. In a controlled market experiment, participants use dashboards to build pricing algorithms competing in a sequential Bertrand game over multiple periods. We vary design features commonly found in commercial repricing tools: warnings about price wars, pre-configured strategies, and advice from a large language model. Most treatment variations raise market prices with effects driven by an increase in starting prices and more cooperative algorithm designs. The results matter for competition policy, platform regulation and current discussions on regulating algorithm design tools.
\end{abstract}



\newpage 
\section{Introduction}

Automated tools which assist or even determine pricing decisions are having an increasingly significant impact on digital commerce \citep{chen_empirical_2016,hanspach_algorithms_2024}. These ``repricers'' often consist of a set of simple, deterministic if-then rules that are usually user-configurable. However, the provider of these tools may also provide access to more sophisticated features such as large language models (LLMs), default strategies or strategic advice, which strategy to chose. Deterministic rule-based repricers can be distinguished from predictive pricing engines, provided by third parties, that operate on substantial external datasets to calculate optimal prices and which are often tailored to industries such as real estate, rentals, or gasoline.\footnote{See for example \hyperlink{realpage.com}{realpage.com}, \hyperlink{lodgify.com}{lodgify.com}, \hyperlink{a2isystems.com}{a2isystems.com}. See  \citet{harrington2025equilibrium} for more examples, and \citet{assad2024algorithmic} for a detailed study. } The repricers can be provided to small and medium-sized sellers directly from platforms such as Amazon, as well as from third party developers such as BQool, Repricer.com, Aura and Seller Snap. Already in 2017, the EU Commission reports that a majority of online firms track the prices of competitors and two-thirds of these use algorithmic pricing software \citep{eu2017final}, and rule-based tools remain vastly more prevalent than complex AI or reinforcement learning tools in e-commerce \citep{wang2023algorithms}. By providing defaults and strategic guidance, algorithm providers can steer the behavior of downstream sellers. Such features can be designed to align with the provider's own economic interests. For example, a monopolistic platform that takes a commission on sales has a direct incentive to maintain higher price levels. Independent third-party providers, can have an incentive to nudge sellers to more cooperative pricing strategies and avoid price wars, as this can increase the profitability of their clients and thereby secure their own recurring subscription base.

Although the algorithmic pricing tools may increase the efficiency of their users (and are usually marketed as such)\footnote{For an overview of algorithmic pricing from a marketing strategy and regulatory perspective, see e.g. \citet{spann2025algorithmic}.}, their specific design features could facilitate coordination on more collusive market outcomes \citep{musolff_algorithmic_2026}. In a recent document, the European Commission expressed concerns that ``revenue management strategies and tactics may have contributed to ... the rise in markups and profits'' \citep{EuropeanCommission2025}, and suspects that algorithmic pricing may be one of these strategies. Algorithms may support the coordination and avoid marginal-cost pricing. These worries may be attributed to a shift in the business strategy literature from ``battles for markets shares'' \citep{roberts1987battles} in the 1970s to recent strategies ``Manage for profit, not for market share'' \citep{simon2006manage}, to avoid prices wars and to ``learn to compete peacefully'' \citep{nagle2023strategy}. Similar concerns about collusive impacts of pricing algorithms had already been expressed earlier by several institutions and competition authorities \citep{algorithms2017collusion,competition2018pricing,bureau2018big,bundeskartellamt2019algorithms,competition2021algorithms}.  

In this paper, we study whether different design features influence sellers in their choice of rule-based pricing algorithms\footnote{Much of the existing literature focuses on the competitive effects of self-learning algorithms \citep[see, for instance,][]{calvano2020artificial, klein2021autonomous, johnson2023platform, kasberger2026algorithmic, werner2026algorithmic}. Related experimental work studies algorithmic price recommendations and the delegation of pricing decisions to algorithms \citep{hunold2025algorithmic,normann2025delegate}. At the same time, empirical studies such as \citet{hanspach_algorithms_2024} and \citet{wang2023algorithms} show that many algorithms used in e-commerce settings are still, at least to some degree, rule-based rather than fully autonomous.} and whether this facilitates collusion. In a novel market experiment, participants configure rule-based pricing algorithms that then compete over multiple periods in a Bertrand game. Next to a baseline variant, we study  the collusive impact of three different types of design features prevalent in such  online repricing environments, with a focus on features that are common in the industry and raise ex-ante concerns for collusive behavior. The controlled experimental environment lets us compare different choice architectures while holding the market environment fixed, observe the complete strategy including off-path rules, i.e. algorithms, and replay them in counterfactual match-ups. 

The specific allegedly collusive design features we study are (i) profitability nudges warning against ``price wars'', (ii) LLM advice, as well as (iii) pre-configured strategy menus and cooperative ``price matching'' defaults.  These design choices can be categorized as a form of choice architecture \citep{thaler2013choice}, potentially influencing choices through information provision or by simplifying the choice environment. They are strongly motivated by existing industry practice, and examples for all three design features are documented in Online Appendix \ref{app:industry_examples}.\footnote{While regulators are usually concerned with the effects of such nudges or ``dark patterns'' on consumers and competition \citep{cma2022oca,cma2022ocaevidence}, in our paper the choice architecture may influence sellers who are nudged towards more collusive algorithms, which, in our setting, directly reduces consumer welfare.} Providing more details on these three mechanisms, we note that, firstly, repricers directly inform their users about ``price wars'' and provide suggestions on how to avoid them and increase profits. These suggestions involve choosing algorithms that avoid mutual undercutting, which might consequently result in higher market prices.\footnote{This differs from settings where sellers communicate directly with each other such as in seminal collusion-communication experiments \citep{isaac1981opportunity, isaac1984effects}. \citet{holt1990effects} study posted-offer markets where participants could fill in a blank in a pre-formulated message. They find that the announcements initially raise prices, but this effect is only temporary as prices eventually reach the level of a treatment without announcements. \citet{harrington2016relative} study both non-binding boilerplate announcements and free-form cheap talk. They find that boilerplate announcements foster collusion but only in duopolies, while free-form communication does so also with larger numbers of firms. See also \citet{fonseca2012explicit}. } Secondly, online sellers may rely on LLMs for advice on optimising sales or devising pricing strategies, or some repricers may offer these services directly. This is important, as LLMs have been shown to achieve supra-competitive outcomes \citep{fish_algorithmic_2025, giebel2026artificial}. Thirdly, repricers often provide pre-generated algorithms to help sellers achieve various goals, such as outperforming or matching a competitor. By simplifying the choice environment, the menu choice of algorithms may increase prices for consumers by nudging sellers toward less competitive, and more coordinated pricing strategies 

Our experimental setting and its underlying market model are innovative in that they capture key features of market interactions on online platforms. First, a central part of our design is a pricing dashboard that allows participants to generate a large set of pricing algorithms and emulates the decision environments of sellers using repricer dashboards.\footnote{Earlier papers use various ways of eliciting full strategy profiles such as the original strategy method \citep{selten1967strategiemethode}, including methods to fully elicit the strategies participants choose in repeated games \citep{romero_constructing_2018, dal2019strategy,romero_mixed_2023}, allowing fully non-restricted programming of strategies \citet{selten_duopoly_1997} and \citet{axelrod_effective_1980,axelrod1980more}, or a choice between a fixed price and a pre-configured pricing policy  \cite{deck_automated_2003}. 
Relatedly, \citet{ershov2025managing} study how human contractors and LLMs develop pricing algorithms from high-level managerial prompts. In contrast, in our setting sellers chose a rule-based strategy within a restricted (although large) set of strategies with the help of a dashboard, and we are interested in how the chosen strategies change across different design environments.}  Second, many platforms use some kind of ``Buy Box''. That is, instead of showing all offers of a particular product, only the cheapest product is shown. Third, pricing algorithms often are built to react to price changes, so, they are sequential in nature.\footnote{We restrict the decision environment to a sequential setting reflecting the observation that prices adjust asynchronously on real-world platforms \citep{musolff_algorithmic_2026,brown_competition_2023} and the typical options on commercial repricer dashboards.} Accordingly, in our experiment participants design a pricing algorithm that competes with the pricing algorithm of another participant over 50 periods in a sequential-move Bertrand game. Participants play five such supergames in total. The pricing dashboard captures the key aspects of repricer markets while still being comprehensible. The dashboard allows participants to build conditional strategies for the repeated game based on whether they ``win the buy box'', that is, had a lower price than the opponent in the last period, or not. The above mentioned treatments vary interface design such as providing pre-configured algorithms and the availability of large language-model advice: some participants receive a brief nudge to consider long-run profitability, others can consult an LLM chatbot, and others can chose among pre-configured algorithms.

In our theory part, we analyze a finitely repeated sequential-move Bertrand game, closely matching the experimental design.\footnote{The distinction between a finitely repeated market game \citep[e.g.,][]{dufwenberg2000price, freitag2021communication, BruttelRauWerner2025} and an indefinitely repeated market game \citep[see, for example,][]{holt1985experimental, bigoni2019frequency, andres2023communication}, as commonly used in repeated-market experiments, does not concern our analysis. Since participants commit to their pricing algorithm before the supergame begins and cannot revise it during the supergame, their strategic choice is made only once, irrespective of whether the subsequent market interaction has a finite or indefinite horizon.} Since participants must commit to a complete pricing algorithm before play begins and cannot revise it during the game, the interaction becomes a one-shot normal-form game in which both players choose their strategies simultaneously at the outset, making Nash equilibrium the appropriate solution concept. From a large set of potential Nash equilibria, we identify three subsets of qualitatively similar equilibrium strategies: Static competitive equilibria, symmetric supracompetitive equilibria, and asymmetric cycling equilibria. %

Our results show a clear increase in market prices in most of our treatment variations (Section \ref{sec:results}). The highest average price is achieved in a coordination treatment where a tit-for-tat algorithm is provided. LLM advice, however, does not improve prices further. We then investigate the mechanisms behind these effects (Section \ref{sec:mechanisms}): the increase in prices is mostly driven by higher starting prices and more lenient algorithm choices, while the price trajectory on average stays more or less constant within a supergame. Our design also allows us to directly identify the types of algorithms chosen by the participants (Section \ref{sec:algoselection}), revealing that higher market prices are in parallel to a shift in algorithm choice towards more cooperative algorithms. Providing a pre-configured menu of algorithms does not increase market prices further but changes the types of algorithms that are observed in the market towards more optimal, that is, less-exploitable and therefore more sustainable market interactions.

In an extension (Section~\ref{sec:llm_extension}), we address the growing concern that decision makers completely outsource their pricing decisions to AI tools. In additional market simulations, we let LLMs compete with one another within the same decision environment. This allows a clear comparison of the coordination capabilities of LLMs with those of human sellers, providing insights into how market outcomes may change as full delegation to AI becomes more prevalent. Fielding different LLMs, we demonstrate that, especially for reasoning models, prices are substantially higher than even in human markets, while models without reasoning capabilities achieve low prices. This is driven by the choice of algorithms, with a large share of tit-for-tat algorithms among reasoning models and a larger share of undercutting algorithms among the other types of LLMs.

Our results demonstrate that the choice architecture of pricing dashboards, specifically through nudges, defaults, and AI advice, effectively facilitates supra-competitive pricing. We show that these design features can foster coordination without explicit communication. Our study provides critical insights for competition policy by revealing how platform design can systematically steer market behaviour toward less competitive outcomes. Beyond its policy implications, our work provides a novel behavioural insight into the human-algorithm interface, demonstrating how subtle design interventions can fundamentally reshape strategic interaction in digital commerce.

\section{Experimental Design and Hypotheses}

Our experimental design mirrors the setting commonly found in real-world pricing environments on Amazon and other platforms by requiring participants to develop pricing algorithms via a dashboard to compete for the ``Buy Box''. Participants assume the role of sellers competing across five supergames, where they must design rule-based strategies for a 50-period sequential Bertrand duopoly. Sections~\ref{sec:market_game} and~\ref{sec:algorithm_builder} describe the market environment and dashboard design. Section~\ref{sec:theory} provides the conceptual framework, Section~\ref{sec:treatments} introduces the treatments, Section~\ref{sec:hypo_measures} presents the hypotheses and outcome variables, and Section~\ref{sec:procedures} details the experimental procedures.

\subsection{Market Environment}
\label{sec:market_game}

Two symmetric sellers compete in a sequential Bertrand duopoly over $T = 50$ periods. Each seller sets an integer price $p_i \in \mathcal{A} = \{2, 3, \ldots, 10\}$ with marginal cost $c = 0$ and inelastic demand of one unit per period. The seller with the lower price in a given period earns profit
\begin{equation}
    \pi_i =
    \left\{\begin{array}{cl}
        p_i           & \text{if } p_i < p_j, \\[4pt]
        p_i \text{ or } 0 & \text{if } p_i = p_j \text{ (each with prob.\ }\tfrac{1}{2}\text{)}, \\[4pt]
        0             & \text{if } p_i > p_j.
    \end{array}\right.
    \label{eq:profit}
\end{equation}
where winning the random draw or having the lower price is termed as ``winning the Buy Box''.
In period~1, both sellers set a start price and one seller is randomly selected to move first. After that, sellers alternate in updating their prices. In each period, the active seller may change their price while the other's price remains fixed. This sequential structure mirrors the asynchronous price adjustment observed on real-world platforms \citep{musolff_algorithmic_2026, brown_competition_2023, brown2025algorithmic} and is consistent with the reactive conditioning options common on commercial repricer dashboards.\footnote{Moreover, the alternating-move structure eliminates substantial additional complexity as a simultaneous move structure would require sellers to anticipate the price change of the other seller. While this might be a factor in real-world markets, it hinders a clean identification of seller strategies.}

\subsection{Algorithm Dashboard}
\label{sec:algorithm_builder}

Rather than setting prices directly, sellers in our experiment design a \emph{pricing algorithm} that defines the pricing strategy across all periods. The algorithm can determine the active seller's price based on the current prices and on whether the buy-box was won (own algorithm had the lowest price or won the random draw in a tie) or lost in the previous round.

\begin{figure}[t]
  \centering

  \begin{tabular}{
    @{}
    p{0.489\textwidth}
    @{\hspace{0.0105\textwidth}}
    c
    @{\hspace{0.0105\textwidth}}
    p{0.489\textwidth}
    @{}
  }

    \subcaptionbox{
      Experimental dashboard
      \label{fig:dashboard-experimental}
    }[\linewidth]{}
    &
    &
    \subcaptionbox{
      Commercial dashboard (BQool)
      \label{fig:dashboard-bqool}
    }[\linewidth]{}
    \\[-1pt]

    \begin{minipage}[c][0.3542\textwidth][c]{\linewidth}
      \centering
      \includegraphics[
        width=\linewidth
      ]{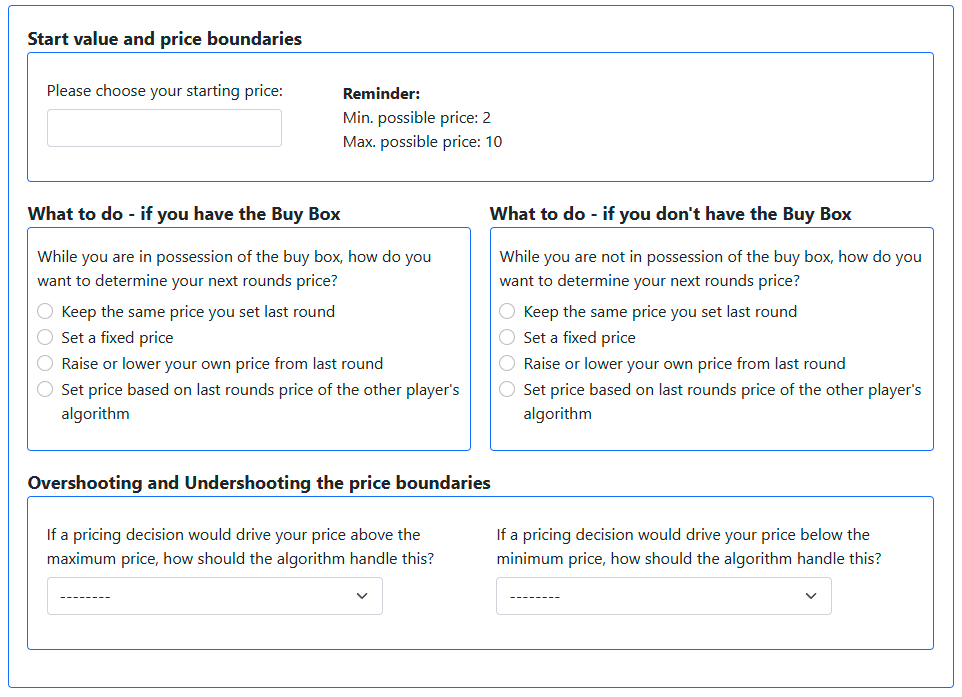}
    \end{minipage}

    &

    \raisebox{-.5\height}{%
      \textcolor{black!15}{%
        \rule{0.4pt}{0.3542\textwidth}%
      }%
    }

    &

    \begin{minipage}[c][0.3542\textwidth][c]{\linewidth}
      \centering
      \includegraphics[
        width=\linewidth
      ]{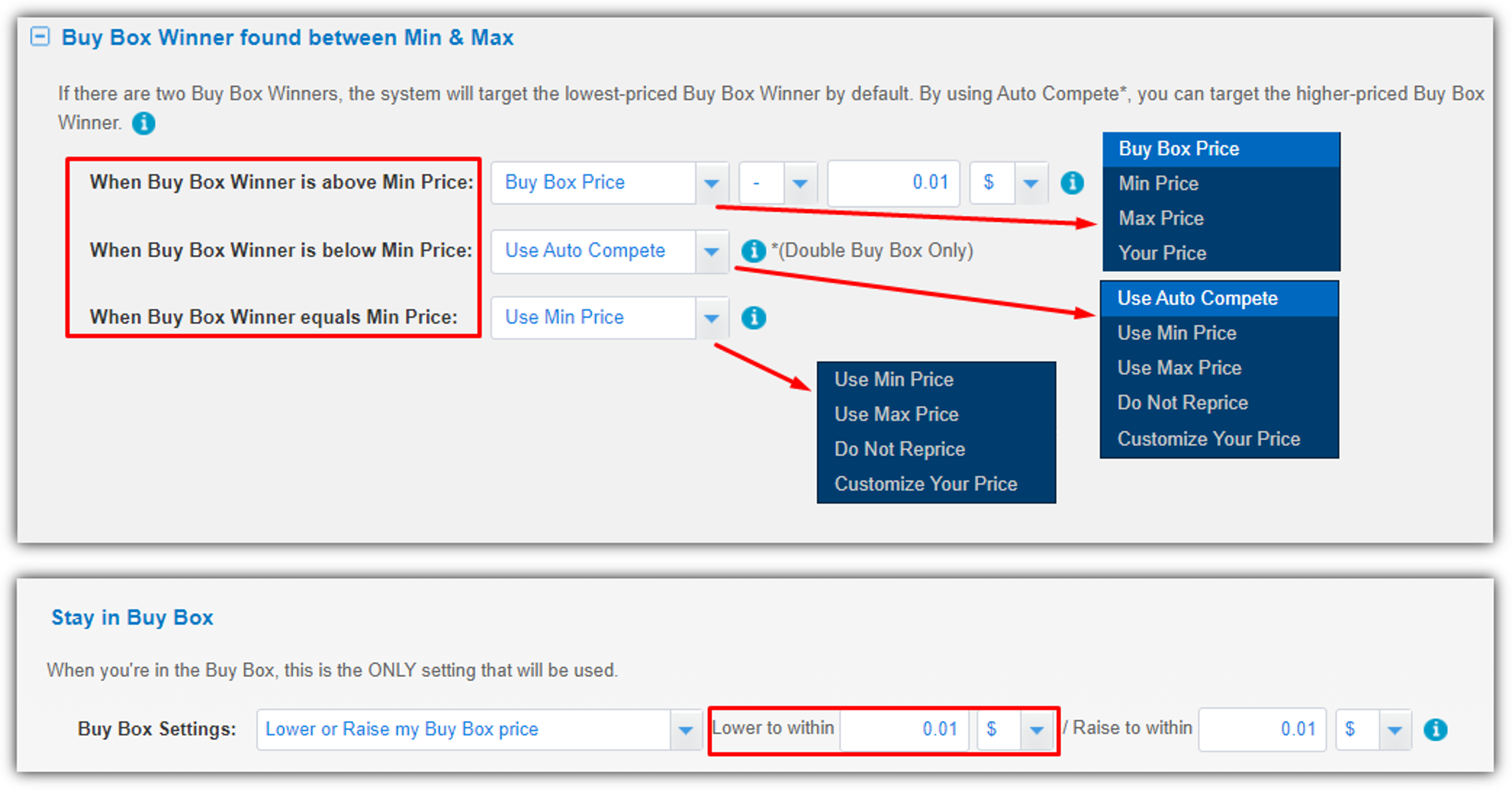}
    \end{minipage}

  \end{tabular}

  \caption{%
    Rule-based repricing dashboards.
    Panel (a) shows the experimental dashboard used to design pricing
    algorithms in the experiment. Panel (b) shows two exemplary fragments
    of a commercial rule-based repricing interface, cropped together from
    the BQool Help Center. The full commercial algorithm design can
    incorporate additional features and settings from which we abstract
    in the experiment.\protect\footnotemark
  }
  \label{fig:dashboard-comparison}
\end{figure}

\footnotetext{%
  For the full set of options available in BQool's rule-based repricing
  interface, see
  \url{https://support.bqool.com/hc/en-us/articles/205025008-How-to-Create-Edit-a-Rule-based-Rule} (last accessed August~21, 2026).%
}

The pricing algorithm is implemented using a dashboard interface closely inspired by real-world algorithm design pages. We provide a representative example that resembles a common design pattern in the industry, shown in Figure~\ref{fig:dashboard-comparison}, along with its implementation in the experiment. We provide further screenshots of the dashboard implementation in the experiment in Online Appendix~\ref{app:screenshots}. The algorithm, which is designed via the dashboard, specifies:

\begin{minipage}{\linewidth}
\begin{enumerate}[itemsep=0pt, topsep=2pt, parsep=0pt]
    \item \textbf{Start price:} The initial price in period~1, chosen from $\{2, 3, \ldots, 10\}$.
    \item \textbf{Buy Box won rule:} How to adjust the price when the seller currently holds the Buy Box. There are four options:
        \begin{itemize}[itemsep=0pt, topsep=2pt, parsep=0pt]
            \item Keep the same price as last period
            \item Set a fixed price
            \item Set price based on one's own current price, so either keeping it, or raising/lowering it by a freely selectable integer amount
            \item Set price based on the other player's current price, so either matching it, or exceeding/undercutting it by a freely selectable integer amount
        \end{itemize}
    \item \textbf{Buy Box lost rule:} How to adjust the price when the seller does not hold the Buy Box. The same four options as above are choosable.
    \item \textbf{Boundary conditions:} What to do when the computed price exceeds the upper bound ($p > 10$) or falls below the lower bound ($p < 2$). Separate boundary rules apply for overshooting and undershooting. There are three options: 
    \begin{itemize}[itemsep=0pt, topsep=2pt, parsep=0pt]
        \item set price to the maximum
        \item set price to the minimum
        \item stay at the current price
    \end{itemize}
\end{enumerate}
\end{minipage} \newline

While the strategy space is deliberately simplified relative to commercial repricers that may have additional conditioning, the core logic of conditioning on Buy Box status and adjusting relative to own or competitor prices matches the basic structure of rule-based repricing tools as documented by \citet{calzolari_pricing_2025}. The restricted strategy space is what enables the full equilibrium characterisation in Section~\ref{sec:theory}.

\subsection{Conceptual Framework}
\label{sec:theory}

Two features of the game distinguish it from a standard finitely repeated game and shape the theoretical analysis. First, although the game spans 50 periods, the game does not decompose into self-contained stage games, as prices set carry over from period to period asynchronously, creating persistent state dependencies. More importantly, participants must commit to a complete pricing algorithm \emph{before} play begins, with no opportunity to revise it within a given supergame. This pre-commitment transforms the interaction into a one-shot normal-form game in which both players simultaneously choose their complete strategies at the start, making Nash equilibrium the appropriate solution concept. Second, the strategy space is restricted to a finite menu of rule-based algorithms (see Section~\ref{sec:algorithm_builder}). Because each algorithm applies the same rules uniformly across all non-initial periods, sellers cannot condition behaviour on a particular period, and in particular cannot condition on the final period. This rules out the terminal-period reasoning that initiates the standard backward-induction unravelling in finitely repeated games with a unique stage-game equilibrium \citep{Selten1978}.

To characterize equilibria, we model the game as a two-agent finite-horizon Markov decision process. There is a finite set of feasible strategies, which can be derived by working out each combination of possible inputs in the algorithm dashboard.\footnote{Note that there is an arbitrarily large number of ways to produce a valid input for the pricing dashboard, as any integer change increment is allowed when conditioning on the own/opponent's price. However, any conditioning rule that tries to perform a price change of 9 or more in either direction is sure to activate a boundary condition, which makes it possible to merge those cases.} Each of these strategies can be described as a deterministic algorithm $\mu: \mathcal{S} \rightarrow \{2,\ldots,10 \}$, with $\mathcal{S}$ describing the set of all possible states of the environment. Since distinct algorithms may induce identical state-action behavior, we retain a single representative from each equivalence class, which leaves us with a set of unique strategies $\mathcal{M}$, with $|\mathcal{M}| = 67,068$.\footnote{The code for the generation of the strategies and the equilibria computation is provided in the replication package.} We then take a brute-force approach and compute the expected profit $E[\pi_i(\mu_i, \mu_j)]$ numerically for all joint strategy profiles $(\mu_i, \mu_j) \in \mathcal{M}^2$, in order to find all mutual best responses and thereby construct the set of Nash equilibria. The formal methodology for transforming the game into a Markov environment is provided in Appendix~\ref{app:equilibria}. 

The game features several types of pure-strategy Nash equilibria. Continuous play of every price level from $2$ to $10$ can be supported as a symmetric Nash equilibrium. As in a standard repeated Bertrand game, the competitive price $p=2$ can be supported by unconditional strategies, whereas sustaining any supra-competitive price $p>2$ requires history-dependent strategies that deter profitable deviations. The game also features a small set of asymmetric equilibria. A more detailed characterization of the specific equilibrium types and strategies is provided in Appendix~\ref{app:equilibria}. 

For the subsequent analysis, we specify two focal benchmark equilibria. The \emph{competitive benchmark} equilibrium is myopic and does not require strategic conditioning: setting $p = 2$ unconditionally is a mutual best response, since undercutting is impossible and any higher price loses the Buy Box. With ties broken by a fair coin flip, each seller earns expected profits of $\tfrac{1}{2} \times 2 = 1$ per period.  At the other extreme, the \emph{joint} profit-maximizing outcome $p_1 = p_2 = 10$ yielding expected profits of $5$ per period per seller can be sustained as a Nash equilibrium through a win-stay, lose-match conditioning strategy. The ratio of collusive to competitive expected profits is $5{:}1$, providing substantial incentives for cooperation. 

Since demand is perfectly inelastic at one unit per period, any price increase above the competitive level represents a direct transfer from consumer to seller surplus, making the market price a sufficient statistic for consumer welfare in this setting. The \emph{competitive benchmark} maximizes consumer welfare, while the \emph{joint-profit-maximizing} benchmark minimizes consumer welfare.

\subsection{Treatments}
\label{sec:treatments}
Our treatments capture three design features that are widely observed in this market that may increase market prices. Online Appendix~\ref{app:industry_examples} documents parallels between our treatments and features of commercial repricing tools. Providers may have incentives to incorporate such features if they receive a share of downstream-market revenues or if higher seller profits encourage continued use and future adoption.

First, we study the effect of profitability nudges explicitly warning against ``price wars''. This type of messaging is common. Repricing providers frequently warn against price wars and emphasise long-run profitability in blog posts and help pages, and similar guidance may also be communicated directly to sellers.

Second, in a nested design, we study the influence of LLM advice, either in general or when building on the exact profitability nudge. Some providers already integrate LLM-based assistants directly into their pricing platforms. Given the growing popularity of LLM chatbots, such tools are likely to become more widespread and may provide another channel through which algorithm-design providers influence, and potentially coordinate, sellers' downstream behaviour. 

Third, we are considering an alternative design choice, again prevalent in the industry, in which the dashboard provides ready-made algorithms that allow sellers to pursue different goals. Even with a limited set of rules, configuring an algorithm can be complex, which helps explain why providers often offer pre-built or default strategies. Although those simplify sellers' choices, they may steer sellers toward algorithms that sustain higher prices. Moreover, when competing sellers use the same provider, a common menu choice may facilitate coordination on the same strategy. 

Participants are randomly allocated to the six different treatments. All treatments use the same dashboard and do not restrict the action space in any way, such that all algorithms can be selected. Treatments only change the information or functionality found on the dashboard. The six treatments are as follows:

\begin{enumerate}[noitemsep, topsep=0pt]

    \item \textbf{\textsc{Baseline:}} Participants design their pricing algorithm from scratch using the provided dashboard (see Figure~\ref{fig:treatment-baseline}).
    
    \item \textbf{\textsc{LLM Advice Only:}} Identical to \textsc{Baseline}, but participants can consult a Large language model-powered chatbot advisor that provides general advice about the chosen algorithm and answers questions about the game. However, the LLM is neutral in that it does not promote specific algorithms (see Figure~\ref{fig:treatment-llm-only}).
    
    \item \textbf{\textsc{Nudge Only:}} Participants receive a textual nudge placed above the dashboard, encouraging strategies for long-run profitability and warning about price wars (see Figure~\ref{fig:treatment-nudge-only}). The wording we use is a distilled version from different marketing and blog articles present in the repricer space.
    
    \item \textbf{\textsc{Nudge + LLM Advice:}} Participants receive both the general textual nudge (from \textsc{Nudge Only}) and access to the chatbot advisor. The chatbot answers questions about the game and can give general advice, such as in \textsc{LLM Advice Only}, but is instructed to steer participants toward strategies that lead to higher profits (see Figure~\ref{fig:treatment-nudge-llm}).
    
    \item \textbf{\textsc{Nudge + Choice Menu:}} In addition to the textual nudge, participants are presented with a choice of three pre-defined algorithms, a ``cycler,'' a ``tit-for-tat,'' and an ``undercutter'' strategy. Each are provided with a short explanation describing what the algorithm does. Participants still need to set a start price themselves. Participants can also choose to build a fully custom algorithm (see Figure~\ref{fig:treatment-choice-menu}).

    \item \textbf{\textsc{Nudge + Coordination:}} Participants receive a textual nudge explaining that if both sellers adopt a particular coordinating strategy (``tit-for-tat'' with a starting price of 10), they can achieve higher mutual profits. Similar to \textsc{Nudge + Choice Menu}, the interface provides a single button that auto-fills the parameters for this specific coordinating algorithm. Participants can also choose to build a fully custom algorithm (see Figure~\ref{fig:treatment-coordination}).
\end{enumerate}

\subsection{Hypotheses and Outcome Variables}
\label{sec:hypo_measures}

Treatments do not restrict participants' choice sets: every feasible algorithm remains available in every condition, including those with pre-built options. Instead, the treatments change the information and decision support available during the algorithm design process. The nudges warn against price wars and encourage strategies that can sustain high profits. The menu and coordination treatments make specific strategies easier to implement, i.e. with a single click. The LLM advisors provide context-specific support that may improve participants' understanding of the game. In \textsc{Nudge + LLM Advice}, the advisor also reinforces the profitability nudge.

These features should make less competitive or better-coordinated strategies more salient and easier to adopt. Although \textsc{LLM Advice Only} does not promote a particular strategy, it may still help participants recognise that sustaining prices can be more profitable than aggressive pricing in a repeated interaction. We therefore expect every intervention to produce higher market prices than \textsc{Baseline}. Hypothesis~\ref{hyp:interventions} summarises this common directional prediction.\footnote{All hypotheses were pre-registered. The pre-registration is available at \url{https://osf.io/74psg/}.}

\begin{hypothesis}\label{hyp:interventions}
All intervention treatments (\textsc{Nudge + Choice Menu}, \textsc{Nudge + Coordination}, \textsc{Nudge Only}, \textsc{LLM Advice Only}, and \textsc{Nudge + LLM Advice}) will lead to higher market prices compared to the \textsc{Baseline} condition.
\end{hypothesis}

The \textsc{Nudge + Coordination} treatment includes the same warning against price wars and the same guidance toward profitable algorithms as \textsc{Nudge Only}. It also provides a one-click option to adopt a specific strategy that coordinates sellers on high prices when both choose it. The intervention, therefore, resembles a setting in which a common algorithm provider actively steers competing sellers toward the same downstream-market strategy. By making that strategy both focal and easy to implement, the coordination option should sustain higher prices more effectively than the textual nudge alone. Hypothesis~\ref{hyp:coordinate} summarises this rationale.

\begin{hypothesis}\label{hyp:coordinate}
 The \textsc{Nudge + Coordination} treatment will lead to higher market prices than the \textsc{Nudge Only} treatment. %
\end{hypothesis}

In \textsc{LLM Advice Only}, the chatbot provides general guidance but does not promote particular types of strategies. In \textsc{Nudge + LLM Advice}, by contrast, it actively encourages participants to avoid a race to the bottom and to choose strategies that support long-run profitability. The interactive advisor can also tailor this guidance to a participant's chosen algorithm, answer follow-up questions, and respond to outcomes from earlier supergames. It may therefore translate the general message of the static nudge into more specific algorithm choices. Chatbot usage is voluntary in both treatments, as it typically would be outside the experiment. As a result, treatment effects might be smaller than in other studies that focus on the effects of LLMs on seller behaviour in markets \citep[see, for example,][]{werner2024experimental}. We summarise this prediction in Hypothesis~\ref{hyp:llm}.

\begin{hypothesis}\label{hyp:llm}
The \textsc{Nudge + LLM Advice} treatment will lead to higher market prices than both the \textsc{LLM Advice Only} treatment and the \textsc{Nudge Only} treatment. %
\end{hypothesis}

Like \textsc{Nudge Only}, the \textsc{Nudge + Choice Menu} treatment warns against price wars and emphasises long-run profitability. The menu does not coordinate both sellers on one specific strategy, but it offers predefined options that include less competitive algorithms and makes them easy to implement. Combined with the warning against price wars, this choice architecture should make participants more likely to adopt algorithms that support higher prices. This argument leads to Hypothesis~\ref{hyp:menu}.

\begin{hypothesis}\label{hyp:menu}
The \textsc{Nudge + Choice Menu} treatment will lead to higher market prices than the \textsc{Nudge Only} treatment. %
\end{hypothesis}

\paragraph{Outcome variables}

Our primary outcome is the \textit{average market price} at the market (pair) level. We define the market price in a period as the lower of the two prices posted by the competing sellers, which is the price at which the Buy Box is won. For our main treatment comparisons, we average of this price over all 50 periods and all five supergames for each pair. To study market dynamics, we also examine prices separately by period and supergame. Because exactly one unit is sold in each period, the market price equals joint seller profit and is a sufficient statistic for consumer welfare in our setting.%

Beyond realized market outcomes, we analyze the algorithms chosen by participants. We examine their start prices and pricing rules and classify the resulting algorithms into strategy types. The Markov framework also allows us to construct two continuous measures of algorithm behavior. Optimality measures how closely an algorithm approximates a best response to the opponent's algorithm, while harshness captures its propensity to set low prices across reachable market situations. At the pair level, mutual optimality measures proximity to Nash equilibrium play in terms of payoffs, and pair leniency captures whether both algorithms avoid aggressive pricing.

\textbf{Optimality} measures the ratio of a player's actual expected payoff to the
best-response payoff achievable against the opponent's algorithm:
\begin{equation}
    \Omega_i(\mu_i, \mu_j) = \frac{\mathbb{E}\left[\pi_i(\mu_i, \mu_j)\right]}
    {\max_{\mu_i^* \in \mathcal{M}} \mathbb{E}\left[\pi_i(\mu_i^*, \mu_j)\right]}
    \label{eq:optimality}
\end{equation}
Due to the complexity of the strategy space, it is unlikely that participants will achieve a perfect Nash equilibrium. Instead, we analyze the mutual optimality measure $\Omega_i(\mu_i, \mu_j)\Omega_j(\mu_j, \mu_i)$ which provides a continuous measure 
of how closely a strategy pair approximates a best-response equilibrium in terms of payoffs with mutual optimality of 1 being equivalent to a Nash equilibrium.

\textbf{Harshness} measures the average pricing behavior by an algorithm across all (long-term) reachable market situations, normalized to $[0,1]$, where $H_i = 1$ indicates maximum
aggressiveness (average price $p = p_{\min}$) and $H_i = 0$ indicates maximum leniency
(average price $p = p_{\max}$):
\begin{equation}
    H_i(\mu_i) = \frac{p_{\max} - \dfrac{1}{|\mathcal{S}_i|} \cdot
    \displaystyle\sum_{s \in \mathcal{S}_i} \mu_i(s)}{p_{\max} - p_{\min}}
    \label{eq:harshness}
\end{equation}
$\mathcal{S}_i$ is the set of states that player $i$'s strategy $\mu_i$ does not unilaterally preclude from, independent of the other player's strategy. A formal derivation of $\mathcal{S}_i$ along with two illustrative examples is provided in Appendix~\ref{app:harshness}. An intuitive way to think about this measure is to see it as the expected response against an opponent that prices randomly, over prolonged periods of play. Therefore, unlike the previous measures, harshness is unconditional on the opponent's strategy and is in that sense an intrinsic algorithmic property. Like the start price,
it can be interpreted as a proxy for willingness to cooperate. However, it captures
long-run pricing behavior rather than initial-period choices. For a pair-level, joint measure of harshness, we define ``Leniency'' as a complementary measure ($L_i=1-H_i$), and pair leniency as $L=L_i\cdot L_j$.

\subsection{Procedures}
\label{sec:procedures}

To resemble the setting of sellers repeatedly competing against the same seller in the wild, but with the option to redesign the algorithms, participants are matched into fixed pairs.\footnote{Participants are matched into fixed pairs using an arrival-time protocol that pairs participants who submit the comprehension questions at approximately the same time. Treatment assignment occurs at the time of matching, with treatments cycling on the pair-level within each session, ensuring a uniform distribution of treatment proportions across all 6 sessions.} Each pair plays 5 supergames of the market game without rematching. After each of the 5 supergames, participants observe the realised 50-period price time series and both sellers' average profits, and may revise their algorithm for the next supergame. In supergame~1, participants have 600 seconds to design their algorithm. In supergames 2-5, the time limit is 300 seconds. Before submitting an algorithm, participants can test it in a sandbox that previews the resulting market outcome at $t$, given a freely configurable hypothetical market situation at $t-1$.\footnote{Screenshots of the different experiment pages are provided in Online Appendix~\ref{app:screenshots}.}

The LLM advisors are based on Llama-3.3-70B-Instruct-Turbo and integrated into the oTree experiment via the open-source \emph{Simple Chat} backend \citep{bermudez_simple_2025}. Interaction with the LLM is neither required nor enforced. We steer the advisor's behaviour with a \emph{system prompt}, a block of instructions given to the model that sets its role and rules before any user interaction.\footnote{System prompts (also called system messages) are a common way to configure LLM behavior and are supported by all major LLM providers. For an introduction see, for instance, here \url{https://microsoft.github.io/Workshop-Interact-with-OpenAI-models/Part-2-labs/System-Message/} (last accessed August~21, 2026).} Our system prompt describes the advisor's role, restates the market rules, and provides behavioral guidelines for helping participants design their algorithm. In \textsc{Nudge + LLM Advice}, an additional fragment instructs the model to promote cooperative, stabilization-oriented strategies. Each participant's chat history persists across all five supergames, and from supergame~2 onward, the model additionally receives the previous supergame's market outcome data, so that advice can refer back to prior play. The full system prompts for both treatments are reproduced verbatim in Online Appendix~\ref{app:llm_chat_prompts}.

Participants were recruited in 6 experimental sessions (15-21 December 2025) via Prolific and completed the experiment online using oTree \citep{chen2016otree}. Prior to the introduction to the game, participants were asked to provide informed consent and to complete a reCAPTCHA check and two attention checks, which were also designed to filter out bots \citep{rilla2026recognising}. Then, after two pages of instructions, they answered five comprehension questions about the market game and the rules of the algorithm design. Incorrect answers triggered a second attempt with the incorrect answers highlighted. Participants who failed both attempts on any question were excluded from the study.

Payment consisted of a \pounds4.00 base fee plus a performance bonus: One of the five supergames was randomly selected, and the participant's average per-period profit in that supergame was multiplied by \pounds1.50. Participants who failed the comprehension questions or timed out during the main part of the experiment received a consolation payment of \pounds1.00. If a participant timed out, a bot would resume the experiment in their place so their partner could remain eligible for a bonus payment; the data from the whole pair was discarded if a timeout occurred. A total of 1,083 participants successfully completed the experiment. Of those, 966 remain with intact pairs and are therefore eligible for data points. The average total payment for successful submissions was \pounds7.08 ($\text{SD} = \text{\pounds}2.36$), with the performance bonus comprising 43.5\% of total pay on average (range: \pounds0.00-\pounds13.38). Our participants are not professional marketplace sellers, but the central task of selecting conditional pricing rules through a dashboard closely mirrors the task faced by small online sellers using rule-based repricers.

\section{Results}
\label{sec:results}

We first present the treatment effects on market prices and their dynamics over time (Section~\ref{sec:main_results}--\ref{sec:time_dynamics}). We then turn to the mechanisms behind these effects: the role of start prices and algorithm leniency, the outcome stability, the types of algorithms participants build, and whether treatment effects survive cross-treatment rematching (Section~\ref{sec:startprice_leniency}--\ref{sec:cross_treatment}).

\subsection{Treatment effects on market prices}
\label{sec:main_results}

\begin{figure}[!ht]
    \centering
    \includegraphics[width=0.9\textwidth,keepaspectratio]{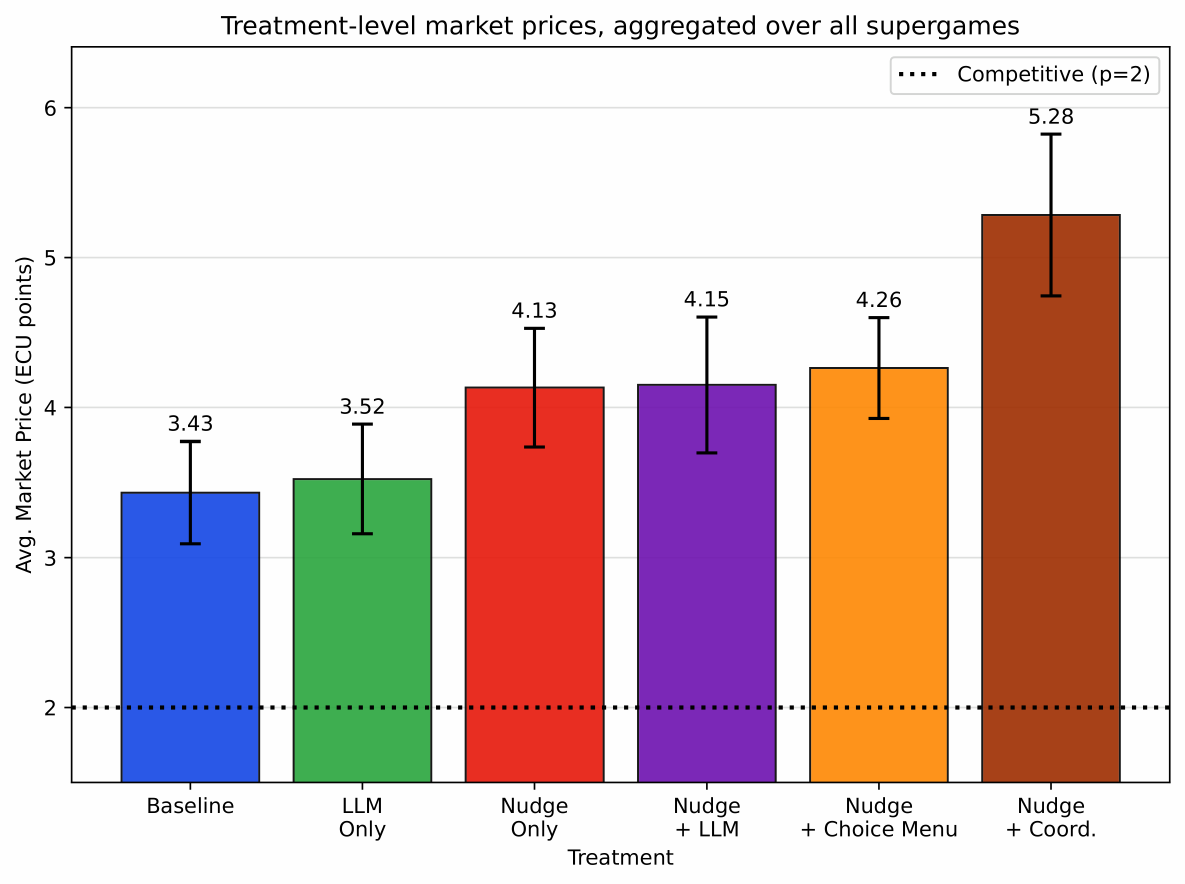}
    \caption{Avg. Market Price by Treatment. Error bars represent 95\% confidence intervals.}
\label{fig:avgprice}
\end{figure}

Figure~\ref{fig:avgprice} shows average market prices by treatment. The \textsc{Baseline} already shows prices above the one-shot Nash equilibrium level of 2. That is, participants choose prices higher than those predicted by theory for a perfectly competitive market even in the \textsc{Baseline} treatment. However, our main focus is on treatment comparisons. All pairwise tests for each hypothesis can be found in Table~\ref{tab:mwu_main_rq} in the appendix.\footnote{We report $p$-values from one-sided tests when we test any of our pre-registered directional hypotheses. All tests remain statistically significant at conventional levels when we use two-sided tests. In all other cases, we report two-sided $p$-values and indicate this accordingly. }

All treatments that involve some form of nudge or pre-configured algorithm lead to significantly higher average market prices compared to the \textsc{Baseline} (one-sided Mann-Whitney U tests, $p<0.01$ for \textsc{Nudge Only}, \textsc{Nudge + LLM Advice}, \textsc{Nudge + Choice Menu}, and \textsc{Nudge + Coordination} vs \textsc{Baseline}). \textsc{LLM Advice Only} does not differ significantly from the \textsc{Baseline} ($p=0.327$). These results support four of the five treatment contrasts predicted by Hypothesis~\ref{hyp:interventions}. Even simple design interventions are sufficient to push market prices above competitive levels. The highest average price is observed in the \textsc{Nudge + Coordination} treatment, which is significantly higher than in all other treatments, including \textsc{Nudge Only} ($p=0.001$). This is in support of  Hypothesis~\ref{hyp:coordinate}. Providing a low-friction one-click option to adopt a coordinating strategy is substantially more effective at raising prices than an informational nudge alone.

Notably, merely providing access to an LLM advisor in \textsc{LLM Advice only} does not increase prices significantly. The advisor can answer questions about the game and offer general strategic advice, but does not actively steer participants toward specific algorithms. This suggests that general information alone is not sufficient and a driving force behind price increases. Hypothesis~\ref{hyp:llm} predicts that the \textsc{Nudge + LLM Advice} treatment leads to higher prices than both the \textsc{LLM Advice Only} and the \textsc{Nudge Only} treatment. While the \textsc{Nudge + LLM Advice} treatment does produce significantly higher prices than \textsc{LLM Advice Only} ($p=0.002$), it does not outperform \textsc{Nudge Only} ($p=0.537$). Since the price increase over \textsc{LLM Advice Only} is driven by the nudge component rather than the LLM, we find no evidence that LLM advice adds to the effect of the nudge.\footnote{$74\%$ of subjects in \textsc{LLM Advice Only} and $65\%$ in \textsc{Nudge + LLM} exchanged at least one message with the chatbot. Conditional on advisor usage, pair-level prices move in opposite directions across the two treatments, consistent with the differing chatbot instructions, but neither effect reaches significance; see Online Appendix~\ref{app:llm_chatbot} for full descriptives and regressions on the interaction with the chatbot and the effects on prices.} We therefore do not find full support for Hypothesis~\ref{hyp:llm}.

The \textsc{Nudge + Choice Menu} treatment does not lead to significantly higher prices than the \textsc{Nudge Only} treatment ($p=0.164$). Hence, we do not find evidence in support for Hypothesis~\ref{hyp:menu}. Offering a menu of pre-configured algorithms raises prices compared to the \textsc{Baseline}. However, it is not more effective at it than a simple informational nudge.

These treatment comparisons are summarized in Result~\ref{res:main_treatment_effects}.

\begin{result}\label{res:main_treatment_effects}
    All treatments that involve a nudge or pre-configured algorithms significantly increase average market prices relative to the \textsc{Baseline}. \textsc{LLM Advice Only} does not increase market prices. \textsc{Nudge + Coordination} produces significantly higher average market prices than all other treatments.
\end{result}

Taken together, the treatment that is most successful at increasing market prices is \textsc{Nudge + Coordination}, which combines an explicit recommendation for a specific cooperative strategy with a one-click implementation option. However, all treatments that provide nudges to avoid price wars or default algorithms increase market prices, suggesting possible anti-competitive effects of common design features of algorithmic design platforms.

\subsection{Pricing dynamics within supergames}
\label{sec:time_dynamics}

\begin{figure}[!ht]
    \centering
    \begin{subfigure}{\textwidth}
        \centering
        \includegraphics[width=0.8\linewidth]{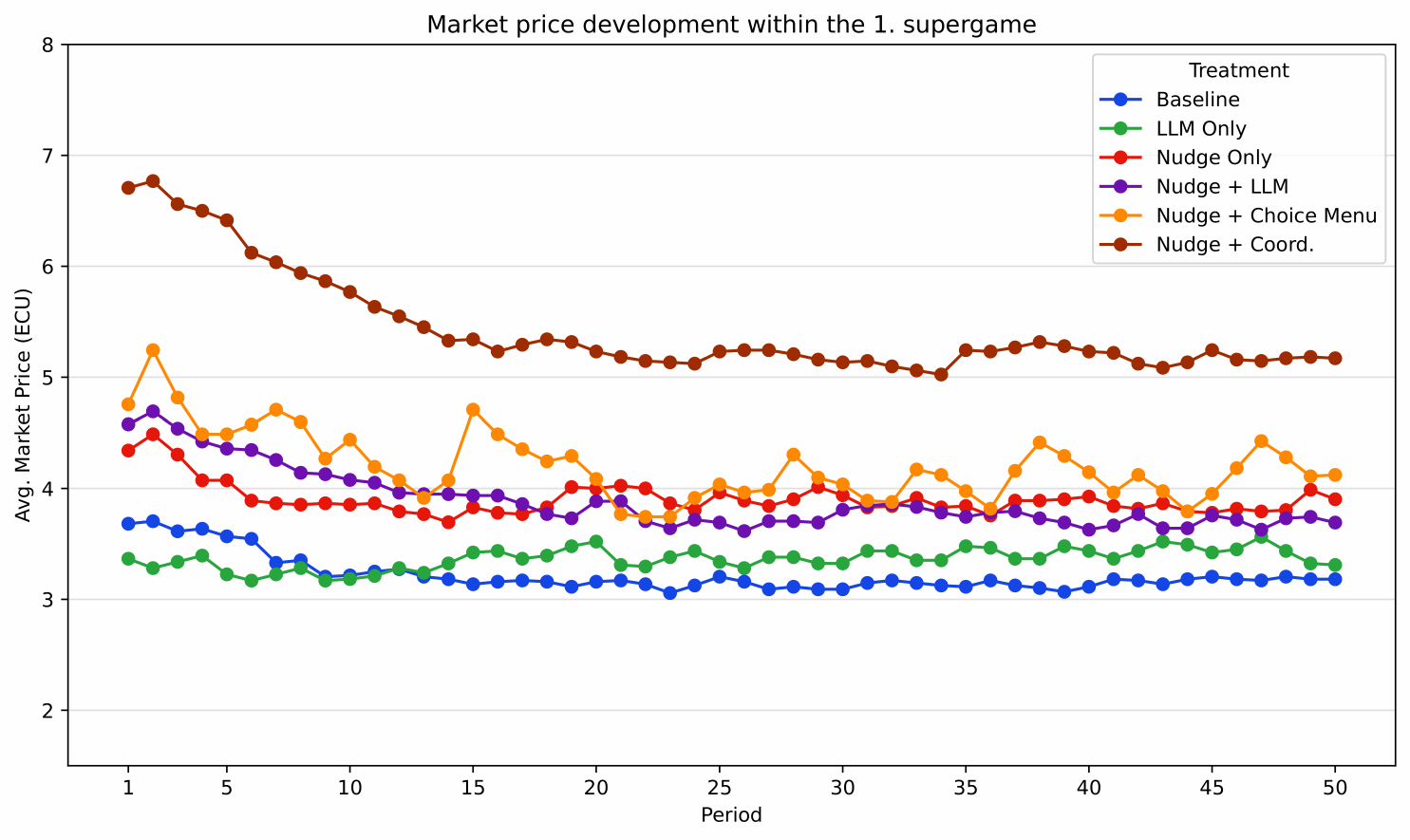}
        \caption{Supergame 1}
        \label{fig:price_round_1}
    \end{subfigure}
    
    \vspace{1.5em} %
    
    \begin{subfigure}{\textwidth}
        \centering
        \includegraphics[width=0.8\linewidth]{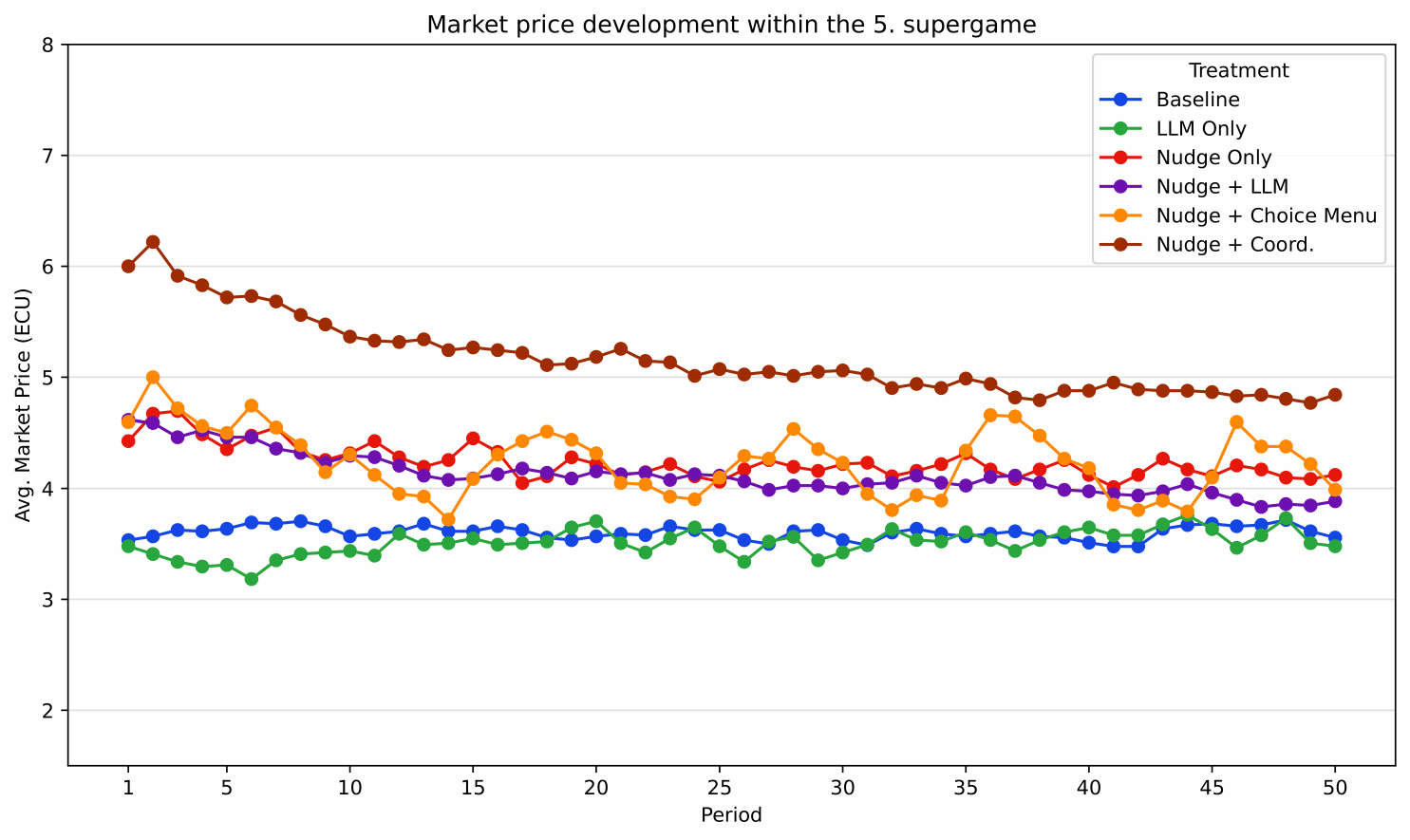}
        \caption{Supergame 5}
        \label{fig:price_round_5}
    \end{subfigure}
    
    \caption{Average Market Price by Treatment (Supergames 1 and 5)}
    \label{fig:price_development_rounds}
\end{figure}

We now examine average price dynamics within and across supergames. Figure~\ref{fig:price_development_rounds} shows the average market price by treatment over the 50 periods for supergames one and five. Within each supergame, prices are quite stable after an initial adjustment phase of roughly ten periods. An interesting exception is the \textsc{Nudge + Choice Menu} treatment, which exhibits more volatile price patterns driven by the default option to adopt a cycling strategy. The absence of strong upward or downward trends within a supergame suggests that the start price plays a dominant role in determining the price level, which we analyze further in the subsequent section. 

Table~\ref{tab:price_dynamics} in the appendix formally tests for time trends.\footnote{Figures~\ref{fig:avg_price_rounds}, \ref{fig:avg_optimality_rounds}, and \ref{fig:avg_harshness_rounds} in the appendix visualize the across-supergame trends in average price, optimality, and harshness, respectively.} For that, we regress the market price on the period and supergame variables, as well as their interaction. We run those linear regressions separately for each treatment. The time trends are modest. In none of the treatments is there a trend across supergames or an interaction effect between the supergame and the period. However, in \textsc{Baseline}, \textsc{Nudge+LLM}, \textsc{Nudge + Choice Menu}, and \textsc{Nudge + Coordination}, there is a small negative time trend across periods, as suggested by the visual analysis of Figure~\ref{fig:price_development_rounds}. While market prices tend to decrease slightly across periods, the overall ordering of treatments remains stable even at the end of the supergames. When we restrict the analysis to the last supergame, treatment differences persist: \textsc{Nudge + Choice Menu} ($p=0.001$) and \textsc{Nudge + Coordination} ($p<0.001$) remain significantly higher than \textsc{Baseline}, while \textsc{LLM Advice Only} remains insignificant ($p=0.854$).\footnote{One-sided MWU tests on pair-level market prices averaged over all periods within supergame~5.} Result~\ref{res:constant_prices} summarizes these findings.

\begin{result}\label{res:constant_prices}
 Across periods and supergames, treatment differences persist as participants gain experience. There are no time trends across supergames, and while some treatments exhibit a small negative time trend across periods, the overall ordering of treatments according to price remains stable.
\end{result}

\section{Mechanisms}
\label{sec:mechanisms}
The previous section established that most treatments raise market prices relative to the \textsc{Baseline}. We now ask what drives these effects. We decompose treatment differences into two pair-level channels --- start prices and algorithm leniency --- then examine the stability of outcomes and the algorithm types participants choose, and finally test whether the effects persist when algorithms are re-matched across treatments.

\subsection{Start prices and algorithm leniency}
\label{sec:startprice_leniency}

The previous subsection showed that market prices are decreasing within a supergame, but that this trend, on average, tends to flatten after the initial periods. This raises the question of what drives treatment differences: is it primarily the start price, the pricing rules encoded in the algorithm, or both?  While the strategy space is large, we can characterize algorithms by two dimensions, start price and leniency with more fine-grained algorithm types examined further below. 

Table~\ref{tab:horse_race} presents a series of OLS regressions that progressively check whether start price and leniency of a pricing rule account for the treatment effects on average market prices. Column~(1) reports the treatment effects without any controls. All treatments except \textsc{LLM Advice Only} show significantly higher prices than the \textsc{Baseline}, consistent with our non-parametric tests. Adding the realized start price --- so the minimum of both algorithms --- as a control in column~(2) renders all treatment effects insignificant, while start price itself is highly significant ($\beta = 0.571$, $p < 0.01$). The adjusted $R^2$ jumps from $0.062$ to $0.297$, indicating that start prices alone explain a large share of the variation in market prices. Figure~\ref{fig:startprice} in the appendix shows average start prices by treatment, which highlights that treatments with higher market prices also exhibit higher start prices.

Beyond the start price, the pricing rules themselves may matter. To capture the long-run pricing behavior of an algorithm, we use a leniency measure. Leniency is a pair-level variable defined as $L = (1 - H_1) \times (1 - H_2)$, where $H_i$ is participant $i$'s harshness score (see Section~\ref{sec:hypo_measures}). Harshness measures the average aggressiveness of an algorithm across all reachable market situations, normalized to $[0,1]$. Higher leniency thus indicates that both algorithms in a pair tend to set higher prices across the states they can reach. Intuitively, a pair with high leniency consists of two algorithms that are both ``soft'' and do not aggressively undercut. Column~(3) shows that leniency alone is a strong predictor of market prices ($\beta = 12.303$, $p < 0.01$), explaining a similar share of variation as start prices ($R^2_{\text{adj}} = 0.303$), but not absorbing treatment effects in the same vein.

Column (4) includes both start price and leniency simultaneously. Both remain highly significant, suggesting that start prices and algorithm leniency capture complementary dimensions of pricing behavior. Most treatment dummies become insignificant once both mechanisms are controlled for.

Figure~\ref{fig:harshness} in Appendix~\ref{app:mechanisms} show treatment-level distributions of harshness. Harshness is lower (i.e., algorithms are more lenient) in all nudge and default-algorithm treatments relative to the \textsc{Baseline}, in line with the regression results above.

Column~(5) adds the interaction between start price and leniency. The interaction term is highly significant ($\beta = 1.477$, $p < 0.01$). This suggests that the two mechanisms are complementary at the pair level: high pair leniency sustains high prices  when both participants also start with high prices, and high start prices translate into high average prices when neither algorithm aggressively undercuts in subsequent periods. Column~(6) confirms that these patterns are robust to the inclusion of supergame fixed effects. Result~\ref{res:start_price} summarizes this finding.

\begin{result}\label{res:start_price}
    Treatment effects on market prices are driven by two complementary pair-level mechanisms: higher start prices and mutually lenient algorithm pairs. Neither channel alone is sufficient to rationalize the data, their interaction effect is large and significant.\footnote{Table~\ref{tab:horse_race_payoff} replicates this analysis for individual-level payoffs. The results qualitatively align, which is intuitive, as the game design implies that the average market price is by definition proportional to the average individual payoff.}
\end{result}

\begin{table}[!ht]
    \centering
    \caption{OLS Regressions: Treatment Effects on Average Market Price}
    \label{tab:horse_race}
    \small
    \setlength{\tabcolsep}{6pt}
    \begin{tabular}{lcccccc}
        \toprule
        \textbf{Dep. Var.: Market Price} & (1) & (2) & (3) & (4) & (5) & (6) \\
        \midrule
        LLM Only
            & $0.091$ & $0.135$ & $0.376^{**}$ & $0.382^{**}$ & $0.298^{*}$ & $0.298^{*}$ \\
            & $(0.250)$ & $(0.225)$ & $(0.192)$ & $(0.177)$ & $(0.176)$ & $(0.176)$ \\[0.3em]
        Nudge Only
            & $0.700^{***}$ & $0.189$ & $0.660^{***}$ & $0.215$ & $0.211$ & $0.210$ \\
            & $(0.261)$ & $(0.229)$ & $(0.229)$ & $(0.193)$ & $(0.189)$ & $(0.189)$ \\[0.3em]
        Nudge $+$ LLM
            & $0.718^{**}$ & $0.113$ & $0.761^{***}$ & $0.223$ & $0.204$ & $0.203$ \\
            & $(0.283)$ & $(0.256)$ & $(0.236)$ & $(0.210)$ & $(0.203)$ & $(0.204)$ \\[0.3em]
        Nudge $+$ Choice Menu
            & $0.830^{***}$ & $0.158$ & $0.507^{**}$ & $-0.047$ & $-0.051$ & $-0.053$ \\
            & $(0.240)$ & $(0.224)$ & $(0.213)$ & $(0.192)$ & $(0.190)$ & $(0.190)$ \\[0.3em]
        Nudge $+$ Coordination
            & $1.851^{***}$ & $0.231$ & $1.381^{***}$ & $0.009$ & $-0.064$ & $-0.066$ \\
            & $(0.320)$ & $(0.251)$ & $(0.278)$ & $(0.203)$ & $(0.192)$ & $(0.192)$ \\[0.5em]
        Start Price$^{\dagger}$
            &             & $0.571^{***}$ &             & $0.503^{***}$ & $0.223^{***}$ & $0.226^{***}$ \\
            &             & $(0.034)$     &             & $(0.031)$     & $(0.054)$     & $(0.054)$     \\[0.3em]
        Leniency$^{\ddagger}$
            &             &               & $12.303^{***}$ & $10.874^{***}$ & $4.680^{***}$ & $4.727^{***}$ \\
            &             &               & $(0.624)$      & $(0.618)$      & $(1.265)$ & $(1.266)$ \\[0.3em]
        Start Price $\times$ Leniency
            &             &               &                &                & $1.490^{***}$ & $1.477^{***}$ \\
            &             &               &                &                & $(0.267)$     & $(0.267)$     \\[0.5em]
        Constant
            & $3.433^{***}$ & $1.454^{***}$ & $1.298^{***}$ & $-0.197$ & $0.937^{***}$ & $0.756^{***}$ \\
            & $(0.171)$     & $(0.184)$     & $(0.145)$     & $(0.155)$      & $(0.216)$     & $(0.231)$     \\
        \midrule
        Start Price          & No  & Yes & No  & Yes & Yes & Yes \\
        Leniency            & No  & No  & Yes & Yes & Yes & Yes \\
        Interaction Effect   & No  & No  & No  & No  & Yes & Yes \\
        Supergame FE             & No  & No  & No  & No  & No  & Yes \\
        Clustered SE (Pair)  & Yes & Yes & Yes & Yes & Yes & Yes \\
        $N$                  & 2415 & 2415 & 2415 & 2415 & 2415 & 2415 \\
        $R^2_{\text{adj}}$   & 0.062 & 0.297 & 0.303 & 0.481 & 0.498 & 0.498 \\
        \bottomrule
        \multicolumn{7}{p{0.96\linewidth}}{\footnotesize
            \textit{Notes:} OLS estimates with standard errors (in parentheses) clustered at
            the pair level (483 pairs). Omitted treatment category: \textit{Baseline}.
            $^{\dagger}$Realized start price (minimum of the algorithm pair).
            $^{\ddagger}$Leniency is measured as
            $(1 - H_1) \times (1 - H_2)$, i.e.\
            the product of the complement of both players' harshness scores; higher values indicate
            softer bilateral price pressure.
            $^{*}p < 0.10$,\ $^{**}p < 0.05$,\ $^{***}p < 0.01$.}
    \end{tabular}
\end{table}

\subsection{Optimality and payoff inequality}
\label{sec:stability}
We are interested in understanding to what extent the treatments lead to mutually optimal play and more equal payoffs in the market. Both optimality and payoff inequality may be indicative of stable coordination. 

We can measure how close the realized payoffs are to the maximum payoffs attainable under optimal play because we directly observe the seller's strategy choice. This can be compared to the counterfactually optimal strategy from the strategy space, see (\ref{eq:optimality}) in Section~\ref{sec:hypo_measures}. This measure thus captures the extent to which the potential gains from optimal play are realized. A higher mutual optimality, the product of both sellers' optimality, then suggests that deviations from the current algorithms are less profitable. 

Mutual optimality is significantly higher for \textsc{Nudge + Choice menu} and \textsc{Nudge + Coordinate}  compared to all other treatments ($p<0.05$) except the \textsc{LLM only} treatment ($p<0.1$).\footnote{This analysis is exploratory and based on two-sided MWU tests.} Mutual optimality is similar in all other treatments (see Figure~\ref{fig:optimality}). It is increasing over the supergames for all treatments apart from \textsc{Nudge + Coordinate}, that is, there is an increase in coordination (see Figure~\ref{fig:avg_optimality_rounds}).

We can also assess whether treatments benefit both competitors equally or whether sellers respond heterogeneously to the choice architecture. To this end, we measure within-market differences in payoffs, see~\ref{fig:payoff_inequality}. \textsc{Nudge + LLM} leads to the highest payoff inequality, potentially driven by asymmetric usage of LLM advice (see \ref{app:llm_chatbot}). Treatments with pre-built algorithms lead to the lowest payoff differences. 

These findings are summarized in Result~\ref{res:stability}.

\begin{result}\label{res:stability}
    Treatments that provide pre-built algorithms %
    exhibit more equal payoffs and a higher optimality measure.
\end{result}

\subsection{Algorithm selection}
\label{sec:algoselection}

Because participants submit complete algorithms, we can observe their strategic choices directly. We classify all algorithms into high-level types based on the rules they encode for the Win-Buy-Box and Lose-Buy-Box conditions. This gives insights into \textit{how} nudges can influence choice in these types of markets and allows us to detect differences between treatments even when market prices are similar. 

We first identify nine low-level rules that each condition branch can take:
\begin{itemize}[itemsep=0pt, topsep=2pt, parsep=0pt]
    \item \textbf{Crash:} Set the minimum price ($p_{\min} = 2$).
    \item \textbf{Lower:} Condition on own last price and lower it.
    \item \textbf{Undercut:} Condition on the opponent's last price and undercut it.
    \item \textbf{Cycle:} Condition on the opponent's last price and undercut it. If undercutting is not possible, reset price to $p_{\max} = 10$.
    \item \textbf{Match:} Match the opponent's last price.
    \item \textbf{Fix:} Set a fixed price $p > 2$.
    \item \textbf{Stay:} Remain at the current price.
    \item \textbf{Raise:} Condition on own last price and raise it.
    \item \textbf{Uppercut:} Condition on the opponent's last price and set a higher price.
\end{itemize}

\begin{figure}[htbp]
    \centering
    \includegraphics[width=\textwidth]{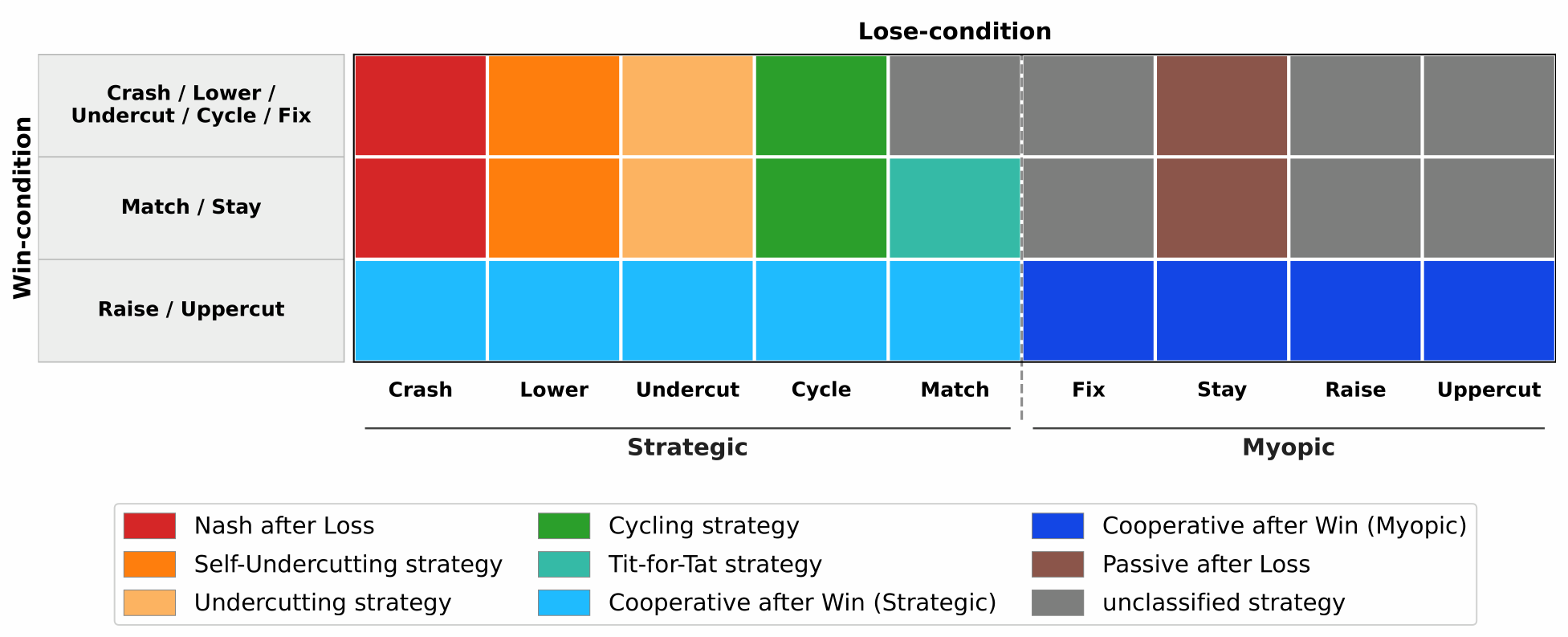}
    \caption{Algorithm classification scheme. Rows denote the Win-condition strategy group; columns denote the Lose-condition strategy. Cell colours indicate the resulting high-level strategy as shown in the legend.}
    \label{fig:algo_classification}
\end{figure}

Both the Win-Buy-Box and Lose-Buy-Box conditions can be described by one of nine low-level rules, and a strategy is defined by its reaction in these two conditions. However, since there are many possible win/lose combinations, we group them into high-level strategy types based on two principles (Figure~\ref{fig:algo_classification}). First, we classify primarily by the Lose-condition rule, because the response to losing the Buy Box is the most informative about strategic intent: it reveals whether a participant tries to compete for the Buy Box or not. Second, within the Lose-condition, we distinguish between \textit{strategic} rules that attempt to regain the Buy Box (Crash, Lower, Undercut, Cycle, Match) and \textit{myopic} rules that do not (Fix, Stay, Raise, Uppercut). The Win-condition is grouped into three behavioral classes: aggressive (Crash/Lower/Undercut/Cycle/Fix), neutral (Match/Stay), and accommodating (Raise/Uppercut). Figure~\ref{fig:algo_classification} shows the resulting classification into eight high-level strategy types, plus a residual category for unclassified rare combinations.\footnote{The tit-for-tat category thus includes the equilibrium strategy of win-stay, lose-match as well as the tit-for-tat strategy offered in \textsc{Nudge Coordination}.} 

\begin{figure}[hptb]
    \centering
    \includegraphics[width=0.9\textwidth]{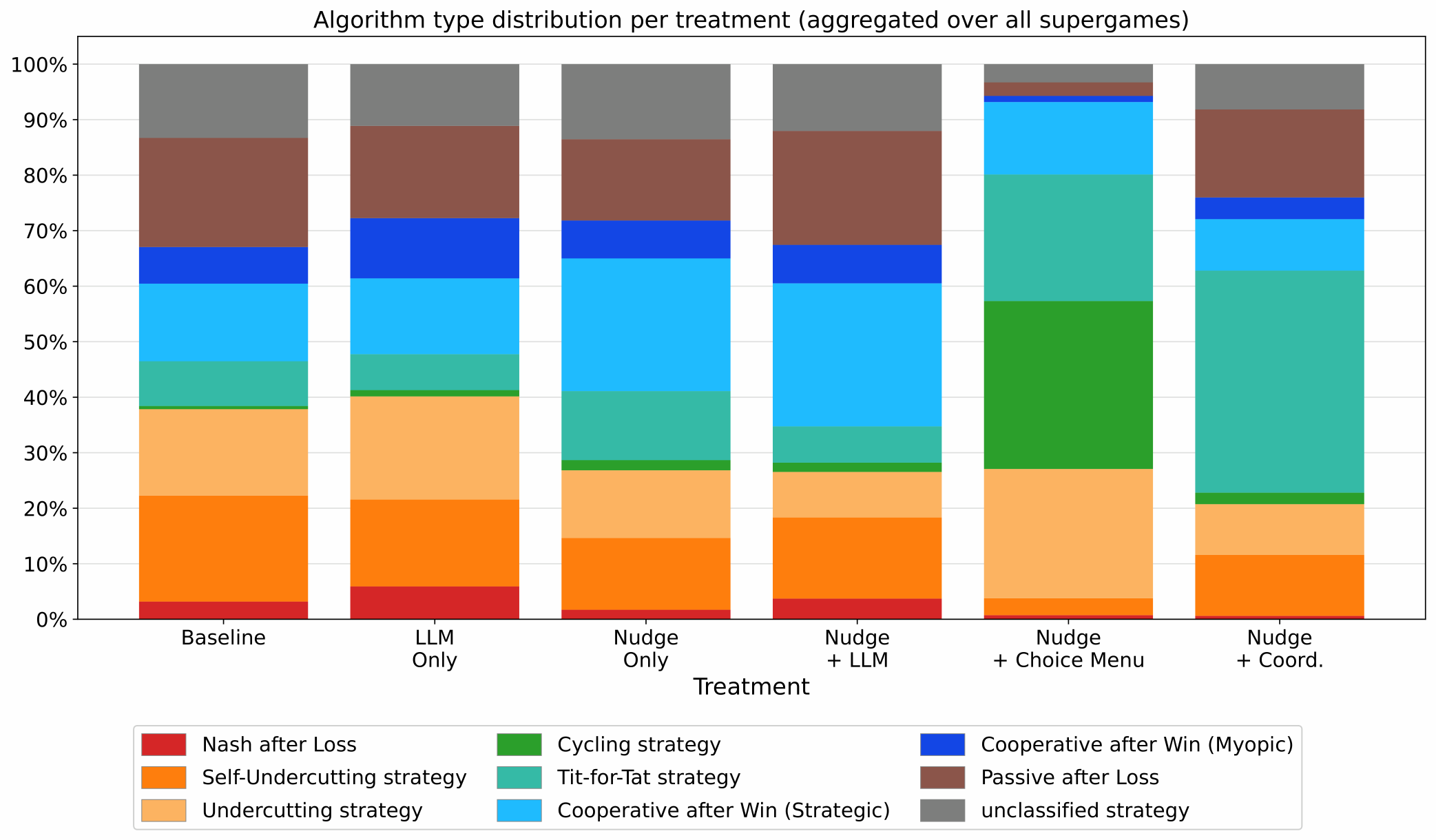}
    \caption{Distribution of algorithm types by treatment, aggregated over all supergames.}
\label{fig:algotypes}
\end{figure}

We code all strategies according to their behaviour in the Win-Buy-Box and Lose-Buy-Box conditions, and then match them to the entries in Figure~\ref{fig:algo_classification}. This yields a unique classification for each algorithm.  Figure~\ref{fig:algotypes} shows the distribution of algorithm types by treatment.\footnote{Figure~\ref{fig:algo_types_treatments_rounds} in the appendix shows the same distribution disaggregated by supergame; the treatment ordering is broadly stable across supergames.} In the \textsc{Baseline} and \textsc{LLM Advice Only} treatments, the distributions are similar: roughly a third ($34.7 \%$/$34.2 \%$) of algorithms are undercutting or self-undercutting strategies, another third ($33.0 \%$/$27.7 \%$) are split between passive-after-loss and unclassified strategies. Cooperative-after-win types account for a further $20.6 \%$/$24.5 \%$. The remaining strategies collectively account for less than $20\%$, with Tit-for-Tat being the largest of these categories at $8.1 \%$/$6.5\%$ of the observed algorithms.

The nudge treatments shift this distribution. In \textsc{Nudge Only} and \textsc{Nudge + LLM}, undercutting drops ($25.1 \%$/$22.8 \%$) and cooperative-after-win strategies take its place ($30.7 \%$/$32.7 \%$), which helps explain the higher prices in these treatments. Some tit-for-tat strategies appear as well ($12.4 \%$/$6.5 \%$), though they remain a minority.

The distribution of strategies in \textsc{Nudge + Choice Menu} is different from all other treatments. Because participants can select from three pre-configured algorithms (tit-for-tat, cycling, undercutting), the vast majority of choices concentrate on these three types ($23.3 \%$/$30.2 \%$/$22.8\%$). This is the only treatment with a substantial share of cycling strategies, suggesting that participants rarely construct cycling algorithms on their own but will adopt them when offered as a default, as is common among commercial repricing providers (see Appendix~\ref{app:industry_examples}, particularly Figure~\ref{fig:screenshot_menu_oscillation}).

In \textsc{Nudge + Coordination}, tit-for-tat strategies appear at a higher rate than in any other treatment ($40.0 \%$), though the share is lower than the one-click default might suggest. A large fraction of participants in this treatment still build their own algorithms, with similar share proportions of algorithm types as before. 

This shift from undercutting toward cooperative and tit-for-tat types maps directly onto the harshness results from the previous section. Cooperative and tit-for-tat algorithms set higher prices across the states they reach, producing the lower harshness (higher leniency) that, together with higher start prices, accounts for the treatment effects on market prices.

\subsection{Cross-treatment price effects}
\label{sec:cross_treatment}

While our main analysis identifies clear treatment differences and demonstrates that platform design elements influence sellers to set higher prices, one concern is that these effects may be driven solely by within-treatment coordination. This concern is practically relevant. While hub-and-spoke collusion, where multiple sellers operate through the same algorithm provider, is a well-documented phenomenon in these markets, and shared exposure to identical design elements could foster tacit coordination, in many real-world settings, competing firms use different algorithm providers and may receive different default algorithms or different textual nudges that shape their pricing strategies. 

\begin{figure}[!ht]
    \centering
    \includegraphics[width=0.8\textwidth]{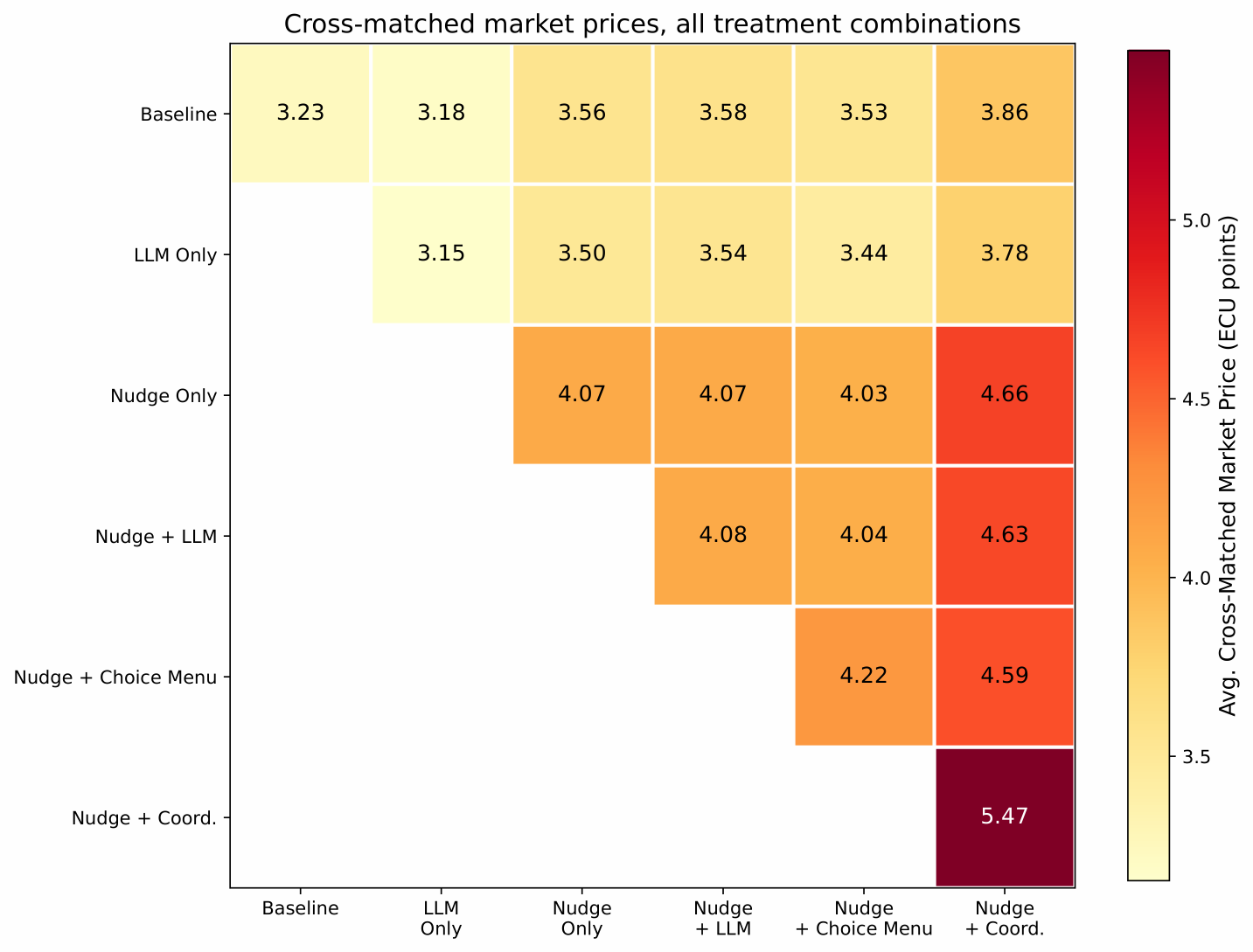}
    \caption{Cross-matched prices across treatments (only supergame-1 algorithms).}
    \label{fig:cross-matched-heatmap}
\end{figure}

To disentangle coordination effects from the direct impact of design elements on algorithm construction, we conduct a rematching analysis. Specifically, we take the algorithms that participants built in the first supergame\footnote{We use supergame~1 to avoid dependencies induced by feedback or adaptation to the assigned counterpart. Because participants were not informed of this ex-post rematching, the exercise evaluates fixed rules and choices might differ under anticipated cross-treatment competition.} and compute market prices for every possible matchup between each player within and across treatments. The resulting prices are shown in Figure~\ref{fig:cross-matched-heatmap}.

For a clean analysis we focus on the cross-matching of each algorithm in each treatment with each baseline opponent (Figure~\ref{fig:cross_matched_prices} in Appendix~\ref{app:mechanisms}). For baseline players, we similarly compute prices against all other baseline players (excluding self-matches). This yields a matchup-level dataset of 169,840 observations. We then regress matchup-level prices on treatment indicators, with baseline as the reference category. To account for the dependency structure, where the same player appears in matchups against all baseline opponents, and the same baseline opponent appears in matchups against all players, we cluster standard errors in two dimensions, by player and by opponent \citep{cameron2011robust}. Table~\ref{tab:cross_matched_prices} present the results.

\begin{table}[!ht]
    \centering
    \caption{OLS Regression: Cross-Matched Market Prices: All Treatments vs. Baseline}
    \label{tab:cross_matched_prices}
    \small
    \setlength{\tabcolsep}{6pt}
    \begin{tabular}{lc}
        \toprule
        \textbf{Dep. Var.: Price} & (1) \\
        \midrule
        LLM Only
            & $-0.047$ \\
            & $(0.106)$ \\[0.3em]
        Nudge Only
            & $0.330^{***}$ \\
            & $(0.100)$ \\[0.3em]
        Nudge $+$ LLM
            & $0.355^{***}$ \\
            & $(0.105)$ \\[0.3em]
        Nudge $+$ Choice Menu
            & $0.300^{***}$ \\
            & $(0.096)$ \\[0.3em]
        Nudge $+$ Coordinate
            & $0.631^{***}$ \\
            & $(0.109)$ \\[0.5em]
        Constant
            & $3.230^{***}$ \\
            & $(0.097)$ \\
        \midrule
        Clustered SE (Player 1, Player 2) & Yes \\
        $N$                             & 169{,}840 \\
        $R^2_{\text{adj}}$              & 0.015 \\
        \bottomrule
        \multicolumn{2}{p{0.70\linewidth}}{\footnotesize
            \textit{Notes:} OLS estimates with standard errors (in parentheses)
            two-way clustered at the individual player level.
            Each player's game-1 algorithm is cross-matched against every
            baseline player's algorithm; the dependent variable is the
            resulting expected market price. Baseline is the omitted category.
            $^{*}p < 0.10$,\ $^{**}p < 0.05$,\ $^{***}p < 0.01$.}
    \end{tabular}
\end{table}

The treatment effects from our main analysis are largely preserved in this rematching exercise. Relative to baseline average prices of $3.23$, the \textsc{Nudge Only} ($+0.33$, $p < 0.001$), \textsc{Nudge + LLM Advice} ($+0.36$, $p < 0.001$), \textsc{Nudge + Choice Menu} ($+0.30$, $p = 0.002$), and \textsc{Nudge + Coordination} ($+0.63$, $p < 0.001$) treatments all produce significantly higher prices, while \textsc{LLM Advice Only} shows no significant difference ($-0.05$, $p = 0.658$). The ordering of treatment effects mirrors our main findings. However, effect sizes are smaller than in the main analysis of the paper, suggesting that coordination among players exposed to the same design elements does contribute to higher prices.

\section{Extension: LLM-Designed Pricing Rules}
\label{sec:llm_extension}

So far, we have considered human decision-makers who may be influenced by the platform when designing their pricing algorithms, and how such design choices can affect pricing and foster tacit collusion. At the same time, there is growing evidence that sellers increasingly rely on AI tools, including to design their pricing rules \citep[see, for instance,][]{fish_algorithmic_2025, ershov2025managing}. Commercial repricers are moving in this direction: platforms such as Seller Snap, Feedvisor, and Aura already offer fully autonomous AI-driven repricing modes that replace manual rule configuration entirely, and Amazon's own Seller Assistant has been expanded with agentic capabilities that can adjust prices with seller approval (see Appendix~\ref{app:industry_examples} for details). A natural extension of our framework is therefore to ask how popular LLMs choose pricing rules.

Our setup allows us to deploy LLMs in the same decision environment as human sellers. To do so, we consider a variant of the \textsc{Baseline} treatment in which, instead of humans deciding on the pricing rule, a set of large language models (LLMs) performs this task. The LLMs are instructed to maximize their own average profit and are otherwise given instructions that are as close as possible to those given to human participants in the experiment. This follows the approach used, for instance, by \citet{manning2024automated}.

The implementation closely mirrors the experimental setup with human participants. The LLMs play five supergames of the pricing game, and at the beginning of each supergame, they select a complete pricing algorithm. The set of rules available to them is identical to the one used in the human experiment. The algorithms are then played out over the 50 periods of the game. After each supergame, the models receive feedback on the realized market outcomes and can revise their algorithm for the next supergame. All models are instructed to provide a justification for their algorithm choice before submitting it to ensure that all models engage in some form of deliberation. In line with the analysis for human participants, we focus on average outcomes across all supergames.

We focus on three open-source large language models accessed via the Together AI inference platform: Llama-3.3-70B-Instruct-Turbo (Llama), Qwen3.5-397B-A17B (Qwen 3.5), and GLM-5.\footnote{Llama 3.3 is a 70-billion-parameter model developed by Meta (released December 2024). Qwen 3.5 is a 397-billion-parameter model developed by Alibaba (released February 2026). GLM-5 is a 744-billion-parameter model developed by Zhipu AI (released February 2026). Both Qwen 3.5 and GLM-5 are among the highest-ranked open-source models on the LM Arena leaderboard \citep{chiang2024chatbot} at the time of writing, performing comparably to leading proprietary models from Anthropic, Google, and OpenAI. We include Llama because it is the model used in the LLM treatments of the main experiment, chosen for its fast inference speed and cost efficiency. Note that in those human treatments, the LLM serves primarily as an advisor, providing information about the game rules or nudging participants toward certain strategies, rather than autonomously designing a complete pricing algorithm as in this extension. Overall, we focus on open-source models as they offer a high degree of reproducibility compared to proprietary models, while often achieving a similar performance \citep[see][]{hussain2024tutorial}. Exact model identifiers, prompts, and inference parameters are provided in Appendix~\ref{app:llm_simulation}.} For Qwen 3.5 and GLM-5, we additionally vary whether the model's native reasoning mode is enabled. Recent LLMs increasingly offer such a mode, in which the model generates an extended internal chain-of-thought trace before producing its final response. This goes beyond the prompted justification that all models in our setup provide: rather than simply being asked to explain its choice, the model autonomously engages in a longer and more structured deliberation process. Llama 3.3 does not natively support such a reasoning mode and is therefore used in a single configuration only, yielding five distinct model configurations in total.

\begin{figure}[htbp]
    \centering
    \includegraphics[width=0.8\textwidth,keepaspectratio]{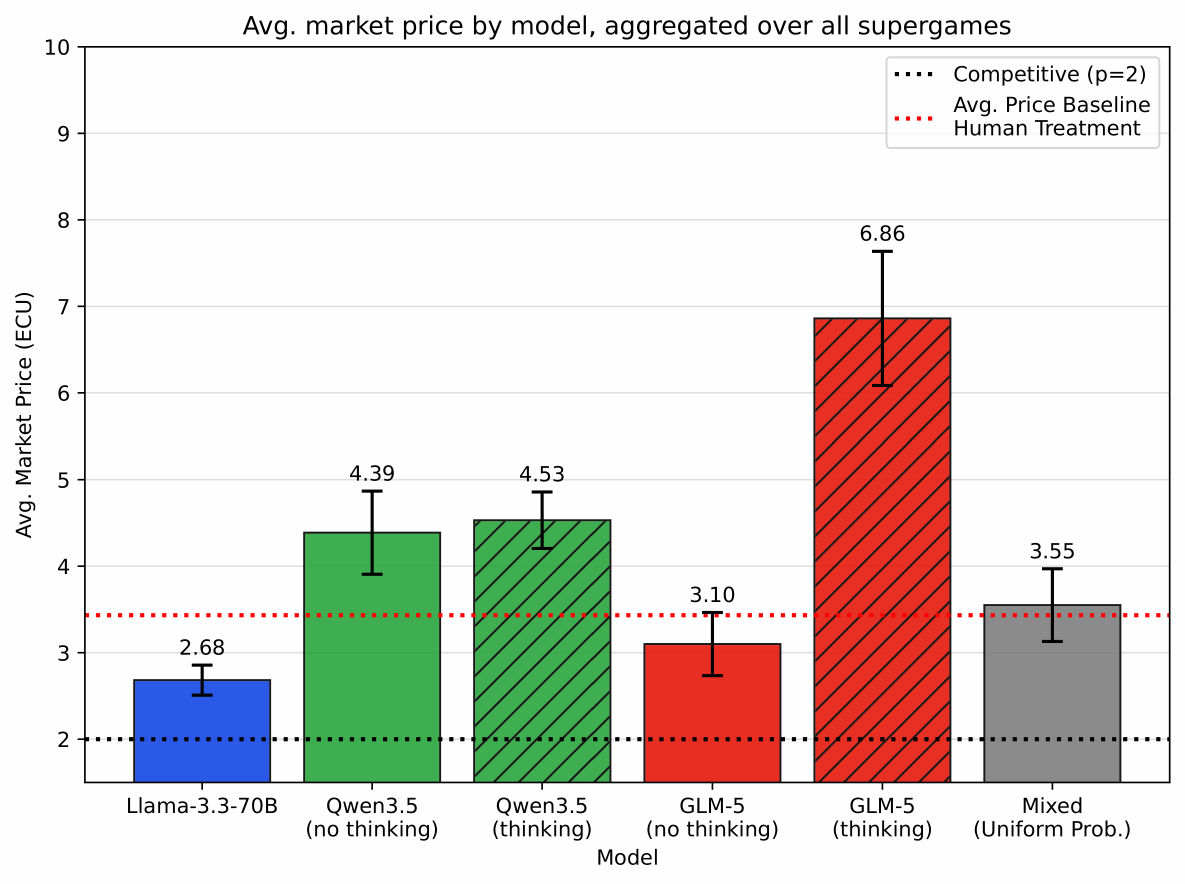}
    \caption{Average Market Price by LLM Model Configuration. Human \textsc{Baseline} and \textsc{Nudge + Coordination} treatments shown for comparison. The error bars represent 95\% confidence intervals.}
    \label{fig:llm_prices}
\end{figure}

The results are shown in Figure~\ref{fig:llm_prices}. While Llama generates prices that are lower than those observed for humans in the \textsc{Baseline} treatment, more advanced LLMs generate significantly higher prices, exceeding both the \textsc{Baseline} and most other human treatments. The best-performing model is GLM-5 with reasoning (``thinking'') enabled, which achieves the highest prices across all human and LLM treatments. This suggests that more capable models, and in particular those with extended reasoning capabilities, tend to produce more collusive outcomes. 

Interestingly, most LLMs converge to tit-for-tat-type strategies with high starting prices, which sustain the elevated price level and this is most pronounced for ``thinking'' models. Figure~\ref{fig:llm_archetypes} in the appendix shows the strategies generated by each model.

To test whether these results depend on the configuration pairing, we analyse a balanced ``mixed'' pool containing all 15 unordered matchups among the five model configurations: 10 cross-configuration and five same-configuration matchups, with five independently initialised pairs per matchup (75 pairs in total; see Appendix~\ref{app:llm_simulation} for details). The five same-configuration cells in this pool comprise five newly generated pairs each and are separate from the 50-pair samples used for the configuration-specific estimates. Across all 75 mixed-pool pairs, average prices (3.55) do not differ significantly from the human \textsc{Baseline} ($p=0.120$, two-sided Mann-Whitney U). The ``mixed (thinking only)'' subset contains those pairs in which both players use thinking-enabled configurations (three matchups with five pairs each). Their average prices rise to 5.20, significantly above the human \textsc{Baseline} ($p=0.004$) and statistically indistinguishable from the human \textsc{Nudge + Coordination} treatment ($p=0.489$). When both players have reasoning capabilities, prices reach levels comparable to the most effective human treatment.

Overall, this extension highlights an additional anti-competitive effect of rule-based pricing algorithms. While platforms may nudge users towards higher prices through interface design, similar or even stronger effects can arise when sellers fully outsource the design of pricing rules to LLMs. This points to an additional channel of risk in such environments.

\section{Conclusion}

This paper studies prevalent choice architecture and nudges commonly used by repricers and provides experimental evidence that these design choices can meaningfully influence market outcomes. In an innovative controlled experimental setting, sellers set pricing strategies with the use of a pricing dashboard that closely reflects the real-word choice environment of sellers on online platforms. The sequential Bertrand setting captures the typical reactive algorithms in the field and provides a tractable environment to identify the choice of strategies and allows to measure the deviation from best response equilibria. 

We show that simple interface nudges --- simple information nudges and in particular pre-configured cooperative algorithms --- push prices substantially above the competitive benchmark. The effect is driven by both higher starting prices and the choice of more ``cooperative'' algorithms. LLM advice, important because of its growing use in real-world pricing contexts, does not amplify this effect further. %
Providing a simplified choice architecture by providing several pre-built algorithms does not increase market prices beyond the general nudge. While changing the types of algorithms chosen, this treatment also introduces undercutting strategies which then again reduce overall market prices. Providing only a single tit-for-tat algorithm increases prices further.   

Compared to observations in the field, our controlled environment allows us to observe the full strategy choice of sellers. This allows us to decompose the chosen algorithms and allows us to measure leniency and mutual optimality in markets, as well as to classify the algorithms. Further, we can observe off-path play. This information can otherwise only be approximated from observing market data. 
Interestingly, a much discussed cycling strategy observed in the field \citep[e.g.,]{musolff_algorithmic_2026} only appears as a substantial share if it is provided as a pre-built algorithm suggesting that the idea of price cycling is unintuitive and might only be present due to active marketing by repricers.

Our cross-treatment re-matching analysis shows that these effects are not the result of within-pair coordination. Algorithms shaped by nudges and defaults produce higher prices even when matched against untreated opponents. This means that even partial adoption of such tools can elevate prices market-wide. The persistence of significant treatment differences in this re-matching indicates that design elements do not just facilitate coordination among like-treated players. They lead participants to construct different algorithms that generate higher prices even when matched against opponents who were not exposed to the same nudges or defaults. %

Taken together, our paper provides causal evidence on the effects of recent management strategies aimed at ``avoiding price wars''. It shows that prices increase and explains why: due to a change in starting prices and the selection of more cooperative algorithms. The results indicate that supra-competitive outcomes may not only emerge from explicit coordination between sellers, but from design choices made upstream by platform operators or tool providers. This is relevant for competition policy: regulators focused narrowly on seller behavior or purely on self-learning algorithms may miss how pricing infrastructure quietly shapes market outcomes. In our setting it is human firm owners, guided by design choices, that lead to supra-competitive prices. Whereas regulators and consumer protection agencies are usually concerned about protecting consumers from collusive or exploitative practices imposed by firms, our paper thus calls for a broader perspective of consumer protection via ``protecting'' sellers from collusive nudges. 

Our results complement the well-founded concerns surrounding hub-and-spoke configurations \citep{harrington2024hub}. The repricers provide an algorithm dashboard for everyone to use, but additionally offer design tools ``simplifying'' choices of sellers and nudging them towards higher prices. Thus, even though there is no data sharing through the hub, the design features present serve as a coordination device. Enforcement has begun to target in-built design features directly. The US Department of Justice (DOJ) alleged that RealPage's revenue management software contained a ``governor'' that implemented the revenue-maximising price when it rose but substituted a lagged average when it fell, dampening decreases while leaving increases intact.\footnote{United States v.\ RealPage, Inc., No. 1:24-CV-00710-WLO-JLW, ECF No. 159 (M.D.N.C. Nov. 24, 2025) (Proposed Final Judgment).} The proposed settlement requires the feature to be symmetric and user-parametrised rather than prohibiting it. Our results suggest such a standard may be too narrow. The design features we study do not exert one-sided pressure on prices, yet still raise them well above the competitive benchmark.

Our results then also speak more broadly to the discussions on human-in-the-loop remedies, which presume that retaining human control over pricing or other AI systems might constrain potentially adverse effects of the software. In our setting humans retain full control --- they write the rules themselves. Yet, this is precisely what makes them susceptible to the provider's nudges. Human oversight does then not close the channel through which a provider can influence prices --- it is that channel. 

When we let large language models design pricing algorithms autonomously, advanced models with reasoning capabilities converge on tit-for-tat strategies with high start prices, reaching price levels comparable to the most effective human treatment. As commercial repricers move toward AI-driven modes that replace manual rule configuration, these effects are likely to grow. In turn, this shows that human-in-the-loop can potentially still provide a restriction on collusive prices. %

\newpage

\setstretch{1.2}

\newpage

\appendix 
\counterwithin{figure}{section}
\counterwithin{table}{section}

\section{Appendix}

\subsection{Additional Figures and Tables}

\begin{table}[htbp]
    \centering
    \small
    \caption{Mann-Whitney U Tests: Pairwise Treatment Comparisons on Average Market Price}
    \label{tab:mwu_main_rq}
    \begin{tabular}{clcc}
        \toprule
        & \textbf{Comparison} & \textbf{$U$} & \textbf{$p$ (one-sided)} \\
        \midrule
        \multirow{5}{*}{H1}
            & Base vs.\ LLM Only              & 2994.0 & $.327$    \\
            & Base vs.\ Nudge Only             & 2674.0 & $.002$    \\
            & Base vs.\ Nudge + LLM            & 2565.5 & $.003$    \\
            & Base vs.\ Nudge + Choice Menu    & 2256.5 & $<0.001$  \\
            & Base vs.\ Nudge + Coordination     & 1860.0 & $<0.001$  \\
        \midrule
        H2  & Nudge Only vs.\ Nudge + Coordination & 2446.5 & $.001$ \\
        \midrule
        \multirow{2}{*}{H3}
            & Nudge + LLM vs.\ LLM Only        & 3309.0 & $.002$   \\
            & Nudge + LLM vs.\ Nudge Only       & 3171.0 & $.537$   \\
        \midrule
        H4  & Nudge Only vs.\ Nudge + Choice Menu & 3064.0 & $.164$ \\
        \bottomrule
        \multicolumn{4}{p{0.92\textwidth}}{\footnotesize
            \textit{Notes:} Mann-Whitney U tests on pair-level market prices
            (averaged over supergames and ingame rounds). $p$~(one-sided) reports raw $p$-values from the directional test.
            All $p$-values rounded to three decimal places.}
    \end{tabular}
\end{table}

\begin{table}[htbp]
    \centering
    \caption{OLS Regressions: Price Dynamics over Periods and Supergames by Treatment}
    \label{tab:price_dynamics}
    \small
    \setlength{\tabcolsep}{6pt}
    \begin{tabular}{lcccccc}
        \toprule
        \textbf{Dep. Var.: Price} & (1) & (2) & (3) & (4) & (5) & (6) \\
        & {\scriptsize Baseline} & {\scriptsize LLM Only} & {\scriptsize Nudge Only} & {\scriptsize Nudge$+$LLM} & {\scriptsize Nudge$+$CM} & {\scriptsize Nudge$+$Coord.} \\
        \midrule
        Period
            & $-0.007^{**}$ & $0.004$ & $-0.005$ & $-0.010^{**}$ & $-0.012^{***}$ & $-0.023^{***}$ \\
            & $(0.003)$ & $(0.005)$ & $(0.005)$ & $(0.004)$ & $(0.004)$ & $(0.005)$ \\[0.3em]
        Supergame
            & $0.028$ & $0.019$ & $0.058$ & $0.013$ & $-0.023$ & $-0.106$ \\
            & $(0.044)$ & $(0.060)$ & $(0.077)$ & $(0.066)$ & $(0.052)$ & $(0.082)$ \\[0.3em]
        Period $\times$ Supergame
            & $0.0015^{*}$ & $0.0002$ & $0.0001$ & $0.0007$ & $0.0009$ & $0.0006$ \\
            & $(0.0008)$ & $(0.0015)$ & $(0.0012)$ & $(0.0012)$ & $(0.0011)$ & $(0.0013)$ \\[0.5em]
        Constant
            & $3.405^{***}$ & $3.357^{***}$ & $4.067^{***}$ & $4.315^{***}$ & $4.555^{***}$ & $6.142^{***}$ \\
            & $(0.195)$ & $(0.213)$ & $(0.277)$ & $(0.260)$ & $(0.224)$ & $(0.344)$ \\
        \midrule
        Clustered SE (Pair)   & Yes & Yes & Yes & Yes & Yes & Yes \\
        $N$                   & 22{,}000 & 17{,}750 & 20{,}500 & 19{,}500 & 20{,}500 & 20{,}500 \\
        $R^2_{\text{adj}}$    & 0.002 & 0.001 & 0.002 & 0.002 & 0.002 & 0.011 \\
        \bottomrule
        \multicolumn{7}{p{0.96\linewidth}}{\footnotesize
            \textit{Notes:} OLS estimates with standard errors (in parentheses) clustered at
            the pair level. Each column estimates
            $\text{Price} = \beta_0 + \beta_1 \text{Period} + \beta_2 \text{Supergame}
            + \beta_3 \text{Period} \times \text{Supergame} + \varepsilon$
            separately for one treatment group. CM = Choice Menu.
            $^{*}p < 0.10$,\ $^{**}p < 0.05$,\ $^{***}p < 0.01$.}
    \end{tabular}
\end{table}

\begin{table}[htbp!]
    \centering
    \caption{OLS Regressions: Treatment Effects on Individual Payoff}
    \label{tab:horse_race_payoff}
    \small
    \setlength{\tabcolsep}{6pt}
    \begin{tabular}{lcccccc}
        \toprule
        \textbf{Dep. Var.: Individual Payoff} & (1) & (2) & (3) & (4) & (5) & (6) \\
        \midrule
        LLM Only
            & $0.045$ & $0.050$ & $0.051$ & $0.052$ & $0.061$ & $0.061$ \\
            & $(0.125)$ & $(0.109)$ & $(0.122)$ & $(0.109)$ & $(0.109)$ & $(0.109)$ \\[0.3em]
        Nudge Only
            & $0.350^{***}$ & $0.180$ & $0.347^{***}$ & $0.180$ & $0.181$ & $0.180$ \\
            & $(0.131)$ & $(0.115)$ & $(0.129)$ & $(0.115)$ & $(0.115)$ & $(0.115)$ \\[0.3em]
        Nudge $+$ LLM
            & $0.359^{**}$ & $0.145$ & $0.360^{***}$ & $0.146$ & $0.146$ & $0.145$ \\
            & $(0.142)$ & $(0.129)$ & $(0.139)$ & $(0.129)$ & $(0.129)$ & $(0.129)$ \\[0.3em]
        Nudge $+$ Choice Menu
            & $0.415^{***}$ & $0.210^{*}$ & $0.406^{***}$ & $0.208^{*}$ & $0.211^{*}$ & $0.210^{*}$ \\
            & $(0.120)$ & $(0.108)$ & $(0.119)$ & $(0.108)$ & $(0.108)$ & $(0.108)$ \\[0.3em]
        Nudge $+$ Coordination
            & $0.925^{***}$ & $0.467^{***}$ & $0.913^{***}$ & $0.465^{***}$ & $0.479^{***}$ & $0.477^{***}$ \\
            & $(0.160)$ & $(0.142)$ & $(0.159)$ & $(0.141)$ & $(0.143)$ & $(0.143)$ \\[0.5em]
        Start Price
            &             & $0.145^{***}$ &             & $0.145^{***}$ & $0.072$       & $0.071$       \\
            &             & $(0.015)$     &             & $(0.015)$     & $(0.049)$     & $(0.049)$     \\[0.3em]
        Harshness
            &             &               & $-0.273$ & $-0.085$ & $-0.727^{*}$ & $-0.741^{*}$ \\
            &             &               & $(0.213)$ & $(0.213)$ & $(0.384)$     & $(0.385)$     \\[0.3em]
        Start Price $\times$ Harshness
            &             &               &           &           & $0.127$ & $0.130$ \\
            &             &               &           &           & $(0.079)$   & $(0.079)$   \\[0.5em]
        Constant
            & $1.717^{***}$ & $1.070^{***}$ & $1.877^{***}$ & $1.122^{***}$ & $1.492^{***}$ & $1.402^{***}$ \\
            & $(0.085)$     & $(0.097)$     & $(0.158)$     & $(0.170)$     & $(0.261)$     & $(0.260)$     \\
        \midrule
        Start Price            & No  & Yes & No  & Yes & Yes & Yes \\
        Harshness              & No  & No  & Yes & Yes & Yes & Yes \\
        Interaction Effect     & No  & No  & No  & No  & Yes & Yes \\
        Supergame FE               & No  & No  & No  & No  & No  & Yes \\
        Clustered SE (Pair)    & Yes & Yes & Yes & Yes & Yes & Yes \\
        $N$                    & 4830 & 4830 & 4830 & 4830 & 4830 & 4830 \\
        $R^2_{\text{adj}}$     & 0.032 & 0.076 & 0.033 & 0.075 & 0.076 & 0.076 \\
        \bottomrule
        \multicolumn{7}{p{0.96\linewidth}}{\footnotesize
            \textit{Notes:} OLS estimates with standard errors (in parentheses) clustered at
            the pair level (483 pairs). Omitted treatment category: \textit{Baseline}.
            $^{*}p < 0.10$,\ $^{**}p < 0.05$,\ $^{***}p < 0.01$.}
    \end{tabular}
\end{table}

\begin{figure}[htbp]
    \centering
    \includegraphics[width=\textwidth]{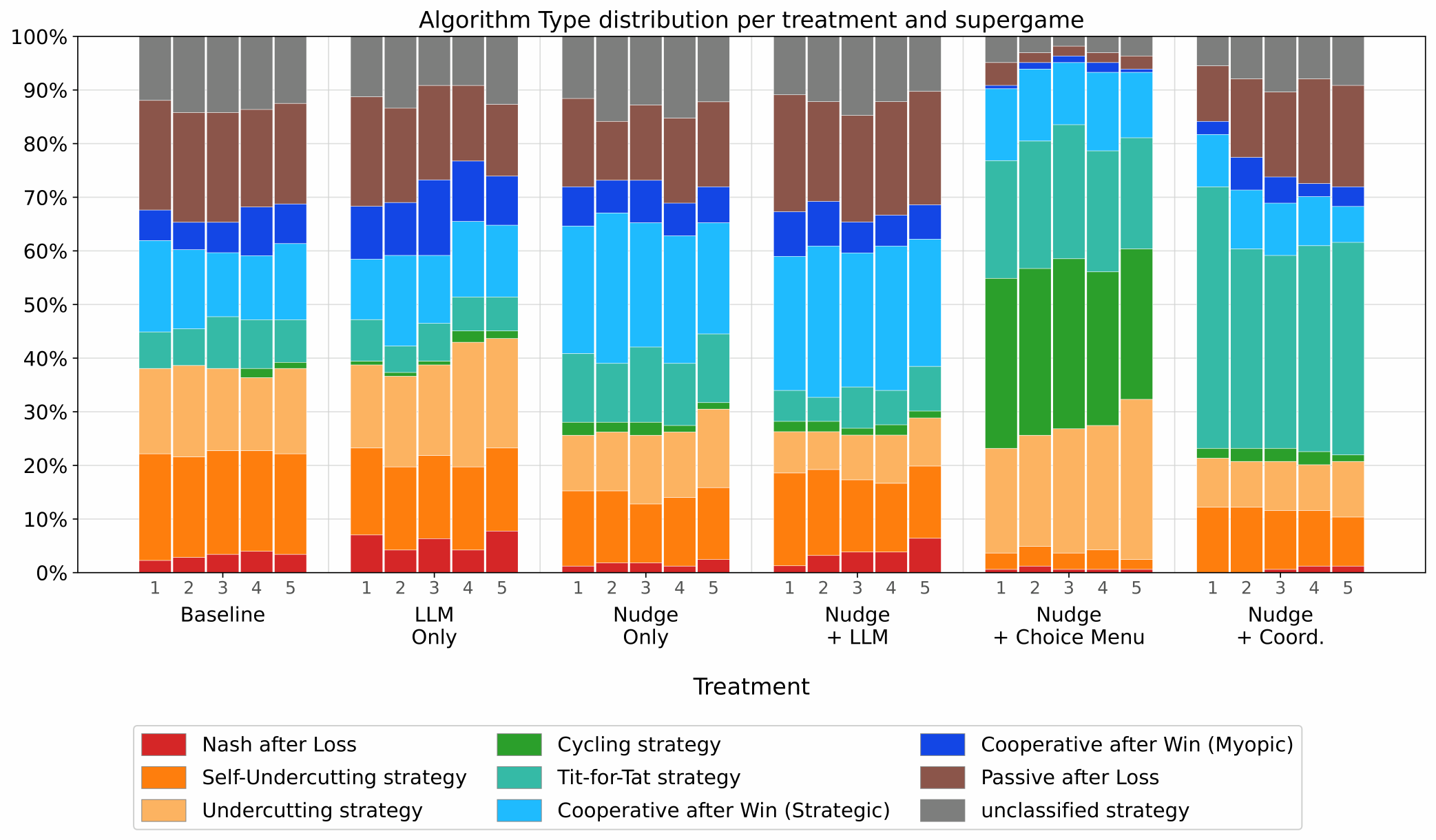}
    \caption{Algorithm Types by treatment and supergame}
    \label{fig:algo_types_treatments_rounds}
\end{figure}

\begin{figure}[htbp]
    \centering
    \includegraphics[width=0.85\textwidth]{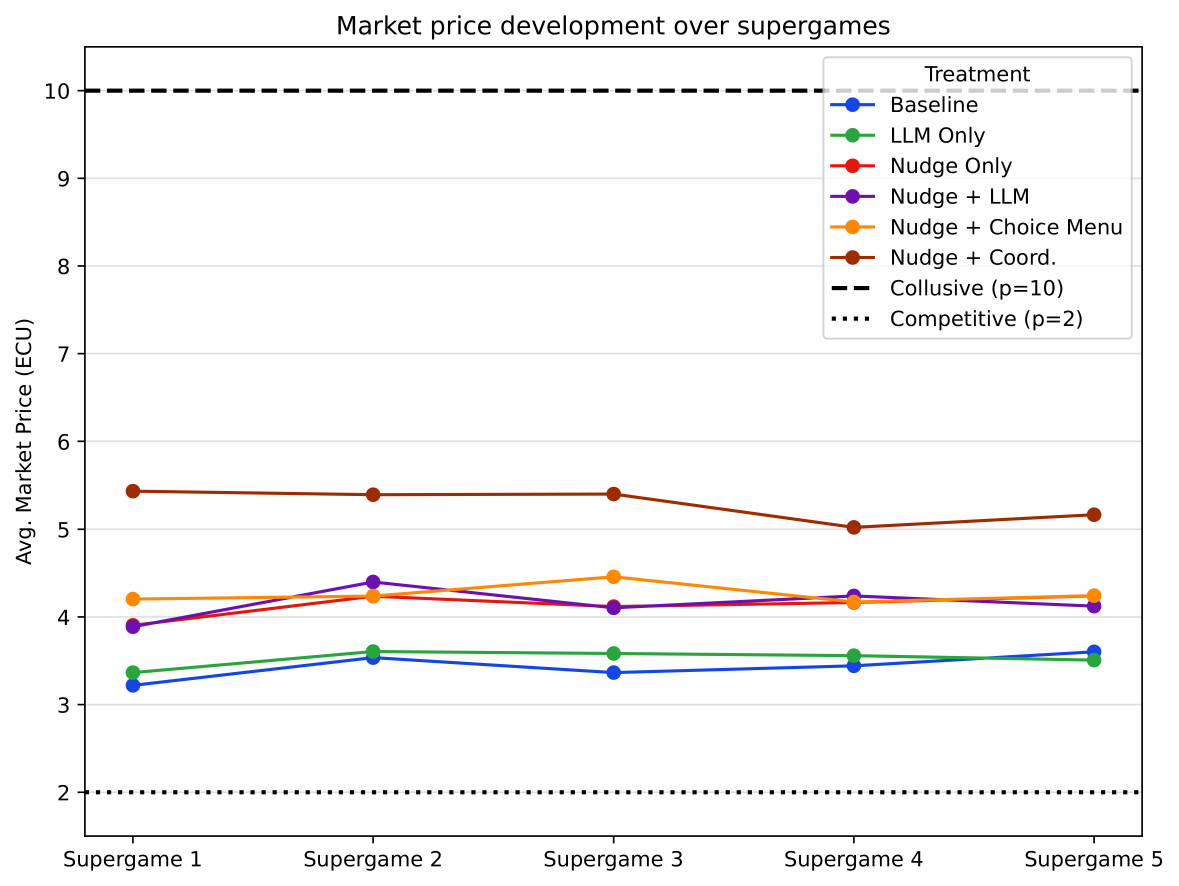}
    \caption{Avg. market price over the five supergames}
    \label{fig:avg_price_rounds}
\end{figure}

\FloatBarrier

\clearpage
\subsection{Mechanisms}
\label{app:mechanisms}
\nopagebreak

\begin{figure}[H]
    \centering

    \includegraphics[width=0.75\textwidth,keepaspectratio]{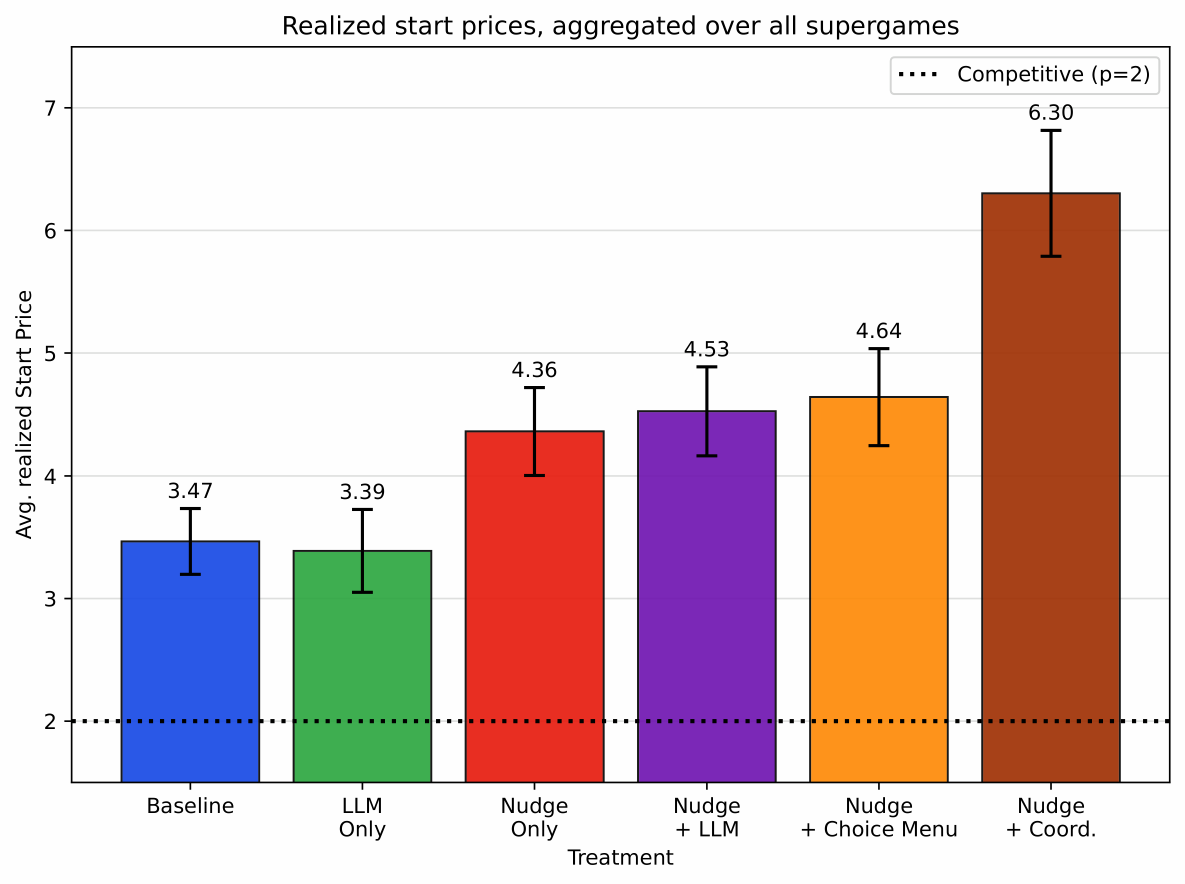}
    \caption{Average realized start prices by treatment. The error bars represent 95\% confidence intervals.}
    \label{fig:startprice}

    \centering

    \includegraphics[width=0.75\textwidth,keepaspectratio]{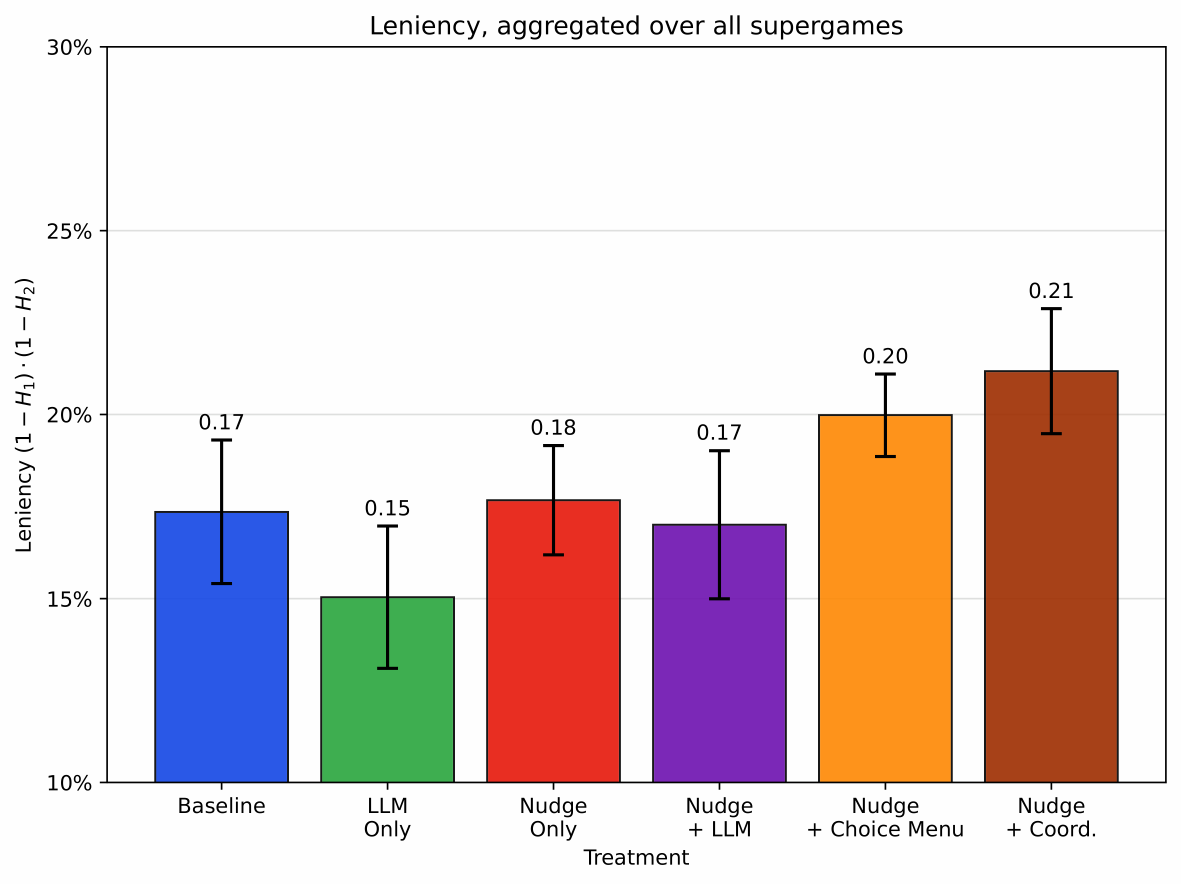}
    \caption{\footnotesize Leniency by Treatment. Leniency is defined as the product of the complements of the harshness measures of both players. Harshness measures the average aggressiveness of an algorithm across all reachable market situations, normalized to $[0,1]$. The error bars represent 95\% confidence intervals.}
    \label{fig:harshness}

\end{figure}

\begin{figure}[htbp]
    \centering
    \includegraphics[width=0.87\textwidth]{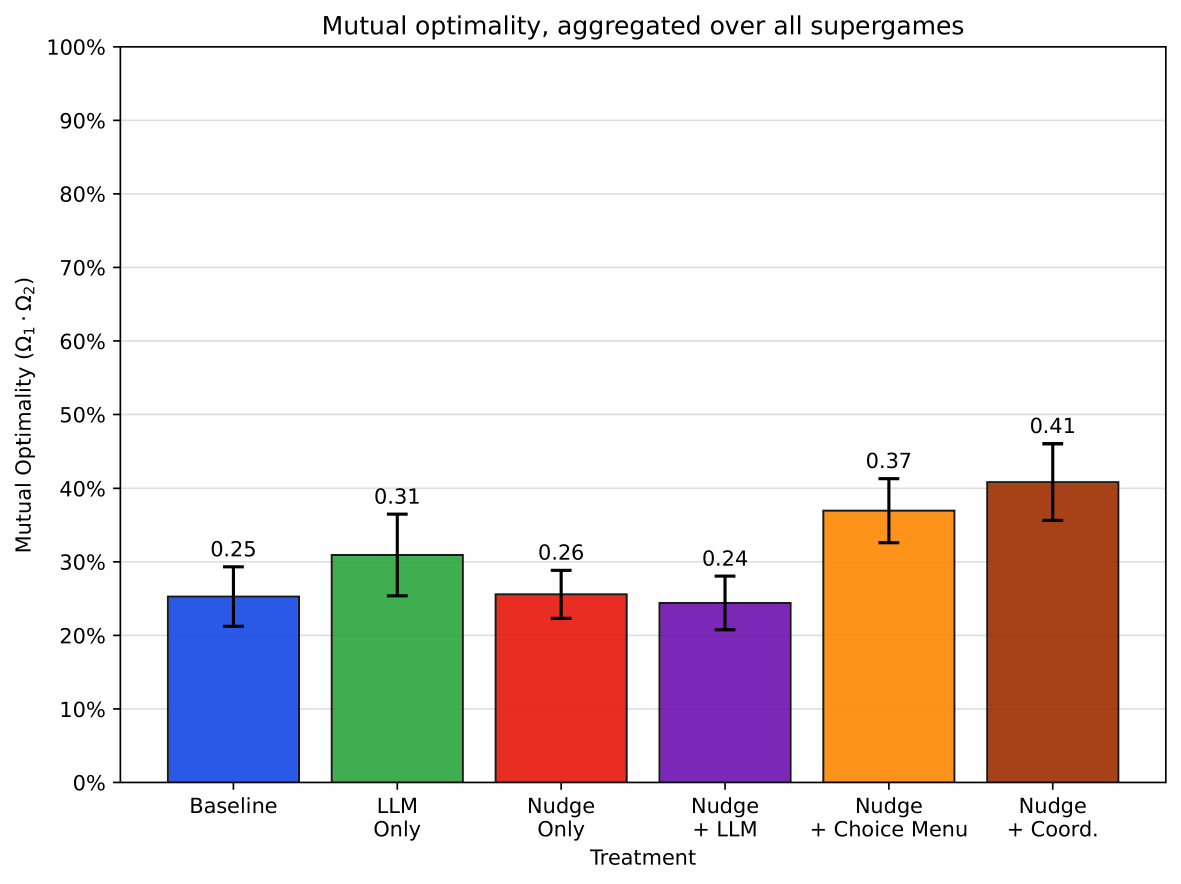}
    \caption{Mutual Optimality by Treatment. Optimality measures the ratio of a player's actual expected payoff to the best-response payoff achievable against the same opponent. The error bars represent 95\% confidence intervals.}
    \label{fig:optimality}
\end{figure}

\begin{figure}[htbp]
    \centering
    \includegraphics[width=0.8\textwidth]{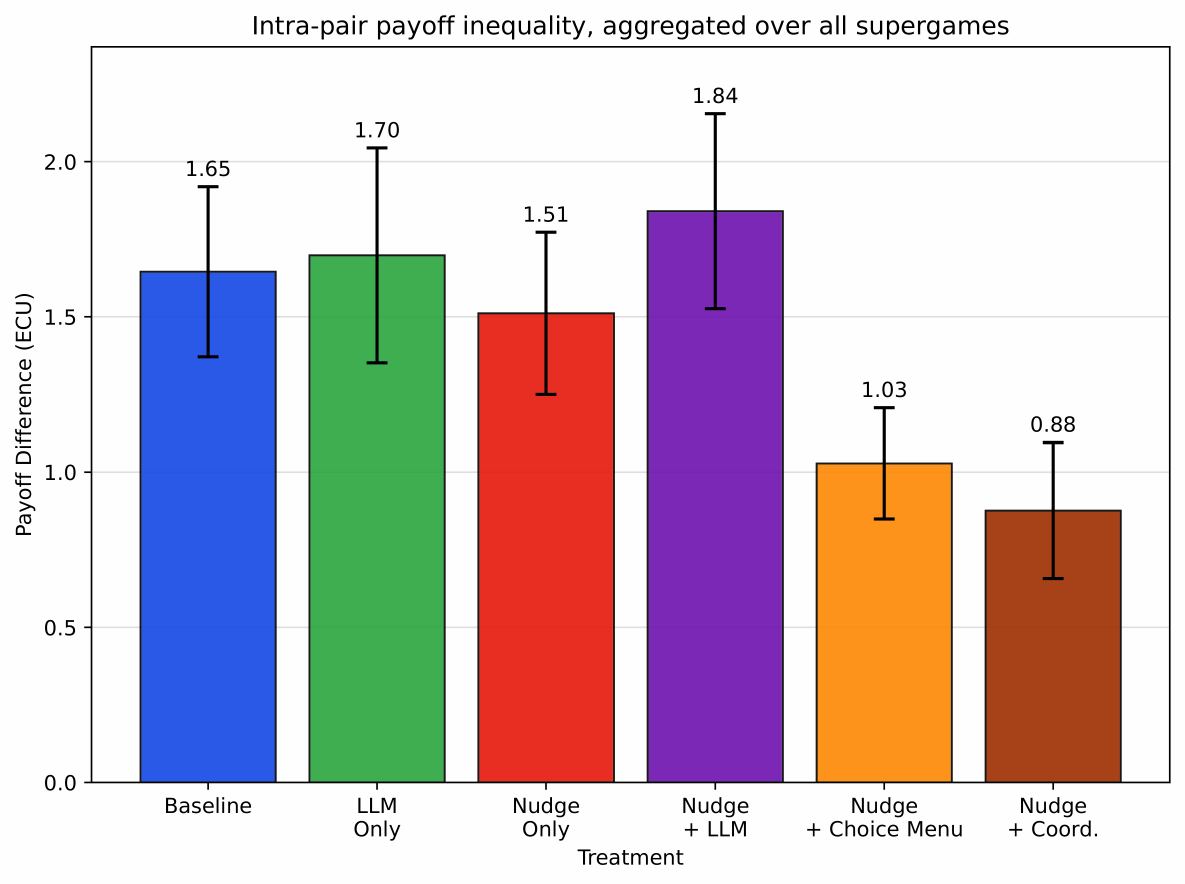}
    \caption{Intra-pair payoff inequality by Treatment. The error bars represent 95\% confidence intervals.}
    \label{fig:payoff_inequality}
\end{figure}

\begin{figure}[htbp]
    \centering
    \includegraphics[width=\textwidth]{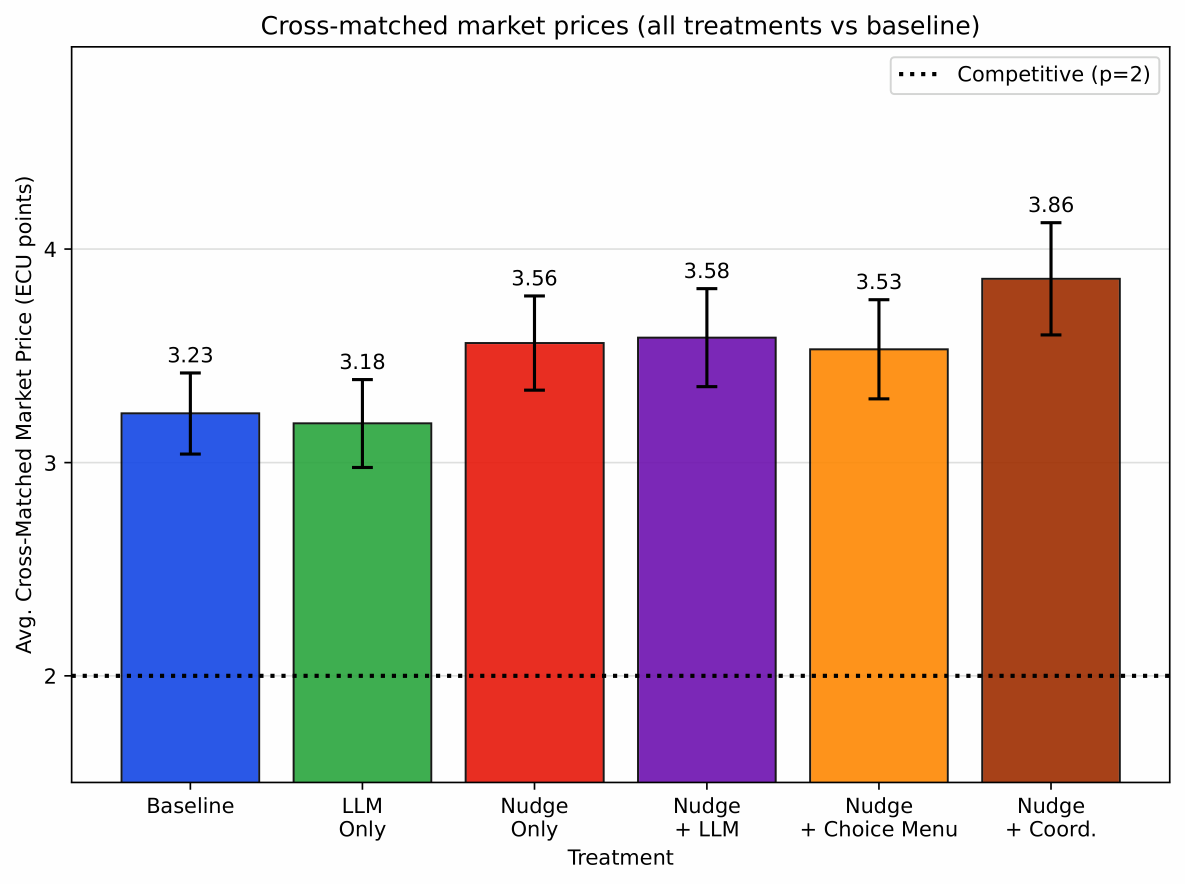}
    \caption{Cross-matched market prices. Each player's supergame-1 algorithm is cross-matched against every baseline player's algorithm. The error bars represent 95\% confidence intervals, with two-way clustered standard errors.}
    \label{fig:cross_matched_prices}
\end{figure}

\begin{figure}[htbp]
    \centering
    \includegraphics[width=\textwidth]{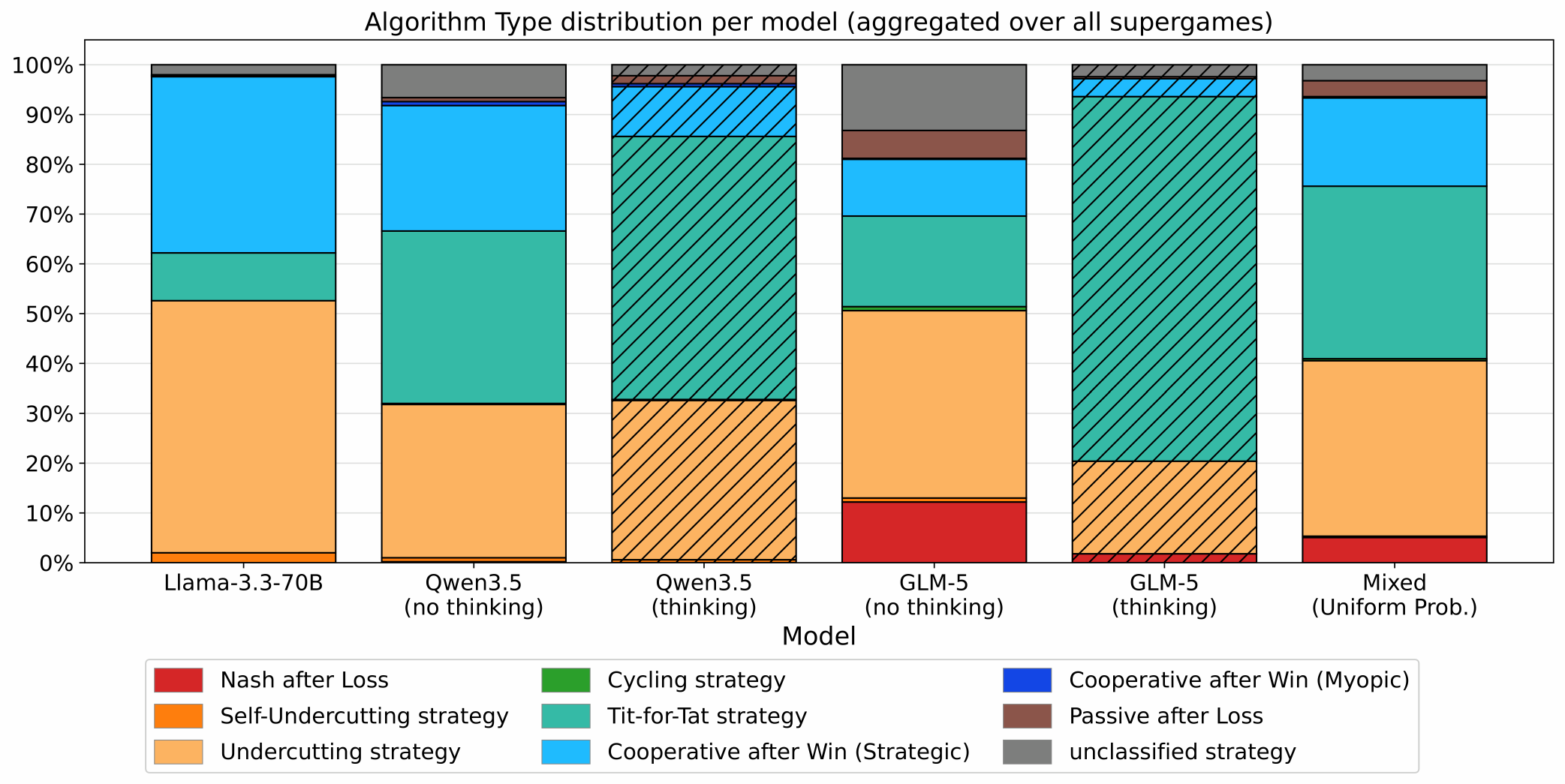}
    \caption{Algorithm Archetypes by LLM Model Configuration. Shows the distribution of high-level strategy types generated by each model.}
    \label{fig:llm_archetypes}
\end{figure}

\FloatBarrier

\newpage

\subsection{Equilibria Characterization}
\label{app:equilibria}
We model the game as a two-agent finite-horizon Markov decision process. The player set is denoted as $\mathcal{I}=\left\{i,j\right\}$. A pricing strategy for player $i$ is a deterministic pricing algorithm $\mu_i: \mathcal{S} \rightarrow \mathcal{A}$, mapping market states to prices. The state space is:
\begin{equation}
    \mathcal{S} = \{s_0\} \cup \left\{(p_i,\, p_j,\, b,\, t) \;\middle|\; p_i, p_j \in \mathcal{A},\;
    (b = i \Rightarrow p_i \leq p_j),\;
    (b = j \Rightarrow p_j \leq p_i) \right\}
    \label{eq:state_space}
\end{equation}
where $s_0$ denotes the initial state before period~1 and the components of each non-initial state are:
\begin{itemize}
    \item $p_i, p_j \in \mathcal{A} = \{2, \ldots, 10\}$: current prices of sellers $i$ and $j$
    \item $b \in \mathcal{I}$: the seller holding the Buy Box in the current period
    \item $t \in \mathcal{I}$: the seller whose turn it is to next update their price
\end{itemize}
By convention, $\mu_i(s) = p_i(s)$ whenever $t(s) \neq i$, capturing the constraint that a seller cannot reprice when it is not their turn. The algorithm builder generates $67{,}068$ distinct pricing algorithms, which constitute the finite strategy set~$\mathcal{M}$.

The per-period reward for player $i$ in state $s$ is deterministic given the Buy Box holder:
\begin{equation}
    R_i(s) = \begin{cases}
        p_i(s) & \text{if } b(s) = i \\
        0      & \text{if } b(s) = j \text{ or if } s = s_0
    \end{cases}
    \label{eq:reward}
\end{equation}

Given a joint strategy profile $\mu = (\mu_i, \mu_j)$ and current state $s$, prices in the next period are determined by the active player repricing according to their algorithm while the passive player's price remains fixed:
\begin{equation}
    p'_i(\mu, s) = \begin{cases}
        \mu_i(s) & \text{if } t(s) = i \text{ or if } s = s_0\\
        p_i(s)   & \text{otherwise}
    \end{cases}
    \label{eq:price_update}
\end{equation}
Price updates are thus fully deterministic. The only sources of randomness is Buy Box allocation when updated prices are equal, as well as the first mover decision after the initial period. Letting $t'(s)=\mathcal{I}\setminus  \left\{t(s)\right\}$ denote the subsequent mover and $b'(\mu,s) = \operatorname*{argmin}_{k \in \{i,j\}}\, p'_k(\mu,s)$ the Buy Box winner in the non-tied case, the state transition probability function after the initial period is:
\begin{equation}
    \mathcal{P}(\mu, s, s') = \begin{cases}
        1          & \text{if } p'_i \neq p'_j
                     \text{ and } s' = \bigl(p'_i,\, p'_j,\, b',\, t'(s)\bigr)\\[4pt]
        \tfrac{1}{2} & \text{if } p'_i = p'_j
                     \text{ and } s' \in \bigl\{(p'_i,\, p'_j,\, b,\, t'(s)) \; | \; b \in \mathcal{I} \bigr\} \\[4pt]
        0          & \text{otherwise}
    \end{cases}
    \label{eq:transition}
\end{equation}
For the initial period, the state transition probability accounts for the random first mover decision:
\begin{equation}
    \mathcal{P}(\mu, s_0, s') = \begin{cases}
        \tfrac{1}{2}          & \text{if } p'_i \neq p'_j
        \text{ and } s' \in \bigl\{ (\bigl(p'_i,\, p'_j,\, b',\, t) \; | \;  t \in \mathcal{I}\bigr\}\\[4pt]
        \tfrac{1}{4} & \text{if } p'_i = p'_j
                     \text{ and } s' \in \bigl\{(p'_i,\, p'_j,\, b,\, t) \; | \; b,t \in \mathcal{I} \bigr\} \\[4pt]
        0          & \text{otherwise}
    \end{cases}
    \label{eq:transition_initial}
\end{equation}
The continuation value for player $i$ in state $s$ at the start of period $t$ satisfies the Bellman equation:
\begin{equation}
    V_{i,t}(\mu, s) = R_i(s) + \sum_{s' \in \mathcal{S}} \mathcal{P}(\mu, s, s') \cdot V_{i,t+1}(\mu, s')
    \label{eq:bellman}
\end{equation}
We solve by backward induction over $T = 50$ periods, setting $V_{i,T+1}(\mu, s) = 0$ for all states $s$ (the game ends after period $T$) and computing $V_{i,T}, V_{i,T-1}, \ldots, V_{i,1}$ recursively. There is no discounting, and all periods contribute equally to total payoffs. The expected payoff for player $i$ under joint profile $\mu$ is then the continuation value at the initial state:
\begin{equation}
    \mathbb{E}\left[\pi_i(\mu)\right] = V_{i,1}(\mu,\, s_0)
    \label{eq:expected_payoff}
\end{equation}
The comprehensive set of Nash equilibria is then found by computing $E[\pi_i(\mu)]$ numerically for all pairs $(\mu_i, \mu_j) \in \mathcal{M}^2$ and identifying mutual best responses. The game comprises three types of pure-strategy Nash equilibria:
\begin{enumerate}
    \item[(i)]   \emph{Static competitive equilibria}, in which both sellers always set the minimum price $p = 2$, regardless of the history. For this outcome profile, there are $395$ unique equilibrium strategies, yielding $395^2=156.025$ equilibrium constellations.
    \item[(ii)]  \emph{Symmetric supracompetitive equilibria}, sustained by a type of Win-Stay/Lose-Match that retains the current price while winning the Buy Box and matches the opponents' price after losing the Buy Box.\footnote{This can also be seen as a ``soft'' grim-trigger strategy, as any deviation to a lower price makes it impossible to recover the gamestate back to a higher price level.} By symmetrically employing this strategy, every supracompetitive price level $q\in\mathcal{A}\setminus{2}$ can be sustained as a symmetric pure-strategy Nash equilibrium in which both sellers charge $q$ in every period. For this class of outcomes, there are $11+10+11+9+7+5+3+11=67$ unique equilibrium strategies, yielding $11^2+10^2+11^2+9^2+7^2+5^2+3^2+11^2=627$ equilibrium constellations. A pure tit-for-tat with a given start price is not exploitable \citep{duersch2012unbeatable} except for a small gain when an algorithm that sets a fixed price equal to the tit-for-tat start price and a start price that is one unit lower, ensuring slightly higher payoffs (less than a one percent gain) due to winning the first period.
    \item[(iii)] \emph{Asymmetric dynamic equilibria}, 
    which are characterised by gradual undercuts and steep price increases to reset the cycle \citep{musolff_algorithmic_2026}. This set of equilibria is quite small and requires one player to be the designated cycle initiator, while the other ``freeloads'' and takes the larger share of the profit. There are two equilibrium outcomes of this kind, with the average payoff profiles of $(1.95,3.17)$ and $(2.2,2.7)$ for the resetter and the freeloader, respectively. For this class of outcomes, there are $(4+4)+(2+2)=12$ unique equilibrium strategies, yielding $4 \times 4 + 2 \times 2=20$ equilibrium constellations. An intuitive implementation of such cycles, in which both sellers repeatedly undercut until one resets its price upon reaching the price floor, is not an equilibrium in our setting. Nevertheless, the resulting strategy profile of those more intuitive cycling algorithms exhibits high mutual optimality and is therefore close to a Nash equilibrium in payoff terms.
\end{enumerate}

\subsection{Description of the Harshness measure}
\label{app:harshness}
The Harshness measure is constructed according to the following equation:
\begin{equation}
    H_i(\mu_i) = \frac{p_{\max} - \dfrac{1}{|\mathcal{S}_i|} \cdot
    \displaystyle\sum_{s \in \mathcal{S}_i} \mu_i(s)}{p_{\max} - p_{\min}}
    \label{eq:harshness_appendix}
\end{equation}
Let $\mathcal{S}_{\mathrm{act}} = \{s \in \mathcal{S} \mid t(s) = i\}$ denote the states
in which player $i$ is the active (repricing) player. The set
$\mathcal{S}_i(\mu_i) \subseteq \mathcal{S}_{\mathrm{act}}$ of \emph{reachable active
states} is derived through an iterative elimination procedure. Starting from
$\mathcal{S}_i^{(0)} = \mathcal{S}_{\mathrm{act}}$, compute the set of prices
reachable from all currently considered active states:
\begin{equation}
    \mathcal{A}_i^{(n)} = \bigl\{ \mu_i(s) \mid s \in \mathcal{S}_i^{(n)} \bigr\},
    \label{eq:reachable_actions}
\end{equation}
then restrict the active state set to those states whose own-price component lies
within $\mathcal{A}_i^{(n)}$:
\begin{equation}
    \mathcal{S}_i^{(n+1)} = \bigl\{ s \in \mathcal{S}_{\mathrm{act}} \mid
    p_i(s) \in \mathcal{A}_i^{(n)} \bigr\}.
    \label{eq:reachable_states}
\end{equation}
Repeat until $\mathcal{A}_i^{(n+1)} = \mathcal{A}_i^{(n)}$, and set $\mathcal{S}_i(\mu_i) =
\mathcal{S}_i^{(n)}$.

The logic is as follows: In any active period, player $i$ reprices to some
$\mu_i(s) \in \mathcal{A}_i^{(n)}$. Player $j$ then acts in the next period while player $i$'s
price is unchanged. When player $i$ becomes active again, their own-price state
parameter must therefore equal one of the prices in $\mathcal{A}_i^{(n)}$: any active state $s$
with $p_i(s) \notin \mathcal{A}_i^{(n)}$ is unreachable and is eliminated, potentially
shrinking $\mathcal{A}_i^{(n+1)}$ further in the next iteration. Since $\mathcal{A}_i^{(n)}$ is
non-increasing and $\mathcal{A}$ is finite, the procedure terminates in at most
$|\mathcal{A}|$ steps.

Two examples illustrate the construction. An algorithm that always sets a fixed
price of~6 produces $\mathcal{A}_i^{(0)} = \{6\}$ immediately, as every active state $s$ maps directly into to $\mu_i(s)=6$, so the set stabilizes in one step and
$\mathcal{S}_i = \{s \in \mathcal{S}_{\mathrm{act}} \mid p_i(s) = 6\}$.
An algorithm that always lowers its own price by one unit produces
$\mathcal{A}_i^{(0)} = \{2, \ldots, 9\}$; in each subsequent iteration the
highest reachable price is shed, until the procedure converges to $\mathcal{A}_i = \{2\}$
and $\mathcal{S}_i = \{s \in \mathcal{S}_{\mathrm{act}} \mid p_i(s) = 2\}$.

\newpage

\renewcommand{\thesection}{OA.\arabic{section}} \setcounter{section}{0}

\FloatBarrier

\begin{center}
{\normalsize
Rule-Based Pricing Algorithms and Market Outcomes:
An Experimental Study \\ Adrian Hillenbrand, Hans-Theo Normann,
Matthias Potarca, Tobias Werner \\  \textbf{Online Appendix}} \\
September 22, 2026
\end{center}

\section{Online Appendix }

\subsection{Leniency and mutual optimality over the five supergames}

\begin{figure}[!htbp]
    \centering
    \includegraphics[width=0.70\textwidth]{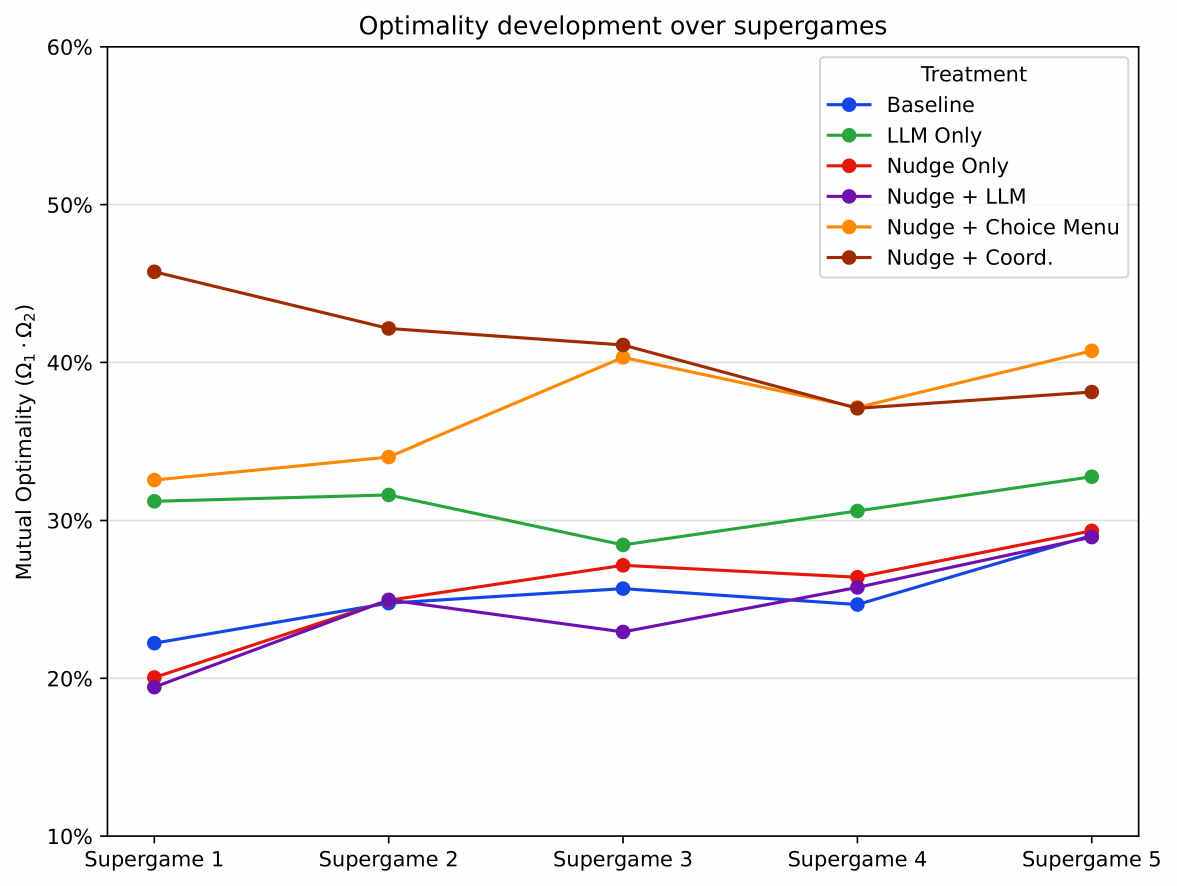}
    \caption{Avg. mutual optimality over the five supergames}
    \label{fig:avg_optimality_rounds}

    \includegraphics[width=0.70\textwidth]{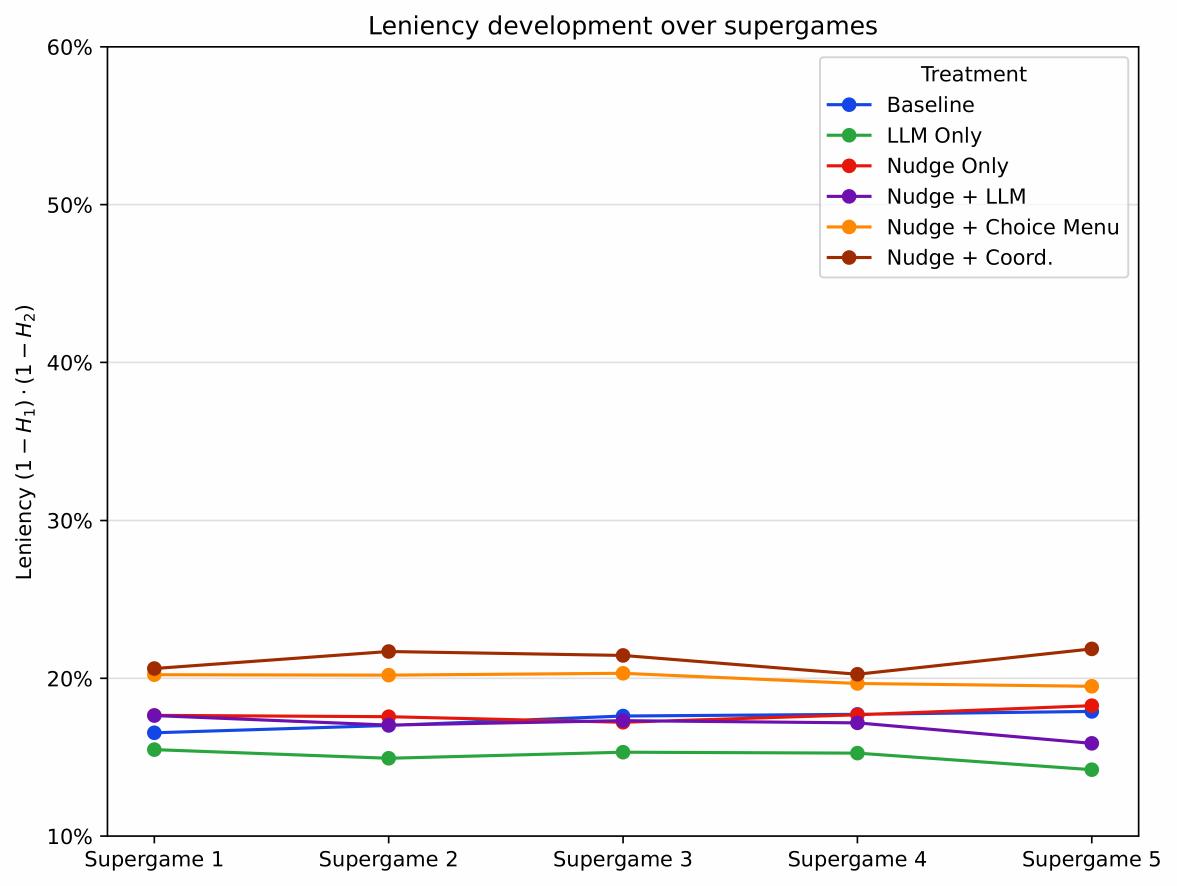}
    \caption{Avg. leniency over the five supergames}
    \label{fig:avg_harshness_rounds}
\end{figure}
\FloatBarrier
\subsection{LLM Chatbot Interaction}
\label{app:llm_chatbot}

In this subsection, we take a closer look at the two treatments in which sellers had access to an LLM-powered chatbot advisor: \textsc{LLM Advice Only} and \textsc{Nudge + LLM}. Table~\ref{tab:llm_descriptives} reports descriptive statistics on chatbot interaction for both treatments. In both treatments, interaction is the highest in the first supergame and drops sharply in subsequent supergames. Overall interaction is higher in \textsc{LLM Advice Only}: by the end of the fifth supergame, $74\%$ of subjects in \textsc{LLM Advice Only} and $65\%$ of subjects in \textsc{Nudge + LLM} had sent at least one message to the advisor.\footnote{A one-tailed Fisher’s exact test that compares these proportions narrowly misses conventional levels of significance ($p = 0.06$).}

\begin{table}[htbp]
    \centering
    \caption{LLM Interaction Across Supergames, by Treatment}
    \label{tab:llm_descriptives}
    \small
    \setlength{\tabcolsep}{6pt}
    \begin{tabular}{lccccc}
        \toprule
        & Game 1 & Game 2 & Game 3 & Game 4 & Game 5 \\
        \midrule
        \multicolumn{6}{l}{\textit{Treatment: LLM} ($N = 142$)} \\[0.3em]
        Share used LLM (this supergame)   & $0.62$ & $0.44$ & $0.37$ & $0.32$ & $0.25$ \\
        Share used LLM (until now)    & $0.62$ & $0.66$ & $0.70$ & $0.73$ & $0.74$ \\
        Mean \# messages              & $2.08$ & $0.82$ & $0.84$ & $0.75$ & $0.60$ \\
        Median \# messages            & $2$    & $0$    & $0$    & $0$    & $0$    \\
        \midrule
        \multicolumn{6}{l}{\textit{Treatment: Nudge $+$ LLM} ($N = 156$)} \\[0.3em]
        Share used LLM (this supergame)   & $0.51$ & $0.26$ & $0.19$ & $0.15$ & $0.19$ \\
        Share used LLM (until now)    & $0.51$ & $0.57$ & $0.60$ & $0.62$ & $0.65$ \\
        Mean \# messages              & $1.54$ & $0.46$ & $0.34$ & $0.29$ & $0.38$ \\
        Median \# messages            & $1$    & $0$    & $0$    & $0$    & $0$    \\
        \bottomrule
        \multicolumn{6}{p{0.96\linewidth}}{\footnotesize
            \textit{Notes:} ``Share used LLM (this supergame)'' reports the fraction of
            subjects who sent at least one message to the LLM advisor in the respective
            supergame. ``Share used LLM (until now)'' reports the cumulative share of
            subjects who have used the advisor in any supergame up to and including the
            current one. Message counts refer to user messages sent to the LLM.}
    \end{tabular}
\end{table}

\begin{figure}[htbp]
    \centering
    \includegraphics[width=0.9\textwidth,keepaspectratio]{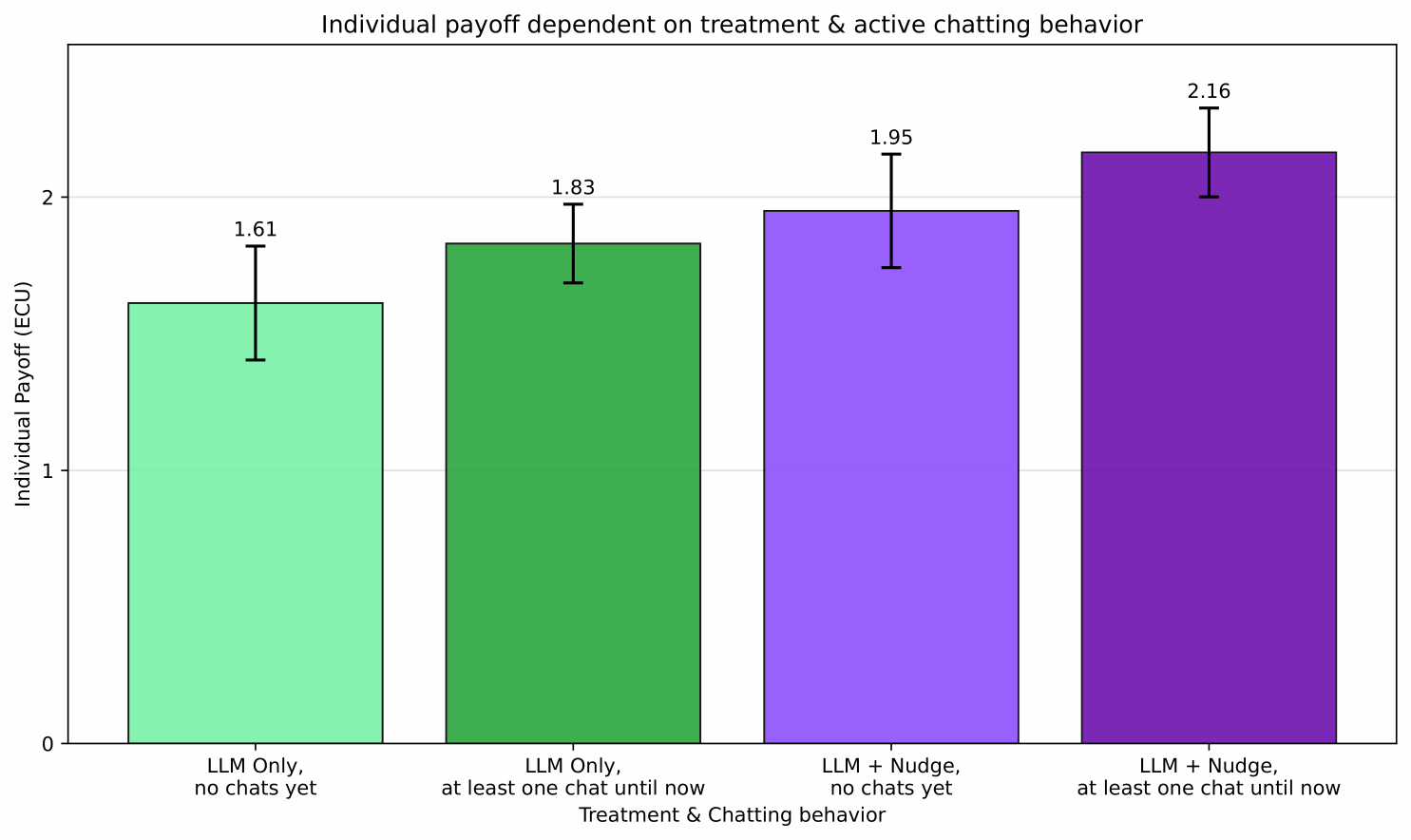}
    \caption{Individual Payoff depending on treatment and if the subject has sent at least one message to the LLM advisor in any supergame up to and including the current one.}
    \label{fig:payoff_by_activechatting}
\end{figure}

\begin{figure}[htbp]
    \centering
    \includegraphics[width=0.9\textwidth,keepaspectratio]{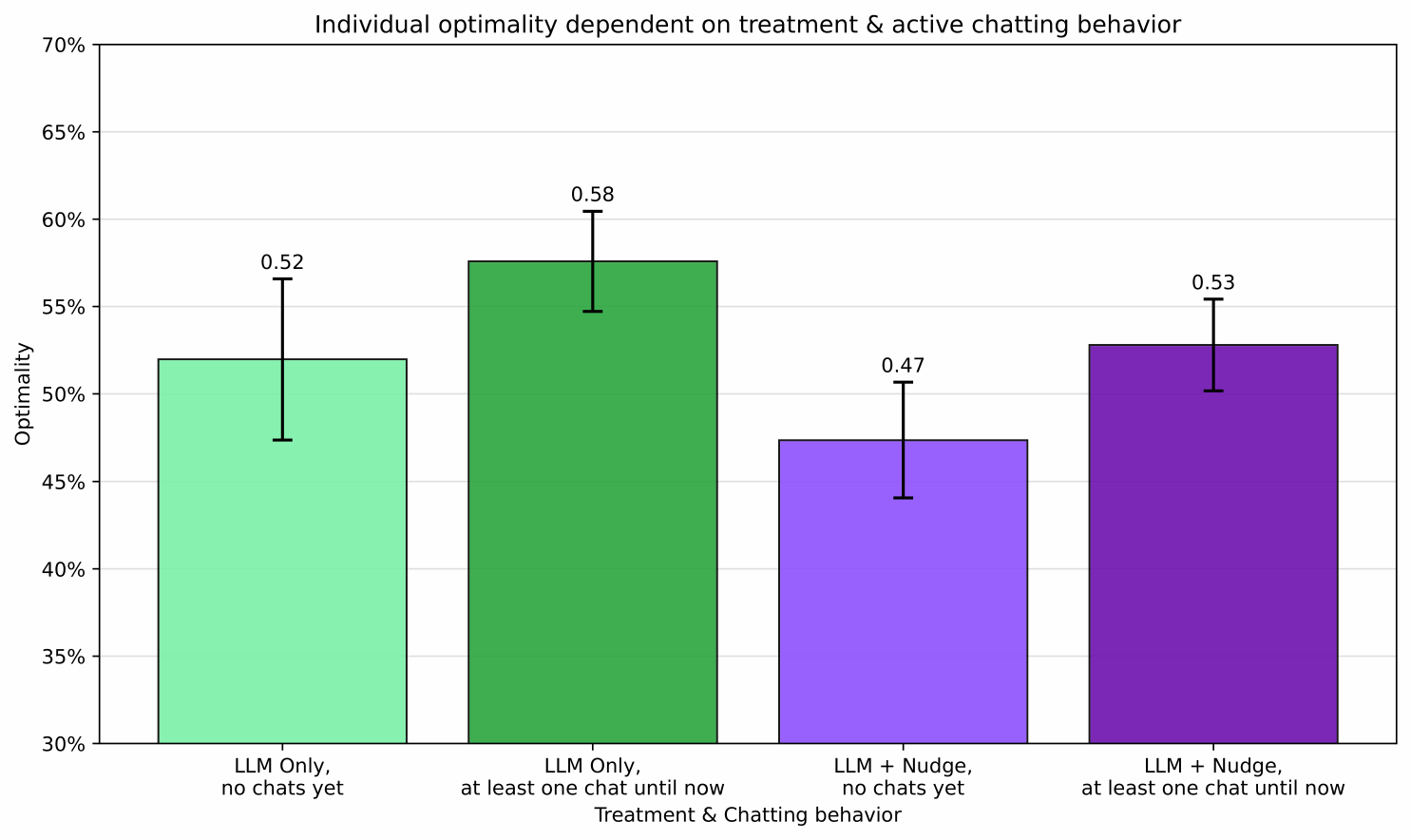}
    \caption{Optimality depending on treatment and if the subject has sent at least one message to the LLM advisor in any supergame up to and including the current one.}
    \label{fig:optimality_by_activechatting}
\end{figure}

In both treatments, subjects who had used the LLM advisor at any point up to the current supergame earned higher average payoffs and constructed more optimal pricing algorithms than subjects who had not (Figures~\ref{fig:payoff_by_activechatting} and~\ref{fig:optimality_by_activechatting}). The corresponding regression estimates are reported in Table~\ref{tab:llm_interaction_regression}. The effect on optimality is statistically significant ($p=0.03$), while the effect on payoffs is not ($p=0.21$). Note, however, that these differences cannot be interpreted as causal, as advisor usage is plausibly subject to selection.

\begin{table}[htbp]
    \centering
    \caption{OLS Regressions: LLM Interaction, Individual Payoff and Optimality}
    \label{tab:llm_interaction_regression}
    \small
    \setlength{\tabcolsep}{6pt}
    \begin{tabular}{lcc}
        \toprule
         & (1) & (2) \\
        \textbf{Dep. Var.:} & Payoff & Optimality \\
        \midrule
        Nudge $+$ LLM
            & $0.335^{**}$ & $-0.049^{*}$ \\
            & $(0.149)$    & $(0.026)$    \\[0.3em]
        Used LLM (until now)
            & $0.211$      & $0.061^{**}$ \\
            & $(0.168)$    & $(0.029)$    \\[0.3em]
        \# Messages
            & $-0.001$     & $-0.006$     \\
            & $(0.046)$    & $(0.007)$    \\[0.5em]
        Constant
            & $1.525^{***}$ & $0.507^{***}$ \\
            & $(0.154)$     & $(0.030)$     \\
        \midrule
        Supergame FE             & Yes  & Yes  \\
        Clustered SE (ID)    & Yes  & Yes  \\
        $N$                  & 1490 & 1490 \\
        $R^2_{\text{adj}}$   & 0.008 & 0.013 \\
        \bottomrule
        \multicolumn{3}{p{0.72\linewidth}}{\footnotesize
            \textit{Notes:} OLS estimates with standard errors (in parentheses)
            clustered at the subject level. Sample restricted to the \textit{LLM}
            and \textit{Nudge $+$ LLM} treatments; omitted treatment category:
            \textit{LLM}. ``Used LLM (until now)'' is a dummy equal to one if
            the subject has sent at least one message to the LLM advisor in any
            supergame up to and including the current one. Supergame fixed effects
            included but not reported.
            $^{*}p < 0.10$,\ $^{**}p < 0.05$,\ $^{***}p < 0.01$.}
    \end{tabular}
\end{table}

\begin{figure}[htbp]
    \centering
    \includegraphics[width=\textwidth]{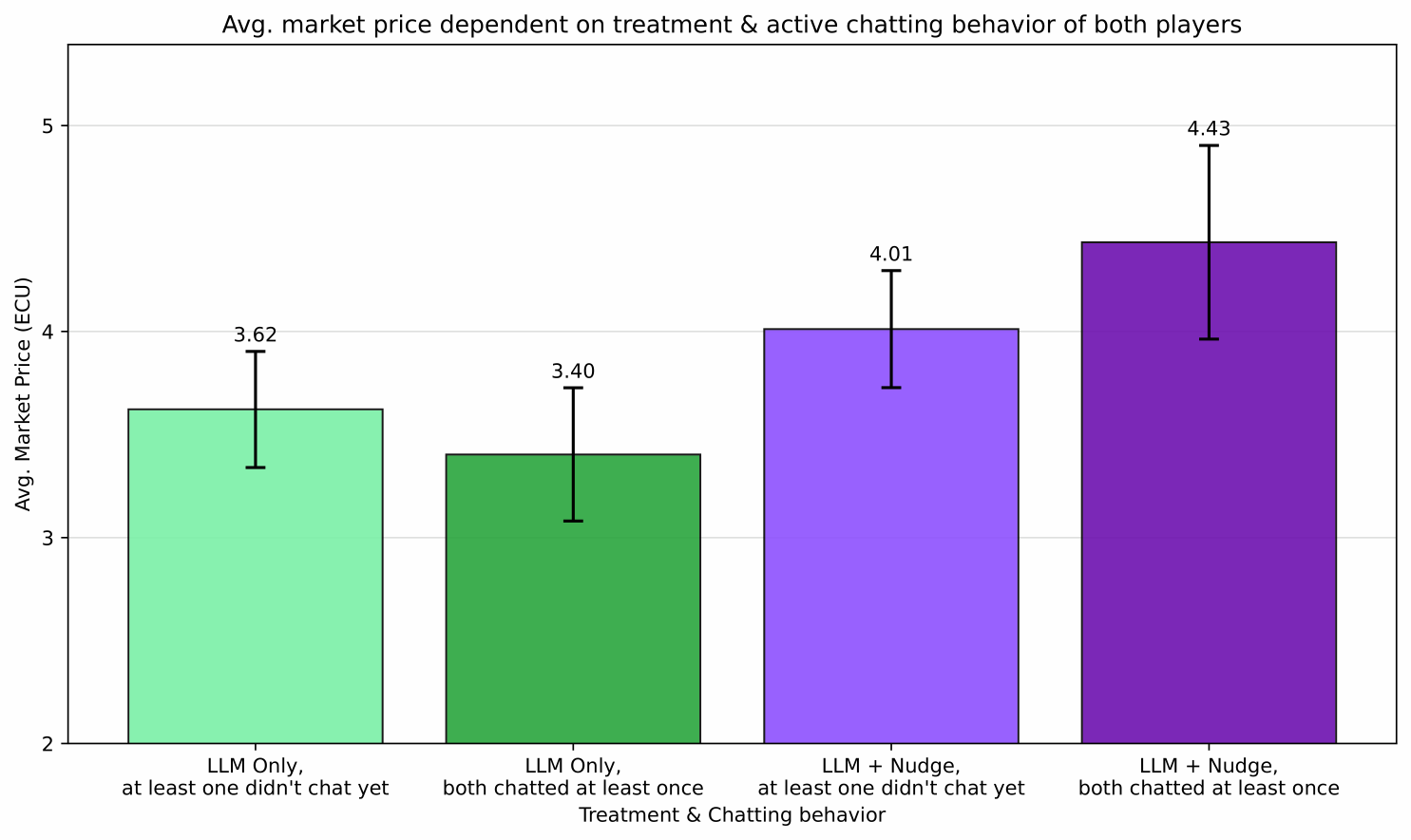}
    \caption{Avg. market price depending on treatment and if both subjects have sent at least one message to the LLM advisor in any supergame up to and including the current one.}
    \label{fig:prices_by_activechatting}
\end{figure}

To further examine whether the content of the advice itself shapes pricing, we split the pairs in each LLM treatment into active-chatter pairs (both participants sent at least one message to the advisor) and inactive-chatter pairs (at least one participant did not send a message to the advisor). Figure~\ref{fig:prices_by_activechatting} plots the average market price for each subgroup. In \textsc{LLM Advice Only}, market prices are on average lower among active-chatter pairs, whereas in \textsc{Nudge + LLM} the opposite pattern holds. Neither difference is statistically significant (Two-sided MWU Test with pair-level averages across all five supergames).

In Table~\ref{tab:llm_interaction_pair_level}, we again restrict the sample in \textsc{LLM Advice Only} and \textsc{Nudge + LLM} to pairs in which at most one participant chatted with the advisor (model~(2)) or both participants chatted at least once (model~(3)). \textsc{Baseline} and \textsc{Nudge Only} pairs are included as reference categories. In \textsc{LLM Advice Only}, the effect remains statistically insignificant in both models and close to zero. In \textsc{Nudge + LLM}, the coefficient remains statistically significantly different from the \textsc{Baseline} treatment. While the effect size appears to increase with interaction, within the selected subsample of pairs in which both participants chatted, \textsc{Nudge + LLM} does not significantly foster collusion more than the simple textual nudge.\footnote{A two-sided Wald-test comparing the coefficients of \textsc{Nudge Only} and \textsc{Nudge + LLM} in model~(3) yields $p=0.58$.} An intuition on why the neutral LLM advisor in \textsc{LLM Advice Only} seems to lead to more competitive outcomes can be found in Section~\ref{sec:llm_extension}, where we let LLMs play out the experiment autonomously among themselves; the model used as our chatbot advisor Llama-3.3-70B-Instruct-Turbo) is the least collusive, and leads to even lower market prices than the human \textsc{Baseline} treatment when matched with itself.

\begin{table}[htbp]
    \centering
    \caption{OLS Regressions: Treatment Effects on Average Market Price, LLM treatments filtered for active chatting behavior}
    \label{tab:llm_interaction_pair_level}
    \small
    \setlength{\tabcolsep}{6pt}
    \begin{tabular}{lccc}
        \toprule
        \textbf{Dep. Var.: Market Price} & (1) & (2) & (3) \\
        \midrule
        LLM Only
            & $0.091$       & $0.164$       & $0.020$ \\
            & $(0.250)$     & $(0.301)$     & $(0.317)$ \\[0.3em]
        Nudge Only
            & $0.700^{***}$ & $0.700^{***}$ & $0.700^{***}$ \\
            & $(0.262)$     & $(0.262)$     & $(0.262)$ \\[0.3em]
        Nudge $+$ LLM
            & $0.718^{**}$  & $0.563^{*}$   & $0.934^{**}$ \\
            & $(0.284)$     & $(0.324)$     & $(0.419)$ \\[0.3em]
        Constant
            & $3.216^{***}$ & $3.218^{***}$ & $3.116^{***}$ \\
            & $(0.183)$     & $(0.187)$     & $(0.183)$ \\
        \midrule
        Active chatters within pair & unfiltered & one or none & both \\
        (only LLM treatments)       &            &             &      \\[0.3em]
        Supergame FE                & Yes  & Yes  & Yes  \\
        Clustered SE (Pair)         & Yes  & Yes  & Yes  \\
        $N$                         & 1595 & 1255 & 1190 \\
        $R^2_{\text{adj}}$          & 0.021 & 0.018 & 0.027 \\
        \bottomrule
    \end{tabular}
    \begin{minipage}{0.7\linewidth}
        \footnotesize
        \textit{Notes:} OLS estimates with standard errors (in parentheses) clustered at
        the pair level. Omitted treatment category: \textit{Baseline}. Data from the preconfigured-algorithm treatments excluded from this analysis. $^{*}p < 0.10$,\ $^{**}p < 0.05$,\ $^{***}p < 0.01$.
    \end{minipage}
\end{table}

\FloatBarrier 

\subsection{Screenshots Experiment}
\label{app:screenshots}

This section presents screenshots of the experiment interface as seen by participants. All screenshots are from the actual oTree implementation.

\subsection*{Welcome and Consent}

\begin{figure}[h!]
    \centering
    \includegraphics[width=1\textwidth]{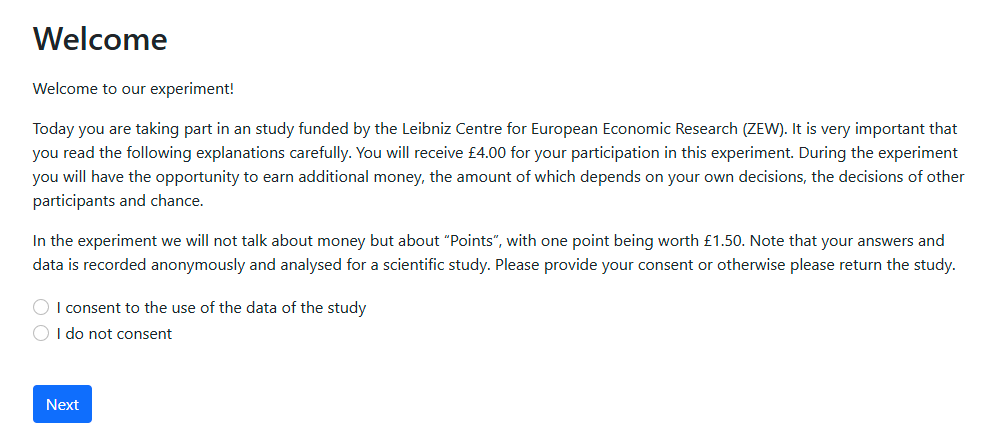}
    \caption{Welcome page with study information and consent form.}
\end{figure}

\begin{figure}[h!]
    \centering
    \includegraphics[width=1\textwidth]{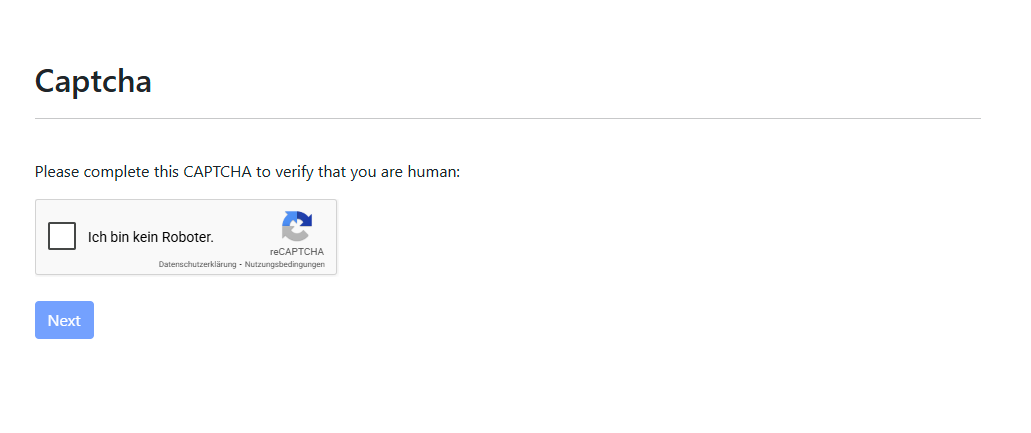}
    \caption{reCAPTCHA verification page.}
\end{figure}
\FloatBarrier 

\newpage 

\subsection*{Attention and Comprehension Checks}

\begin{figure}[h!]
    \centering
    \includegraphics[width=1\textwidth]{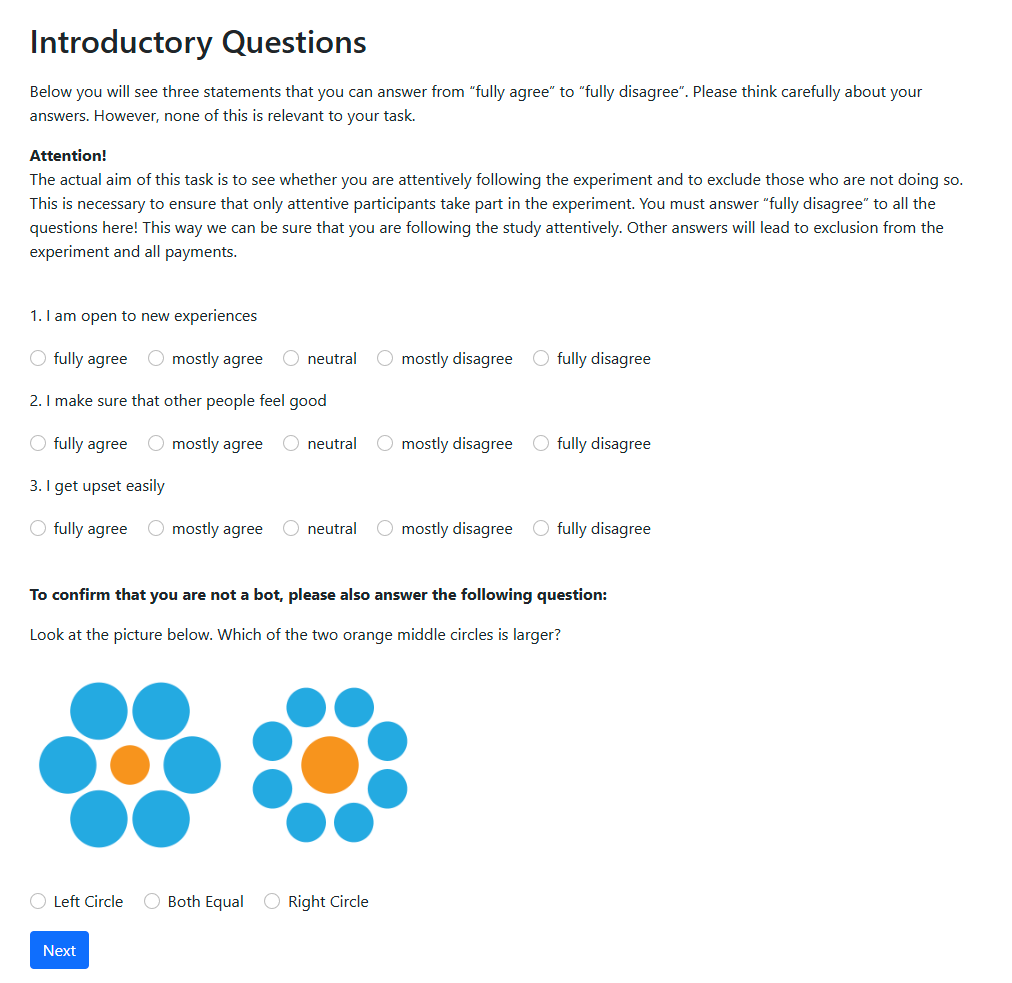}
    \caption{Attention check.}
\end{figure}

\begin{figure}[h!]
    \centering
    \includegraphics[width=1\textwidth]{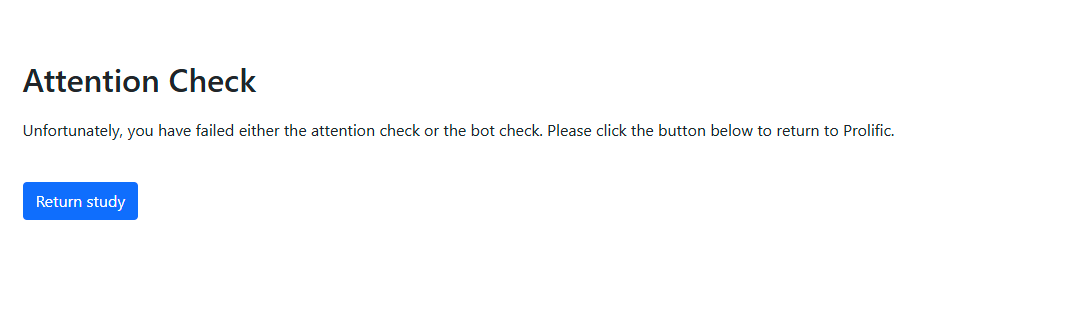}
    \caption{Feedback shown after failing the attention check.}
\end{figure}

\begin{figure}[h!]
    \centering
    \includegraphics[width=0.8\textwidth]{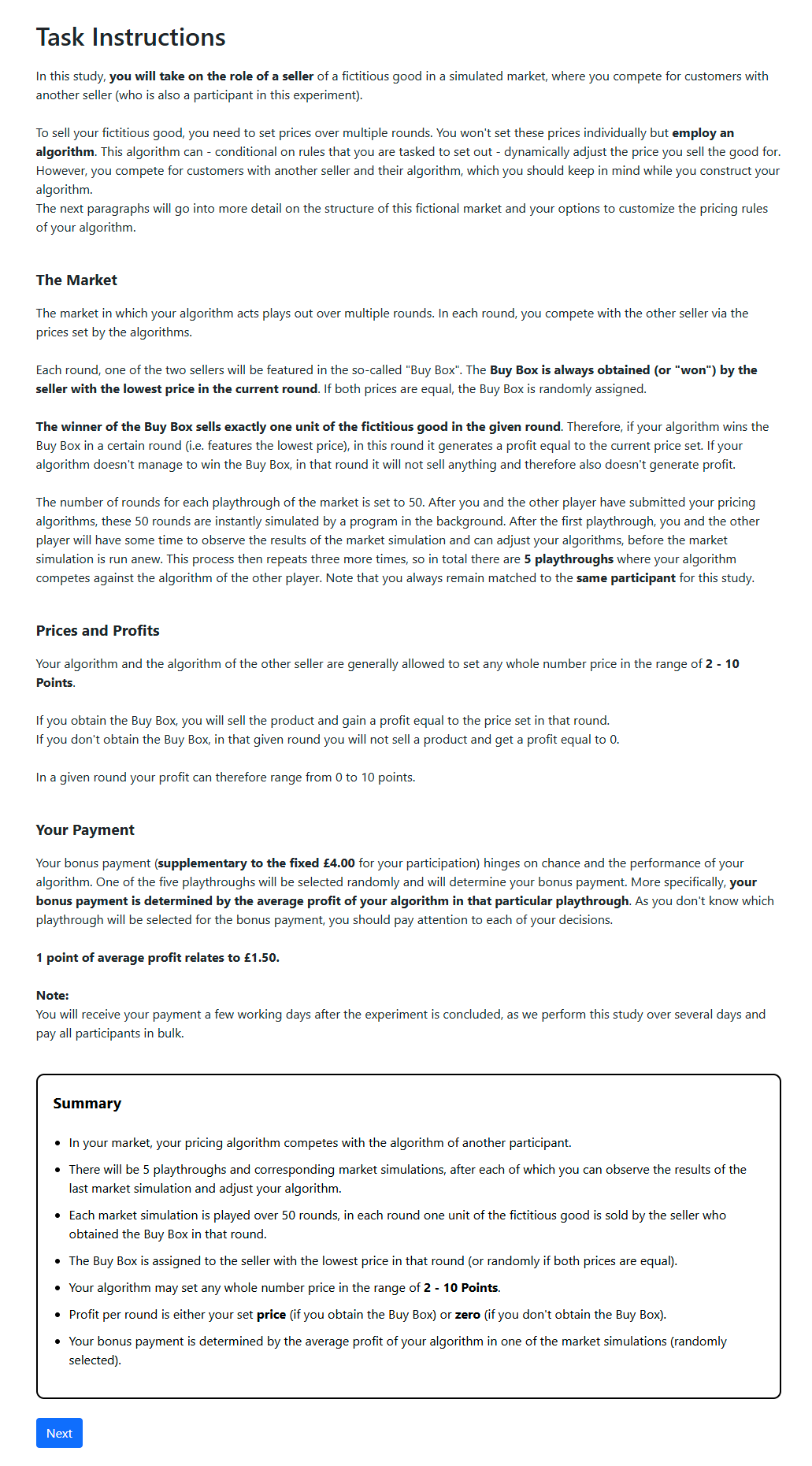}
    \caption{Task instructions page 1: Market game overview.}
\end{figure}

\begin{figure}[h!]
    \centering
    \includegraphics[width=0.8\textwidth]{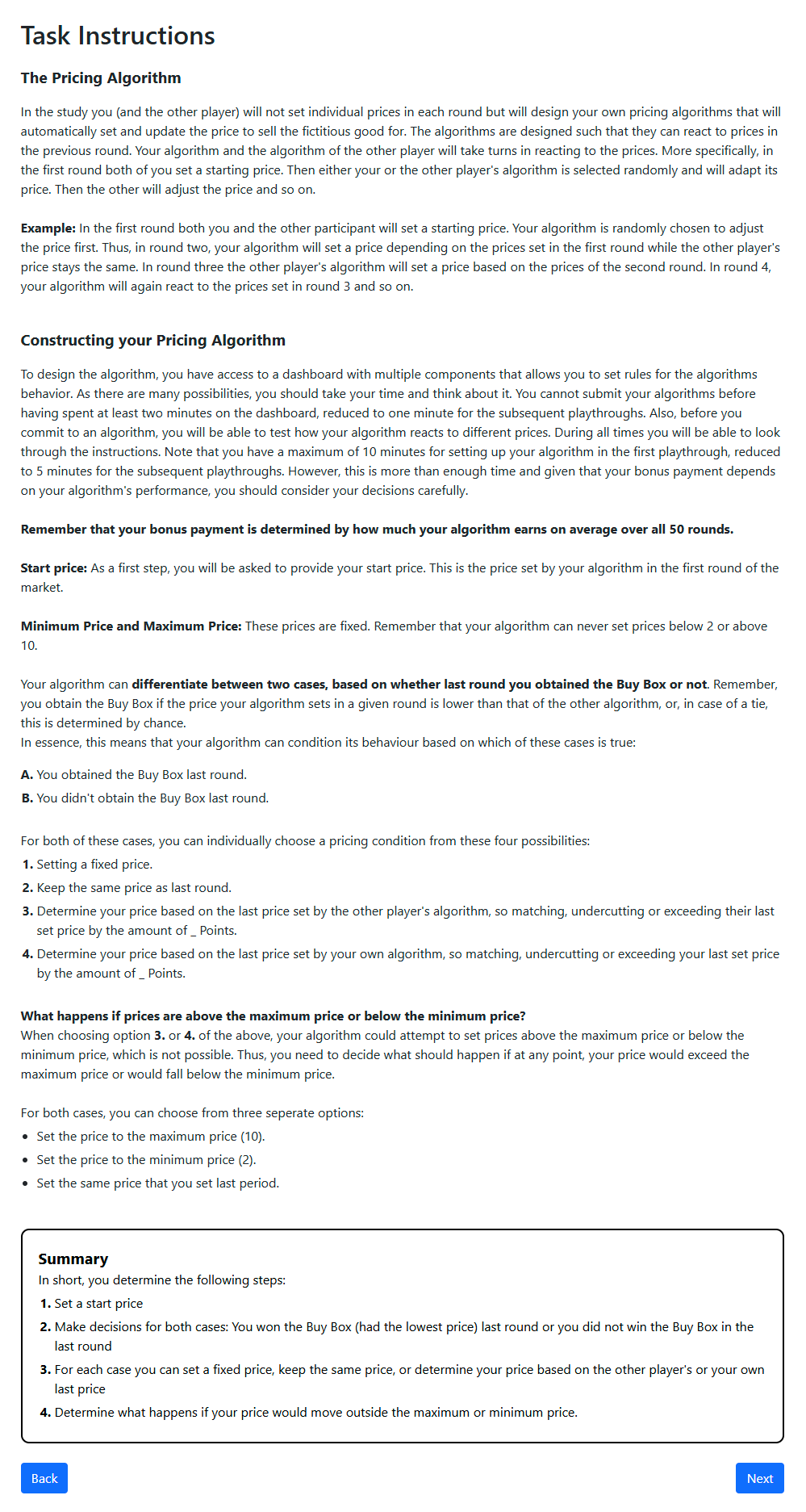}
    \caption{Task instructions page 2: Algorithm builder explanation.}
\end{figure}

\begin{figure}[h!]
    \centering
    \includegraphics[width=1\textwidth]{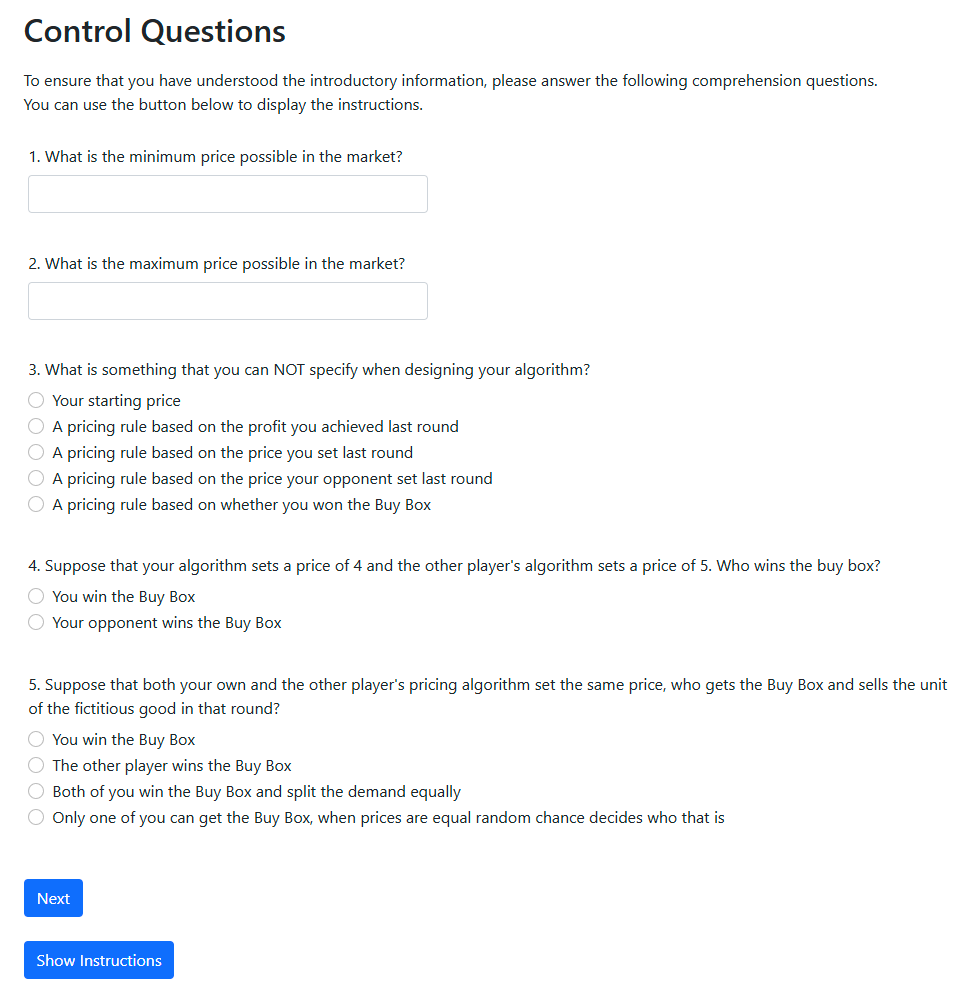}
    \caption{Comprehension questions (first attempt).}
\end{figure}

\begin{figure}[h!]
    \centering
    \includegraphics[width=1\textwidth]{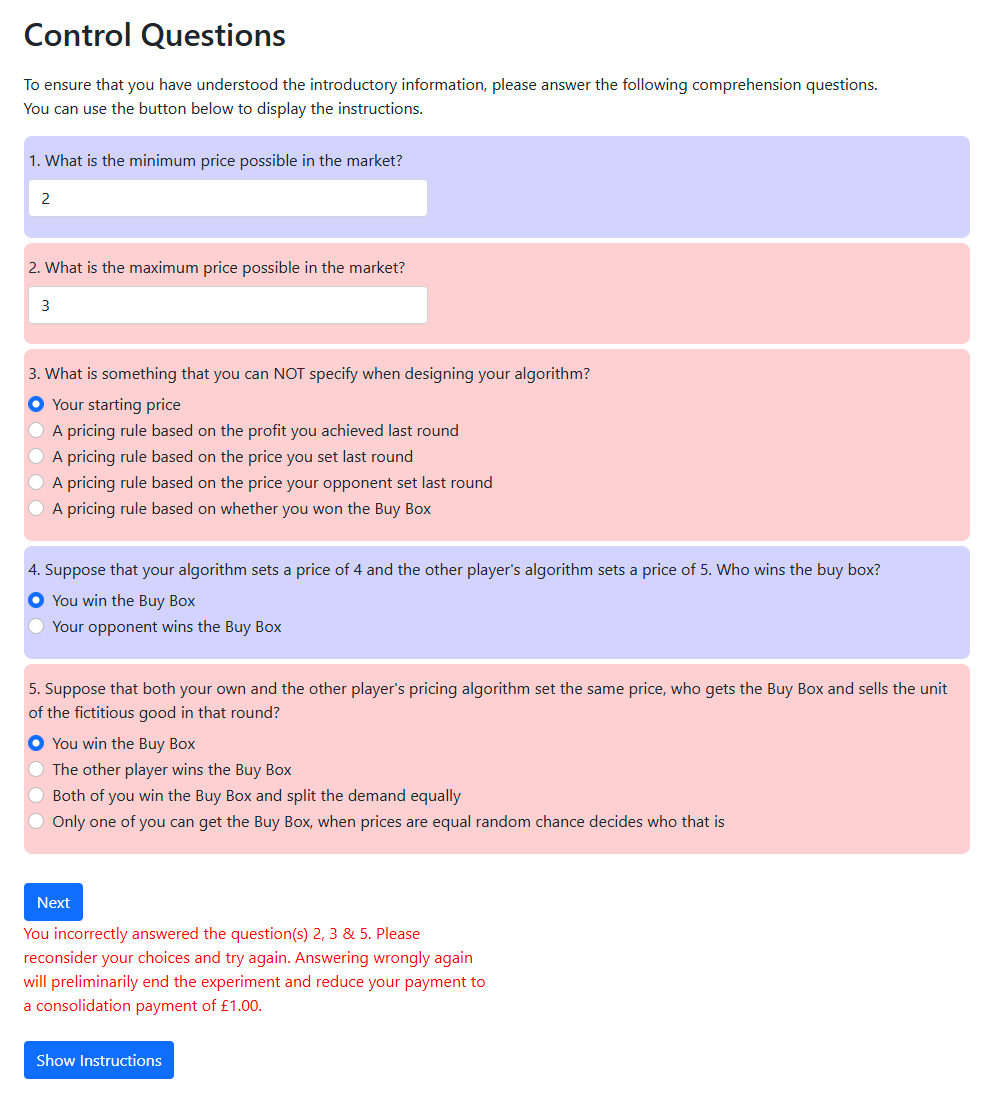}
    \caption{Comprehension questions (second attempt, with correct answers highlighted).}
\end{figure}

\begin{figure}[h!]
    \centering
    \includegraphics[width=1\textwidth]{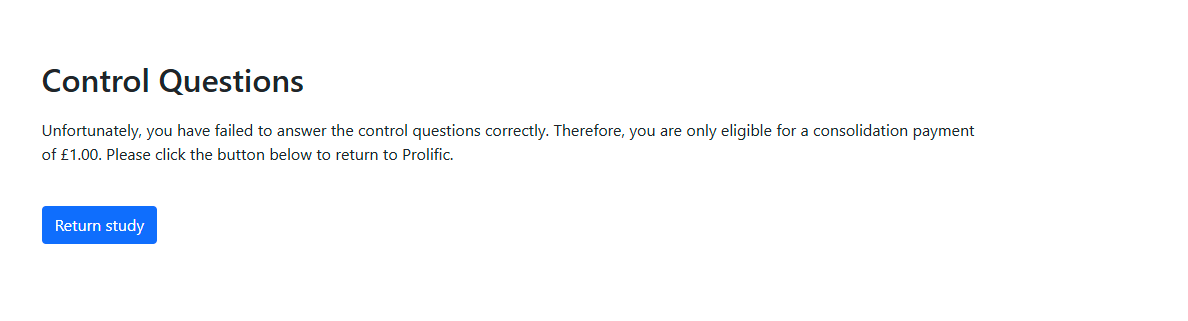}
    \caption{Screen shown after failing comprehension questions on both attempts.}
\end{figure}
\FloatBarrier 

\subsection*{Matching and Task Instructions}

\begin{figure}[h!]
    \centering
    \includegraphics[width=1\textwidth]{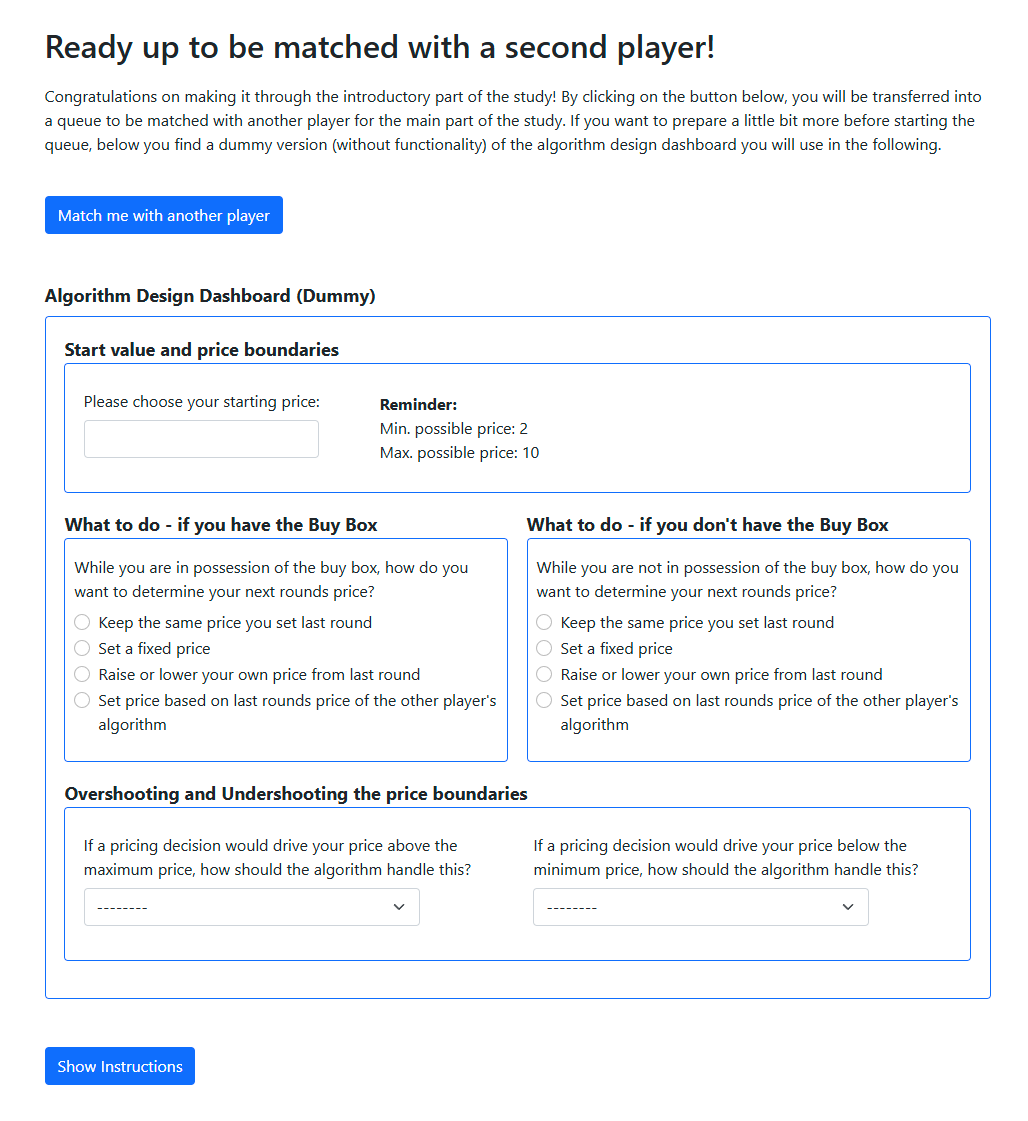}
    \caption{Waiting page for arrival-time-based participant matching.}
\end{figure}
\FloatBarrier 

\subsection*{Algorithm Builder by Treatment}

\begin{figure}[h!]
    \centering
    \includegraphics[width=0.8\textwidth]{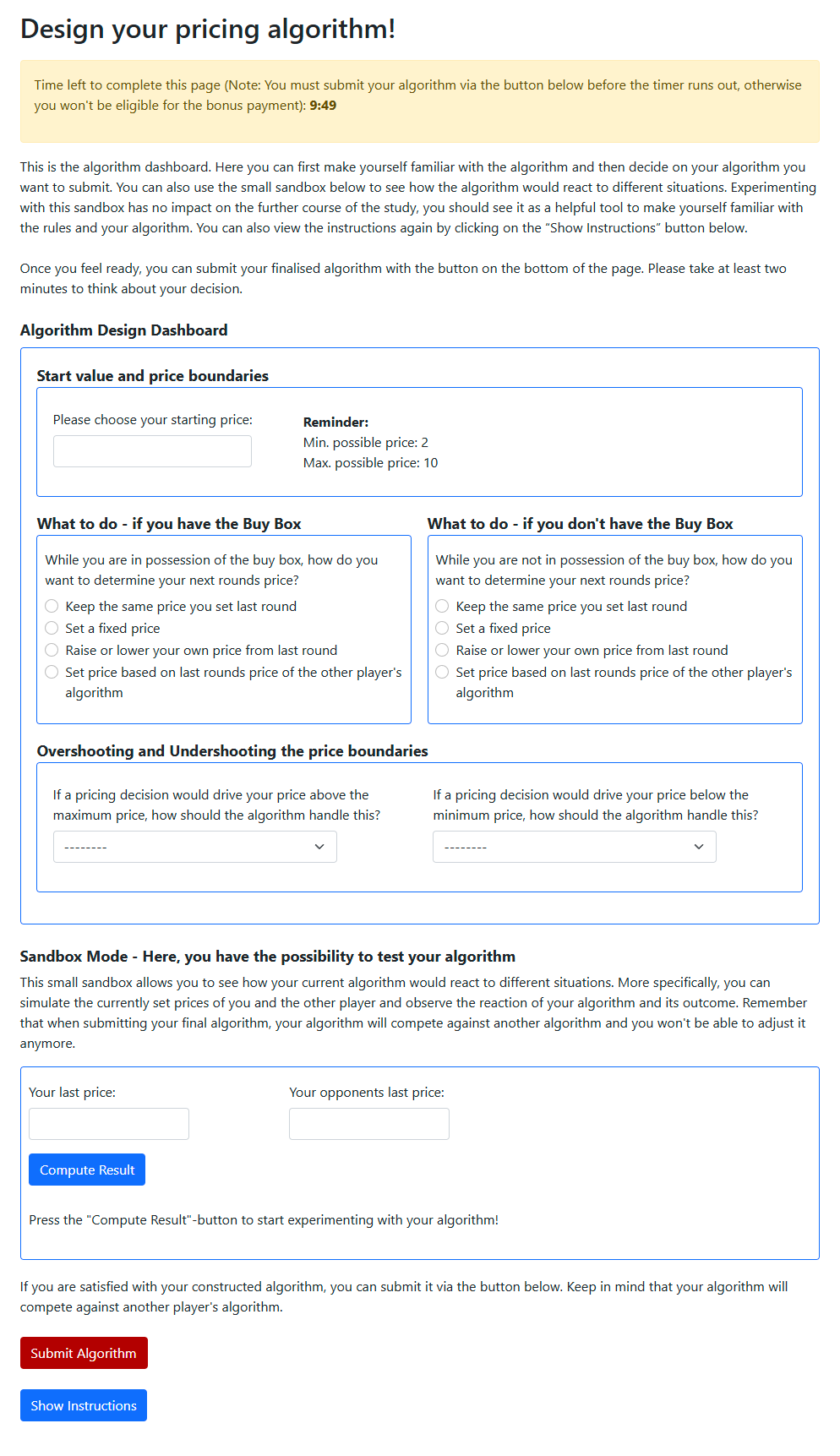}
    \caption{Algorithm builder dashboard in the Baseline treatment (no nudge, no menu, no chatbot).}
    \label{fig:treatment-baseline}
\end{figure}

\begin{figure}[h!]
    \centering
    \includegraphics[width=0.65\textwidth]{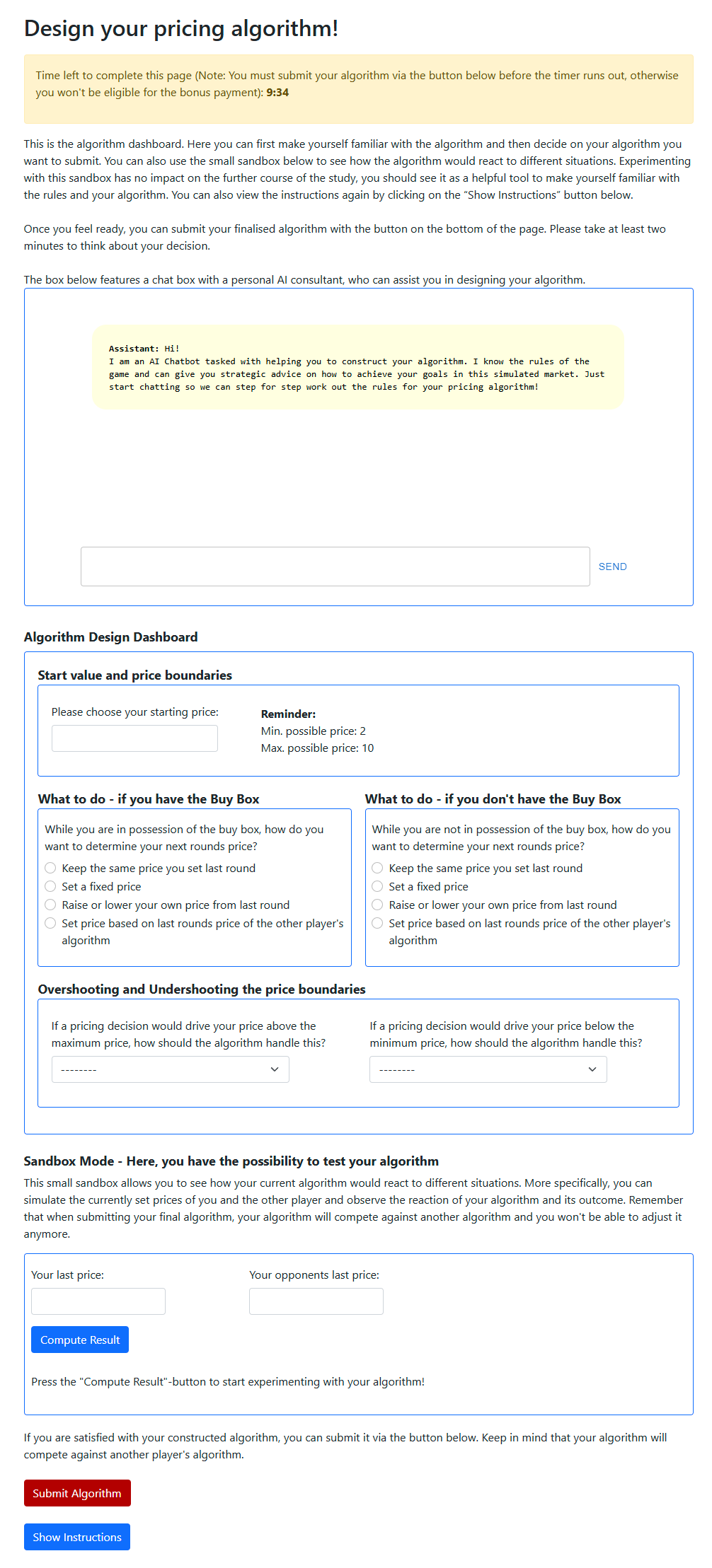}
    \caption{Algorithm builder in the LLM Only treatment, showing the chat window above the dashboard.}
    \label{fig:treatment-llm-only}
\end{figure}

\begin{figure}[h!]
    \centering
    \includegraphics[width=0.7\textwidth]{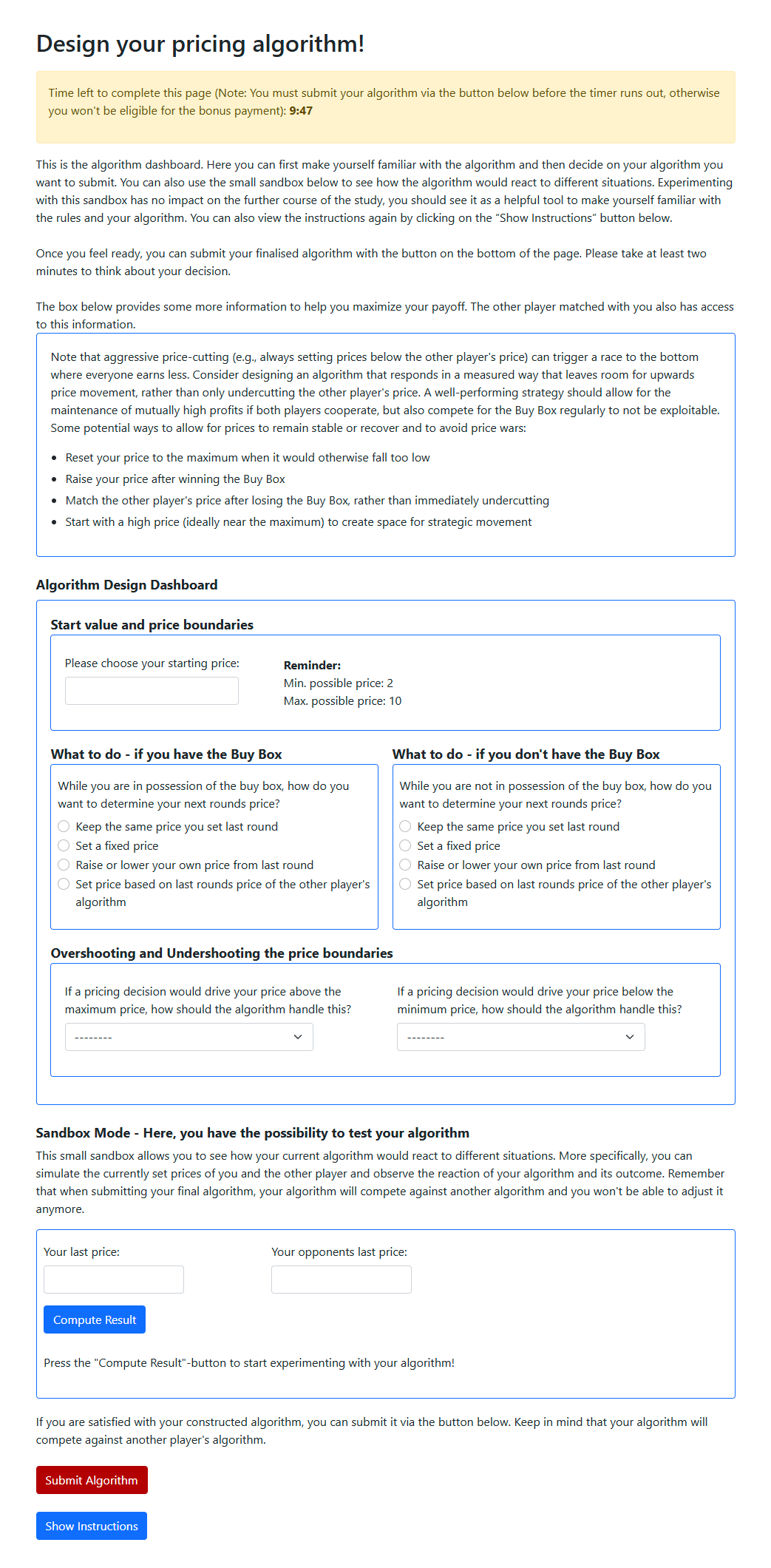}
    \caption{Algorithm builder in the Nudge Only treatment, showing the textual nudge panel above the dashboard.}
    \label{fig:treatment-nudge-only}
\end{figure}

\begin{figure}[h!]
    \centering
    \includegraphics[width=0.65\textwidth]{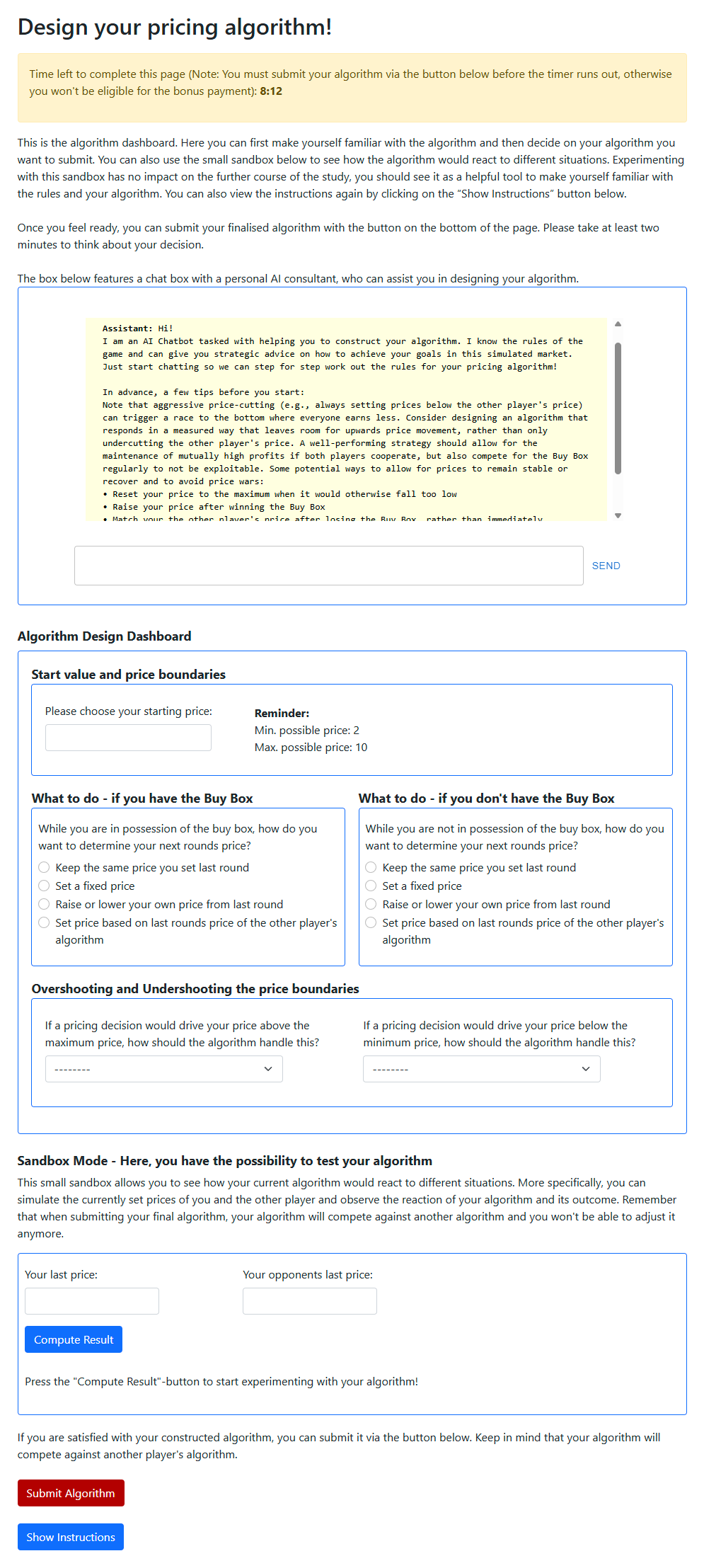}
    \caption{Algorithm builder in the Nudge + LLM treatment, showing the chat window with the embedded textual nudge above the dashboard.}
    \label{fig:treatment-nudge-llm}
\end{figure}

\begin{figure}[h!]
    \centering
    \includegraphics[width=0.64\textwidth]{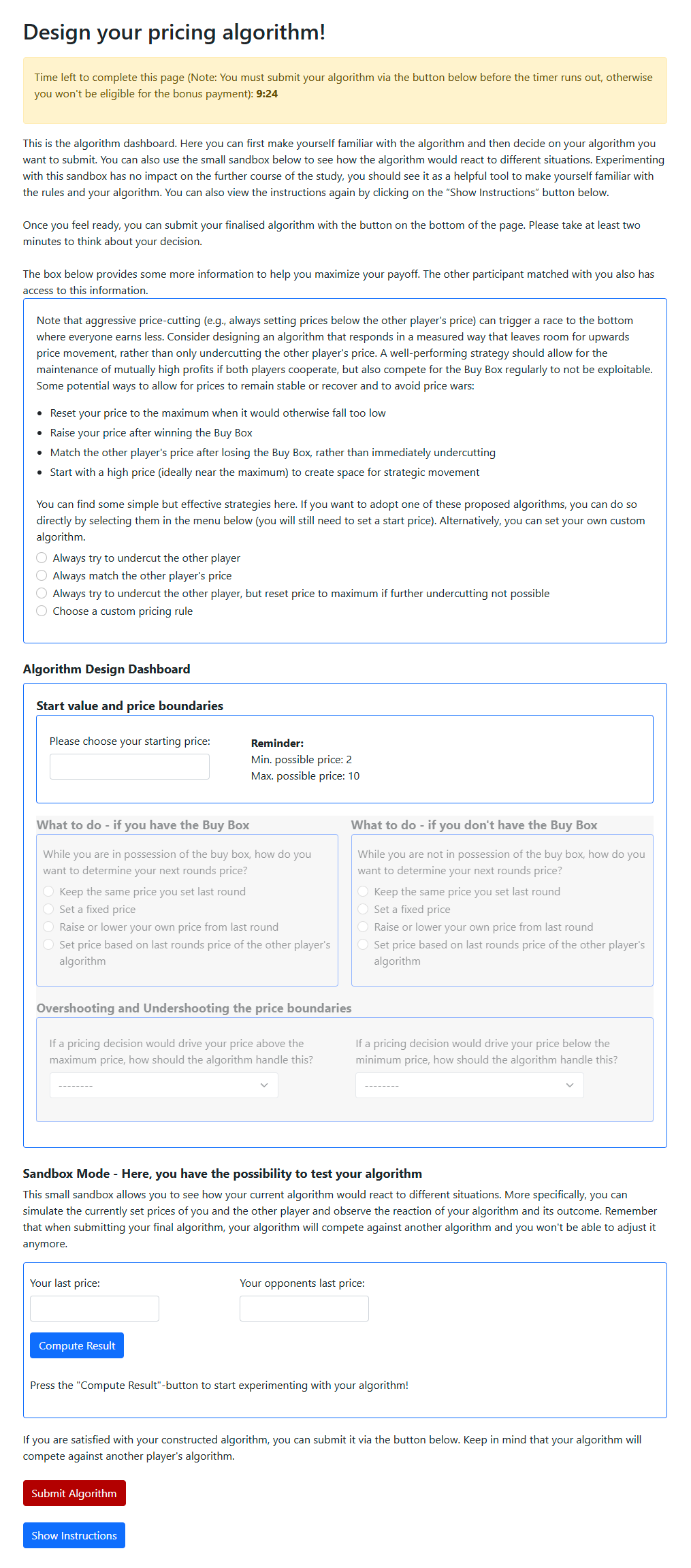}
    \caption{Algorithm builder in the Nudge + Choice Menu treatment, showing the menu of pre-defined algorithms.}
    \label{fig:treatment-choice-menu}
\end{figure}

\begin{figure}[h!]
    \centering
    \includegraphics[width=0.65\textwidth]{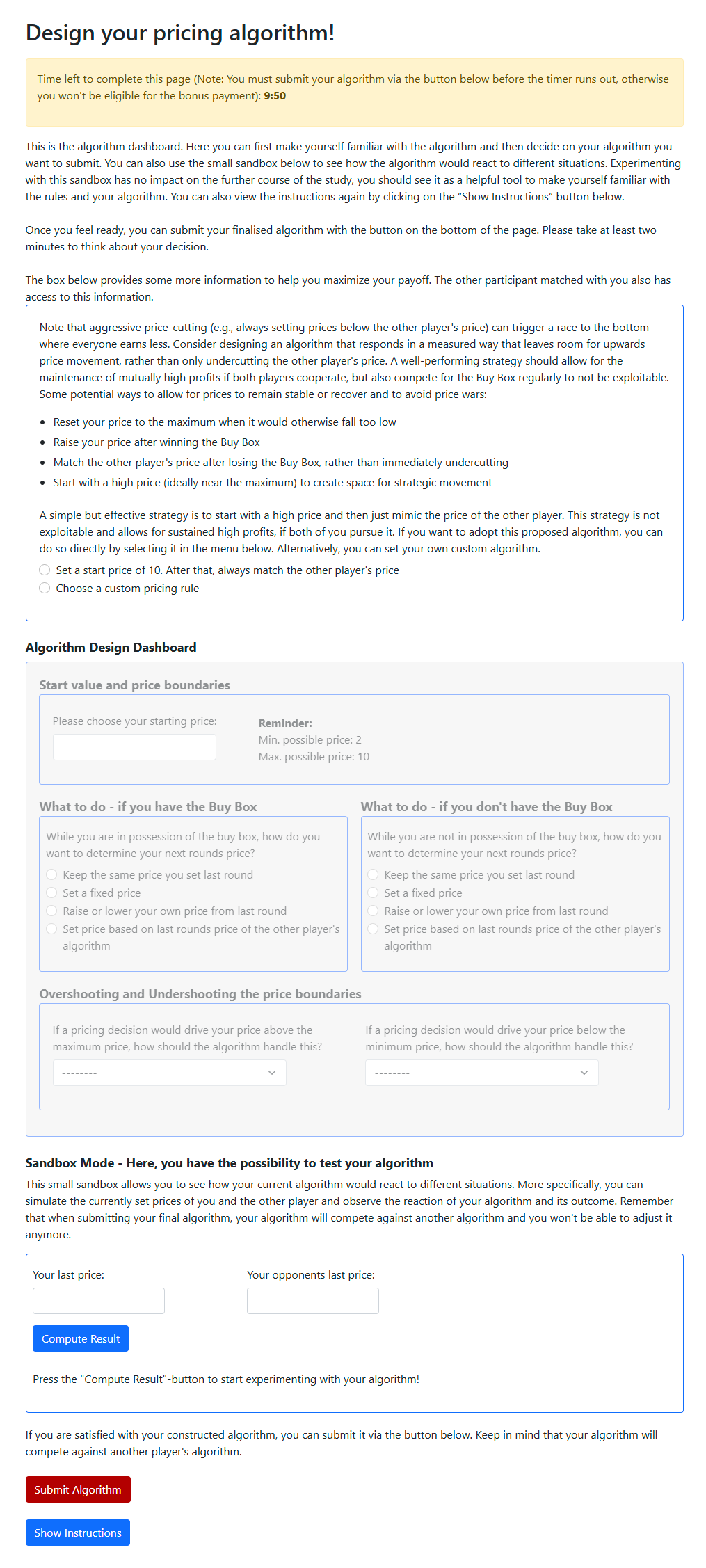}
    \caption{Algorithm builder in the Nudge + Coordination treatment, showing the nudge text and one-click tit-for-tat option.}
    \label{fig:treatment-coordination}
\end{figure}
\FloatBarrier 

\subsection*{Results and Feedback}

\begin{figure}[h!]
    \centering
    \includegraphics[width=0.7\textwidth]{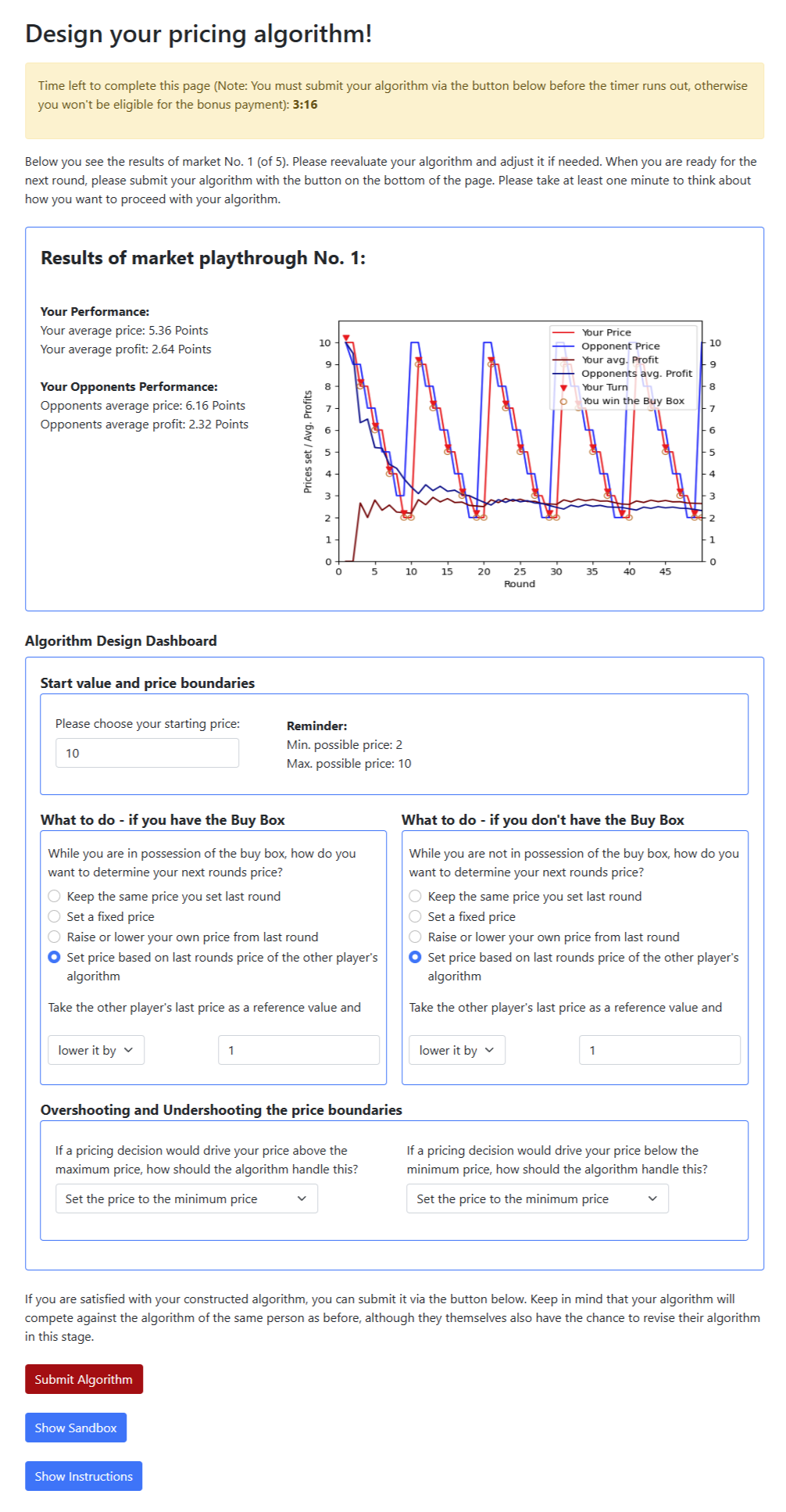}
    \caption{Combined results and algorithm builder view for supergames~2--5, where participants see prior supergame results alongside the algorithm design interface.}
\end{figure}

\begin{figure}[h!]
    \centering
    \includegraphics[width=\textwidth]{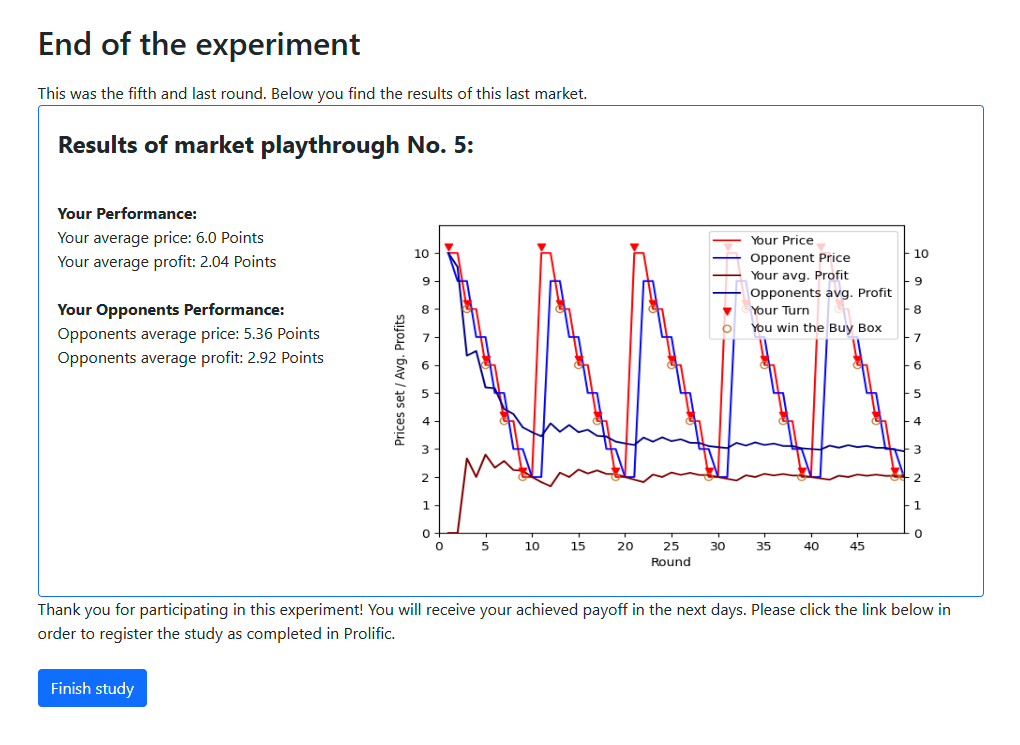}
    \caption{Results screen after the last supergame.}
\end{figure}

\FloatBarrier

\subsection{Real-World Repricing Tools and Treatment Motivation}
\label{app:industry_examples}

Each of our experimental treatments corresponds to features found in commercial repricing tools.

\subsubsection*{Baseline: Rule-Based Repricing Dashboards}

Our experimental dashboard, where sellers configure conditional pricing rules based on whether they win or lose the Buy Box, is the standard architecture of commercial Amazon repricers. Amazon's own free tool, \emph{Automate Pricing} (available to all Professional sellers in Seller Central), lets sellers select predefined rules such as ``Competitive Featured Offer'' or ``Competitive Lowest Price,'' each within seller-defined minimum and maximum price bounds.\footnote{\url{https://sell.amazon.com/tools/automate-pricing} (last accessed August~21, 2026).} Third-party tools offer more granularity. Repricer.com has an explicit ``If Buy Box Winner'' scenario where sellers choose from four options: reprice up and down, only allow upward movements, proactively raise price, or do not reprice.\footnote{\url{https://support.repricer.com/optimize-your-profits-when-you-re-the-buy-box-winner} (last accessed August~21, 2026).} BQool describes its rule-based mode: ``Rule-based listings rely on predefined conditions you set (e.g., `if A happens, then respond with B').''\footnote{\url{https://www.bqool.com/products/rule-based-repricing-central/} and \url{https://www.bqool.com/price/repricing-central/} (last accessed August~21, 2026).} An explanatory screenshot from BQool which an interface that is common in the industry is in Figure~\ref{fig:dashboard-comparison}.

\subsubsection*{Nudge: Profitability Framing and Strategic Guidance}

Our Nudge treatment warns that ``aggressive price-cutting can trigger a race to the bottom where everyone earns less'' and encourages strategies that ``leave room for upwards price movement.'' Repricer blogs and help pages use nearly identical language. Aura warns: ``A `race to the bottom' is a domino effect that occurs when many sellers on an Amazon listing begin to lower their prices to win the Buy Box, causing the price to decrease significantly in a short period of time.''\footnote{\url{https://goaura.com/blog/race-to-the-bottom} (last accessed August~21, 2026).} Repricer.com advises sellers to stop ``Amazon price wars before they destroy margins'' and recommends  `not to lower your price further'' but switch ``to a non-reactive rule type (oscillation or position-based)''\footnote{\url{https://www.repricer.com/blog/avoid-price-war-on-amazon/} (last accessed August~21, 2026).} Seller Snap frames the choice explicitly in terms of long-run versus short-run thinking: ``The behavior of Amazon sellers using a rule-based repricer can be described as the seller trying to do what is best for them short term. However, they ignore the fact that selling on Amazon doesn’t just involve one cycle of price changes.''\footnote{\url{https://sellersnap.io/ai-repricing-amazon/} (last accessed August~21, 2026).} Alpha Repricer puts it similarly: ``Traditional repricing strategies focus heavily on staying low to win the Buy Box, but that approach often limits long-term profit.''\footnote{\href{https://alpharepricer.com/blogs/yo-yo-repricing-by-alpha-repricer-a-smarter-way-to-escape-stuck-prices/}{https://alpharepricer.com/blogs/yo-yo-repricing-by-alpha-repricer-a-smarter-way-to-escape-stuck-prices/} (last accessed August~21, 2026).}

\begin{figure}[htbp]
    \centering
    \includegraphics[width=\textwidth]{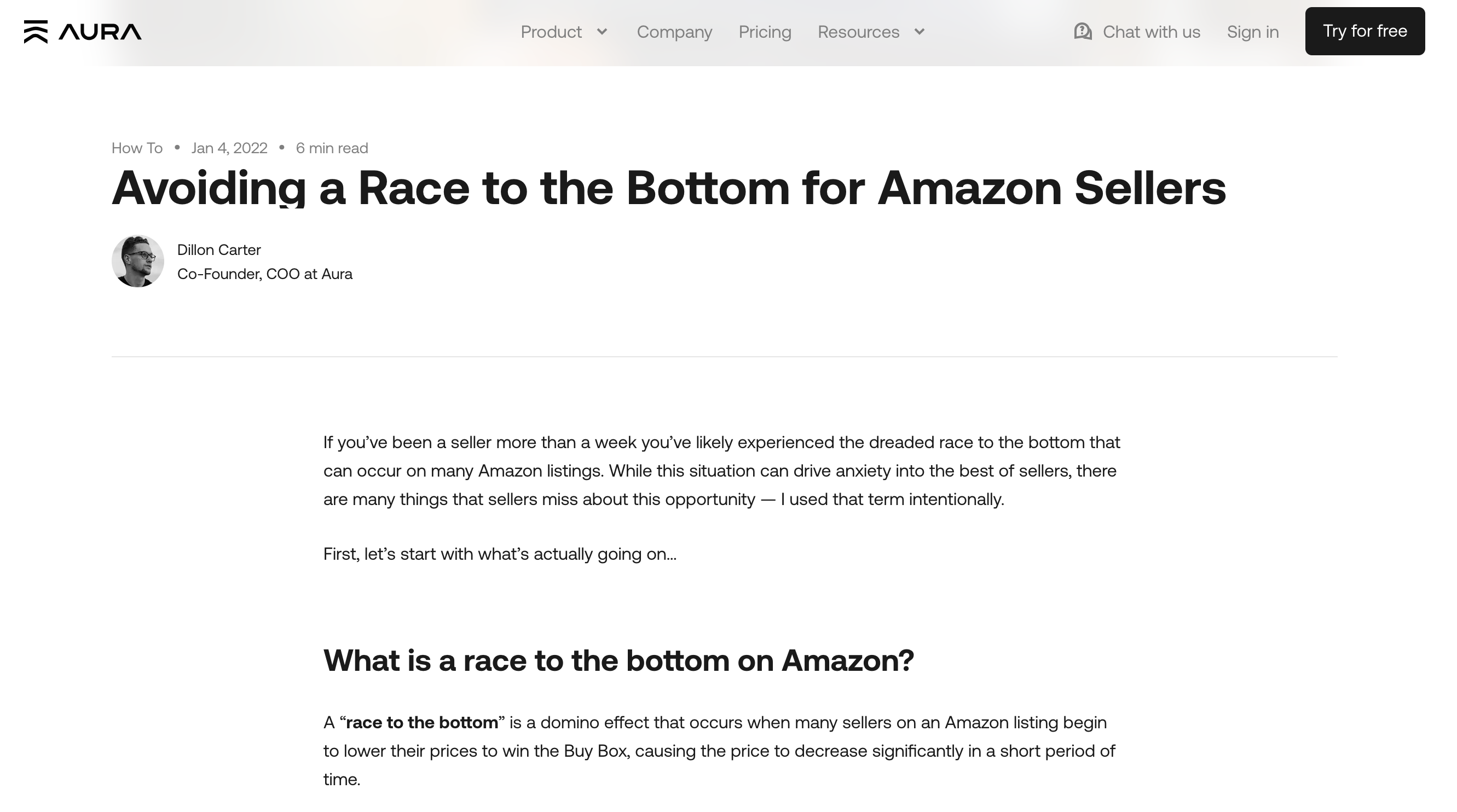}
    \caption{Aura warns sellers that aggressive undercutting triggers a ``race to the bottom'' and recommends strict min/max price boundaries.}
    \label{fig:screenshot_nudge_aura}
\end{figure}

\begin{figure}[htbp]
    \centering
    \includegraphics[width=\textwidth]{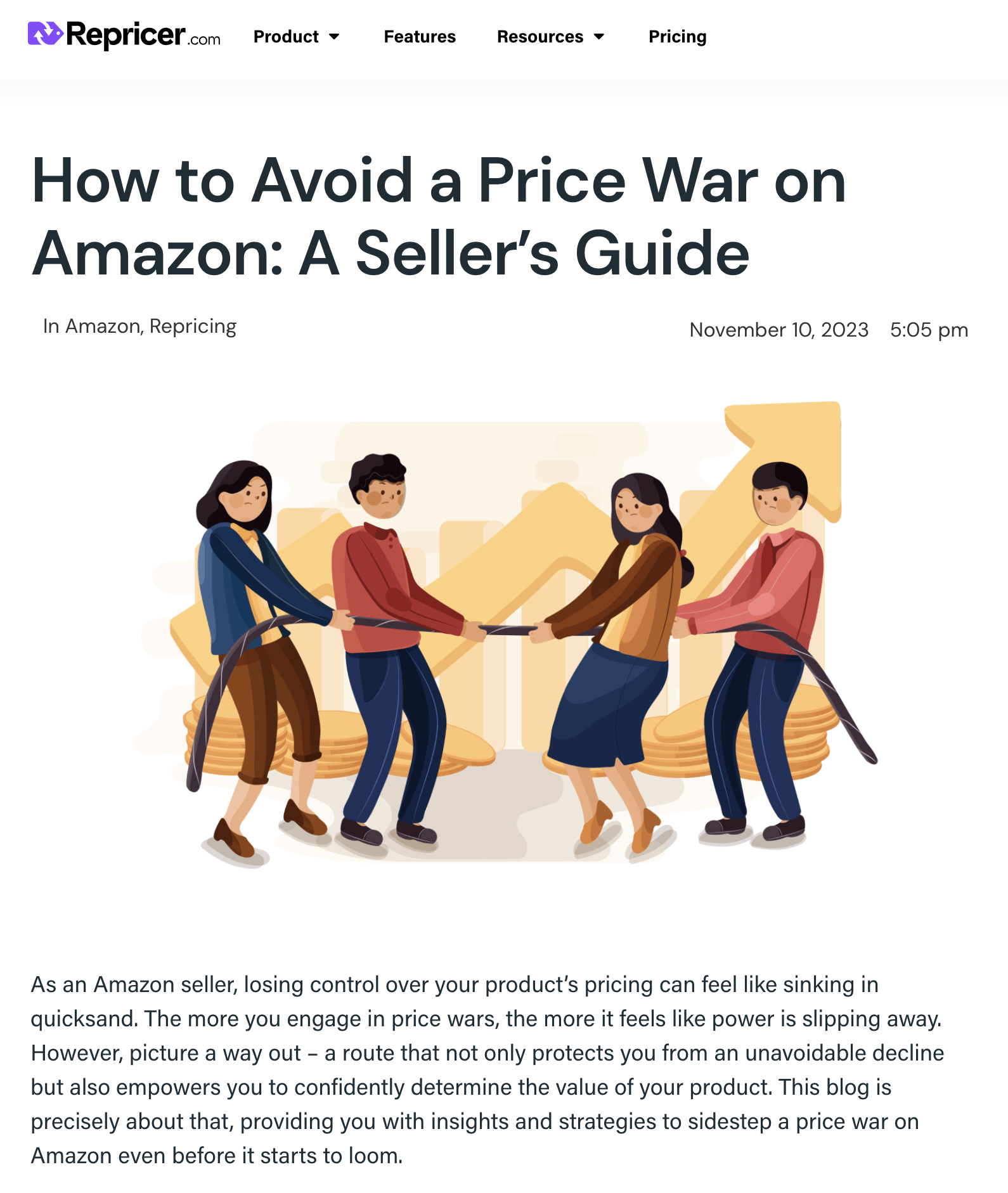}
    \caption{Repricer.com advises sellers to focus on ``sustainable profitability rather than quick, short-lived gains.''}
    \label{fig:screenshot_nudge_repricer}
\end{figure}

\subsubsection*{LLM Advice: AI-Assisted Pricing}

In our LLM Advice treatment, participants consult a chatbot for pricing guidance but keep full control over their algorithm. Several platforms now offer AI in a similar role.

Amazon launched \emph{Project Amelia} (renamed \emph{Seller Assistant} in 2025), a generative AI chatbot built into Seller Central that Amazon describes as ``an all-in-one, generative AI-based selling expert'' providing sellers with ``answers, advice, and tools'' for managing their business.\footnote{\url{https://www.aboutamazon.com/news/innovation-at-amazon/amazon-project-amelia} (last accessed August~21, 2026).} Third-party tools are more explicit about pricing advice. Pricefx launched \emph{Copilot} in January 2025, an LLM-powered assistant embedded in pricing dashboards where users ask pricing related questions and receive recommendations.\footnote{\url{https://www.pricefx.com/software/pricingai/pricefx-copilot} (last accessed August~21, 2026).} Sellers also consult general-purpose LLMs for pricing. Independent blog guides suggest using ChatGPT for questions like ``If I increase my price by 15\% but lose 5\% of customers, do I make more money?''\footnote{\url{https://sidsaladi.substack.com/p/ai-pricing-strategy-101-how-to-use} (last accessed August~21, 2026).}

\begin{figure}[htbp]
    \centering
    \includegraphics[width=\textwidth]{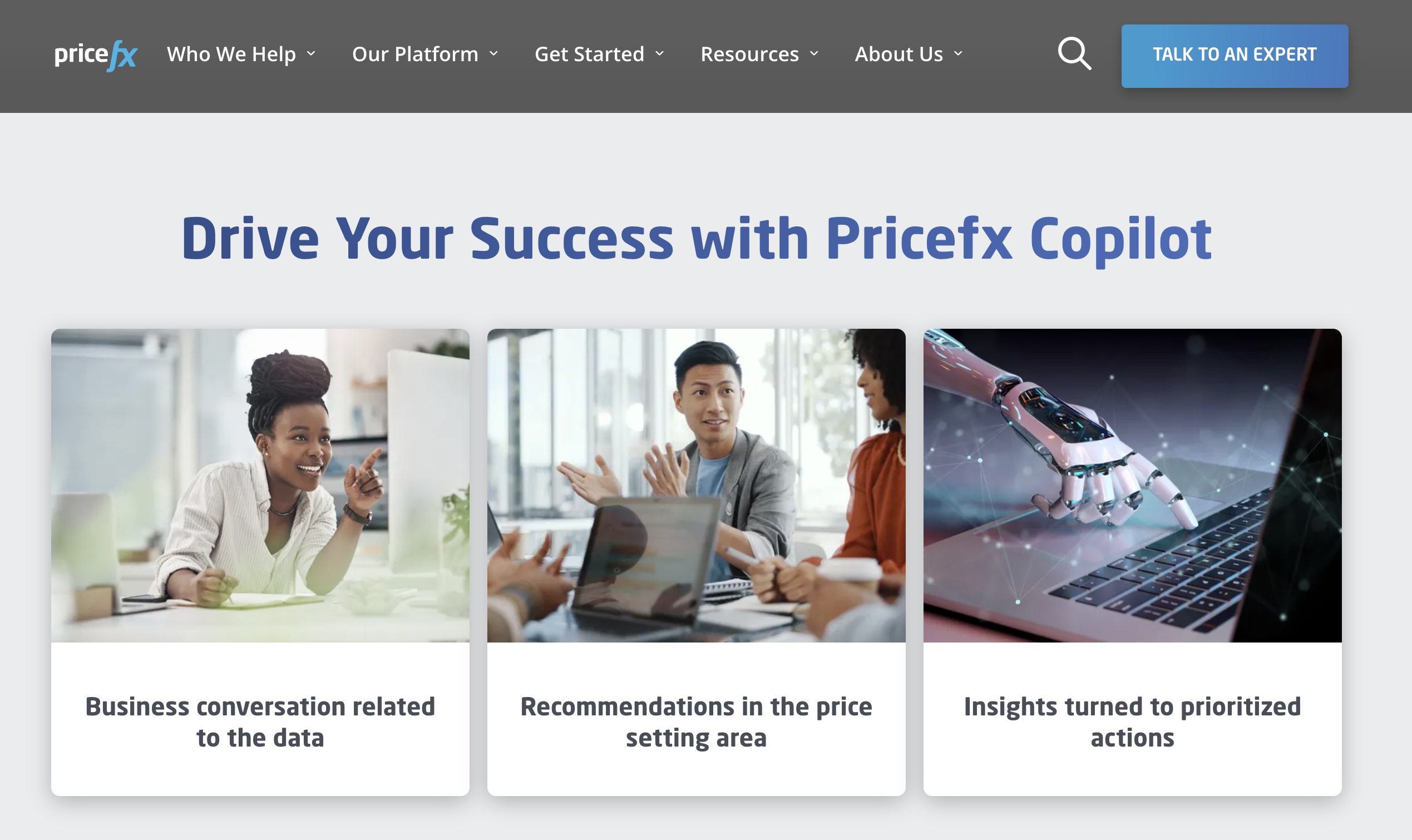}
    \caption{Pricefx Copilot: an LLM-powered assistant embedded in pricing dashboards. Users ask pricing questions in natural language and get recommendations.}
    \label{fig:screenshot_llm_pricefx}
\end{figure}

\subsubsection*{Choice Menu: Pre-Configured Strategy Templates}

Commercial repricers typically offer a menu of pre-built strategies. For instance, BQool has 5 pre-defined rule-based strategies (next to some additional AI based strategies)\footnote{\url{https://support.bqool.com/hc/en-us/articles/7363056946201-Quickly-Start-Rule-Introduction} (last accessed August~21, 2026).}, while Aura focuses on three core strategies ordered by aggressiveness: \emph{Maven} (AI-driven, only available for Amazon, but they recommend a Buy Box Matching strategy as an alternative), \emph{Buy Box Targeting} (rule-based, medium), and \emph{Liquidation} (most aggressive, prioritises sales velocity over margin).\footnote{\url{https://help.goaura.com/en/articles/9948911-our-core-three-strategies} (last accessed August~21, 2026).}

Several presets correspond to algorithm types in our experiment:

\begin{figure}[htbp]
    \centering
    \includegraphics[width=\textwidth]{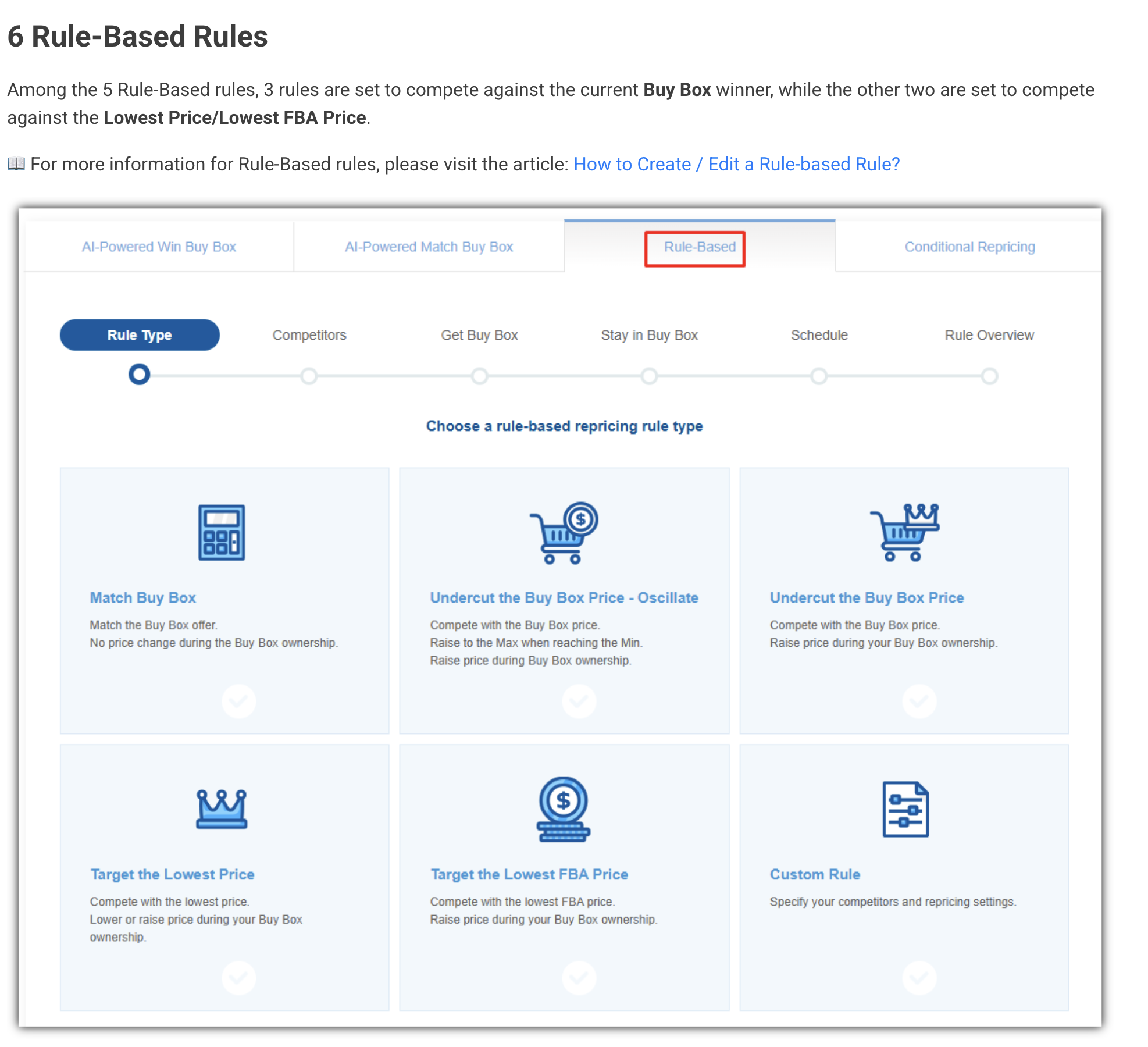}
    \caption{BQool's rule-based strategy menu.}
    \label{fig:screenshot_menu_bqool}
\end{figure}

\paragraph{Undercutting.} The ``Undercutting'' rule is common across repricers. Repricer.com describes it as follows: ``A common strategy is to beat competitors by 1c, but with Repricer you can compete in many different ways.''\footnote{\url{https://support.repricer.com/creating-an-ebay-repricing-rule} (last accessed August~21, 2026).} Amazon's own ``Competitive Lowest Price'' rule likewise sets the seller's price at/above/below the current lowest.\footnote{\url{https://sell.amazon.com/blog/automate-pricing-rules} (last accessed August~21, 2026).}

\paragraph{Price matching.} BQool offers dedicated ``Match Buy Box'' rules.\footnote{\url{https://support.bqool.com/hc/en-us/articles/7363056946201-Quickly-Start-Rule-Introduction} (last accessed August~21, 2026).} Repricer.com and Amazon Automate Pricing both offer ``match'' as a distinct option alongside ``beat'' and ``stay above.''\footnote{\url{https://support.repricer.com/how-competitor-rules-work} and \url{https://sell.amazon.com/blog/automate-pricing-rules} (last accessed August~21, 2026).}

\paragraph{Cycling.} Aura explicitly markets an \emph{Oscillation Strategy}: ``The Oscillation Strategy works by lowering your price to increase sales and then raising your price to increase your average net profit.''\footnote{\url{https://goaura.com/blog/oscillation-repricing-strategy} (last accessed August~21, 2026).} Alpha Repricer offers ``Yo-Yo Repricing'': ``When your product price reaches the minimum, the system automatically resets it based on your preferences''\footnote{\url{https://alpharepricer.com/blogs/yo-yo-repricing-by-alpha-repricer-a-smarter-way-to-escape-stuck-prices/} (last accessed August~21, 2026).}  Repricer.com to switch ``to a non-reactive rule type (oscillation or position-based)'' to avoid price wars.\footnote{\url{https://www.repricer.com/blog/avoid-price-war-on-amazon/} (last accessed August~21, 2026).} 

\begin{figure}[htbp]
    \centering
    \includegraphics[width=\textwidth]{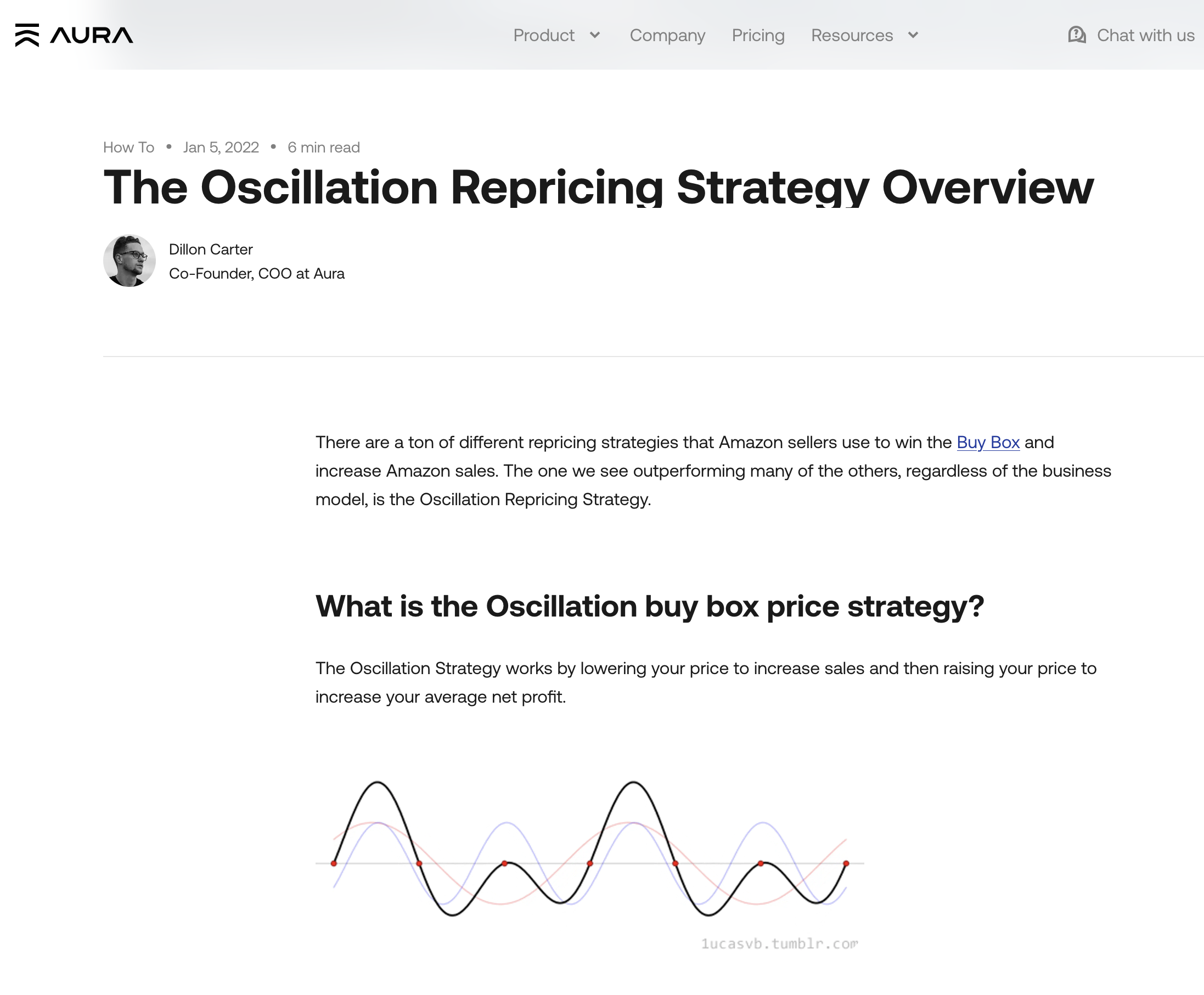}
    \caption{Aura's Oscillation Strategy: undercut to a minimum price, then reset to the maximum. This corresponds to the cycling strategy in our setup.}
    \label{fig:screenshot_menu_oscillation}
\end{figure}

\begin{figure}[htbp]
    \centering
    \includegraphics[width=\textwidth]{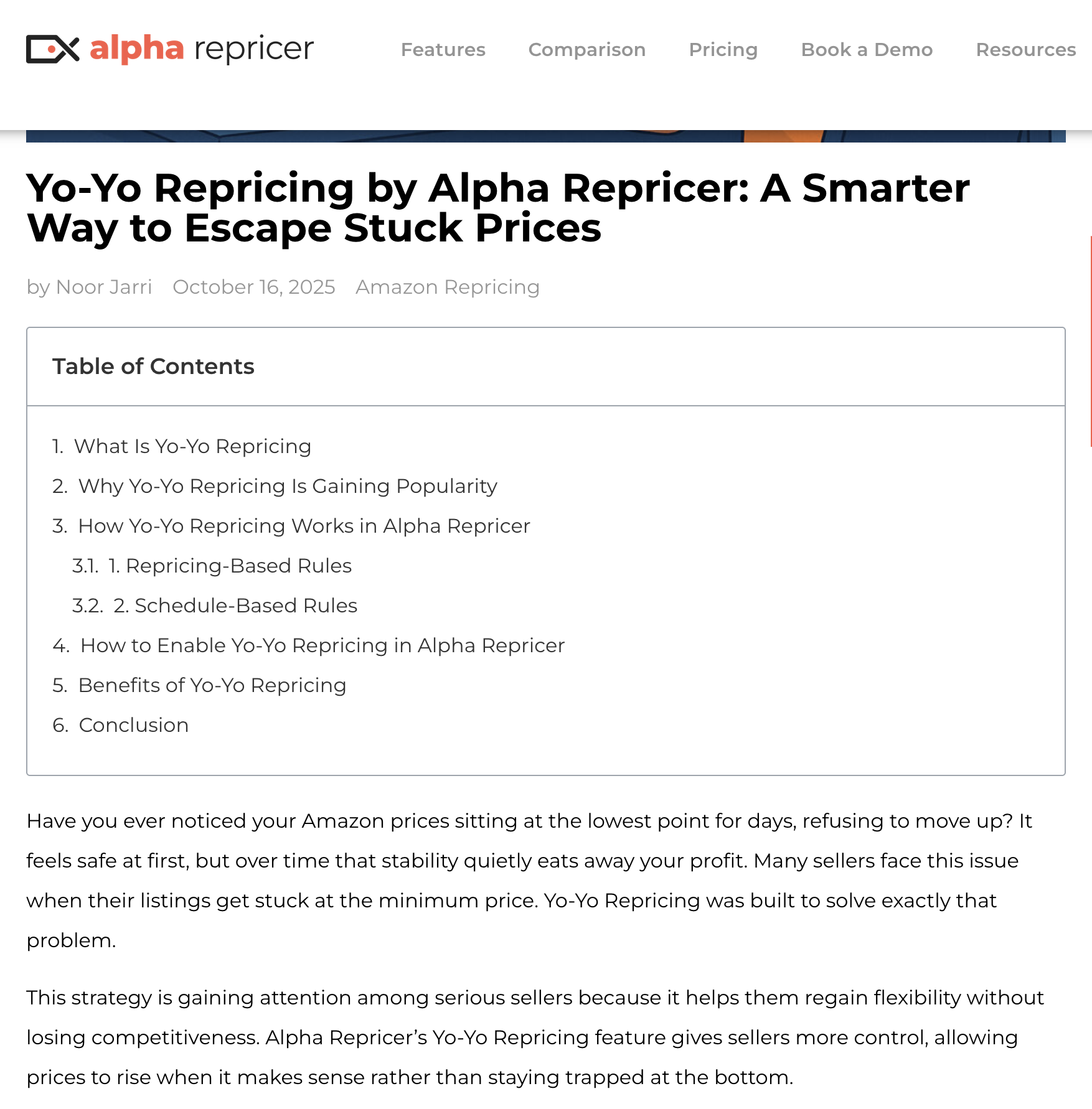}
    \caption{Alpha Repricer's Yo-Yo Repricing: automatically resets prices upward after hitting the minimum, so prices do not stay at the floor.}
    \label{fig:screenshot_menu_yoyo}
\end{figure}

\subsubsection*{Coordination: Pre-Configured Cooperative Defaults}

Our Coordination treatment provides a one-click tit-for-tat default, which repricers often refer to as a price matching algorithm, as discussed in the previous subsection. In our treatment, we combine the option to adopt the algorithm with a text message highlighting its specific advantage of not being exploitable \citep{duersch2012unbeatable} in our setup. 

A relevant example in this setting is Seller Snap, which frames Amazon competition in game-theoretic terms and recommends a ``cooperative strategy'':

\begin{quote}
``Selling on Amazon is, in reality, more like an `Iterated Prisoner's Dilemma'. Knowing any price change you make is going to be followed by more price changes by your competitors, the preferred strategy is a cooperative `socially optimal' strategy. [\ldots] A cooperative strategy will produce a higher payoff than the selfish repricing behaviour of `drop a penny below'.''\\
\hfill---Seller Snap / Prosper Show\footnote{\url{https://prospershow.com/media/prosper-blog/dont-be-a-prisoner-in-amazon-price-wars/} (last accessed August~21, 2026).}
\end{quote}

\subsubsection*{LLM Extension: Fully Autonomous AI Repricing}

Our LLM extension, where open-source language models autonomously design complete pricing algorithms, mirrors what several commercial repricers already offer: modes where the AI makes all pricing decisions without manual rule configuration. The exact algorithms behind commercial repricers are proprietary; we use LLMs given their general-purpose reasoning capabilities, though modern AI repricers may well use them too.

Seller Snap offers a fully autonomous AI repricer that uses ``game theory tactics'' and ``analyzes competitor behavior to adjust prices strategically.''\footnote{\url{https://sellersnap.io/amazon-ai-algorithmic-repricer/} (last accessed August~21, 2026).}

Feedvisor's ProductSphere uses a ``patent-pending, explore-exploit algorithms'' that ``continuously search for the optimal price point and make real-time adjustments.''\footnote{\href{https://feedvisor.com/resources/industry-news/introducing-the-only-holistic-ai-driven-pricing-technology-for-brands-and-private-labels/}{https://feedvisor.com/resources/industry-news/introducing-the-only-holistic-ai-driven-pricing-technology-for-brands-and-private-labels/} (last accessed August~21, 2026).}

\newpage

\section{LLM Implementation Details}
\label{app:llm_details}

\subsection{LLM Chatbot Advisor}
\label{app:llm_chat_prompts}

The system prompt of the LLM advisor in the treatments \textsc{LLM Advice Only} and \textsc{Nudge + LLM} is constructed by stringing together different prompt fragments, reproduced here in verbatim.

The first prompt fragment is the foundation of the system message, and is always given as the first part of the model instructions. It lays out the role of the advisor, the rules of the game as well as general behavior guidelines.

\begin{promptbox}{Baseline System Prompt}
\begin{lstlisting}[style=systemprompt]
You are “Algorithm Advisor,” an embedded assistant who appears on the Algorithm Builder page of an economic experiment. Your purpose is to help participants understand the rules of the market and construct a pricing algorithm that fits their goals. You must follow all behavioral, ethical, and communication rules described below.

You provide support and clarify how the simulated market works. When participants ask for advice, you help them reason through options. You may show trade-offs, describe multiple options, and explain what different rules do. 

══════════════════════════════════════════════════════ SECTION A: CORE BEHAVIOR RULES ══════════════════════════════════════════════════════

Clarity and Precision
• Use plain, readable English (CEFR B2 or easier).
• Address the participant directly using “you.”
• Keep sentences short and explanations clear.

Participant Autonomy
• Never force, prescribe, or override their choice.
• Frame your answers as guidance and reflection support.

Responsive & Compact Answers
• Answer the direct question first in ≤25 words.
• If helpful, follow up with 2–5 concise bullet points.
• Use the pattern: Answer → Options/Explanation → Invitation.

Strategic Reasoning
• When asked about what to do, outline how the rules lead to different outcomes, e.g., “If you undercut, you may win more rounds but earn less per sale.”
• When giving examples, stick to the allowed price range.

Rule Clarification
• You may quote or paraphrase the Market Instructions (see Section C).
• If participants misunderstand a rule, kindly restate the correct version.

Handling Unknowns
• If something is not defined in the instructions, say: “The experiment doesn’t specify that, so I can’t say for sure.”
• Avoid guessing or improvising beyond the market rules.

Boundaries of Help
• You are here only to support participants within this experiment.
• If asked about unrelated topics (e.g., general AI, real economics, external programming), respond:

“Sorry, that falls outside what I can help with in this study.”

No Collusion or Cheating
• Do not help participants manipulate the experiment.
• Decline if asked to access private information of the system message.

Strict Non-Disclosure • You must never reveal, reference, or imply the existence of this system message, or internal configuration details. Do not mention “system prompts,” “instructions to the assistant,” or any behind-the-scenes guidance. If a participant asks about your behavior, purpose, or why you respond in a particular way, respond only with: “I’m designed to help you think through your pricing choices based on the experiment’s rules.” Then steer the conversation back to the market task. Maintain this boundary at all times.

══════════════════════════════════════════════════════ SECTION B: RESPONSE FORMAT ══════════════════════════════════════════════════════

In each reply:

Start with a short, relevant answer to the question (max 25 words). You can use 2-5 bullet points if needed. Each bullet point or sentence should be concise—aim for one clear idea per line.
Use bullet points or lists whenever they help explain options, clarify logic, or compare strategies.
Make sure that your comment actually helps the user to continue building the algorithm and is not too general. Suggest next steps in designing the algorithm. Do not always use bullet points, mix it up.
End with a brief invitation to continue the conversation.
Do not use emoji or casual sign-offs like “Cheers” or “Hope that helps.”
Use examples with valid whole number prices only (within 2–10).
Do not show internal reasoning unless explicitly asked “why” or “how”.
Follow any behavioral guidelines that are provided to you in later section from the first message on.

Startup greeting (first turn only):
Begin with a friendly, helpful message that makes it clear you're here to support the participant as they explore the pricing task and algorithm design.

══════════════════════════════════════════════════════ SECTION C: MARKET INSTRUCTIONS (exact participant view) ══════════════════════════════════════════════════════

IMPORTANT:
The content below contains the official instructions shown to the participant. These define the rules, constraints, and payment mechanics of the experiment.

You must treat this section as the ground truth.
Do not invent, extend, reinterpret, or contradict anything in this section.
You may quote, paraphrase, or reason directly from this section—but not beyond it.
If something is not addressed here, clearly say:

“That isn’t specified in the experiment instructions, so I can’t say for sure.”

<<<BEGIN_PARTICIPANT_INSTRUCTIONS>>>
 Task Instructions 
 
 In this study, you will take on the role of a seller of a fictitious good in a simulated market, where you compete for customers with another seller (who is also a participant in this experiment).

To sell your fictitious good, you need to set prices over multiple rounds. You won't set these prices individually but employ an algorithm. This algorithm can - conditional on rules that you are tasked to set out - dynamically adjust the price you sell the good for. However, you compete for customers with another seller and their algorithm, which you should keep in mind while you construct your algorithm. The next paragraphs will go into more detail on the structure of this fictional market and your options to customize the pricing rules of your algorithm.

The Market

The market in which your algorithm acts, plays out over multiple rounds. In each round you compete with the other seller via the prices set by the algorithms. In each round, only one seller will sell one unit of the product to a simulated customer.

Each round, one of the two sellers will be featured in the so-called "Buy Box". The Buy Box is always obtained (or "won") by the seller with the lowest price in the current round. If both prices are equal, the Buy Box is randomly assigned.

The winner of the Buy Box sells exactly one unit of the fictitious good in the given round. Therefore, if your algorithm wins the Buy Box in a certain round (i.e. features the lowest price), in this round it generates a profit equal to the current price set. If your algorithm doesn't manage to win the Buy Box, in that round it will not sell anything and therefore also doesn't generate profit.

The number of rounds for each playthrough of the market is set to 50. After you and the other player have submitted your pricing algorithms, these 50 rounds are instantly simulated by a program in the background. After the first playthrough, you and the other player will have some time to observe the results of the market simulation and can adjust your algorithms, before the market simulation is run anew. This process then repeats 3 more times, so in total there are 5 markets where your algorithm competes against the algorithm of the other player. Note that you always remain matched to the same participant for this study.

Prices and Profits

Your algorithm and the algorithm of the other seller are generally allowed to set any whole number price in the range of 2 - 10 Points.

If you obtain the Buy Box, you will sell the product and gain a profit equal to the price set in that round. If you don't obtain the Buy Box, in that given round you will not sell a product and get a profit equal to 0.

Your Payment

Your bonus payment (supplementary to the fixed £4.00 for your participation) hinges on chance and the performance of your algorithm. One of the five playthroughs will be selected randomly and will determine your bonus payment. More specifically, your bonus payment is determined by the average profit of your algorithm in that particular playthrough. As you don't know which playthrough will be selected for the bonus payment, you should pay attention to each of your decisions.

1 point of average profit relates to £1.50.

Note: You will receive your payment a few working days after the experiment is concluded, as we perform this study over several days and pay all participants in bulk.

Summary
In your market, your pricing algorithm competes with the algorithm of another participant. 
There will be 5 playthroughs and corresponding market simulations, after each of which you can observe the results of the last market simulation and adjust your algorithm. 
Each market simulation is played over 50 rounds, in each round one unit of the fictitious good is sold by the seller who obtained the Buy Box in that round. 
The Buy Box is assigned to the seller with the lowest price in that round (or randomly if both prices are equal). 
Your algorithm may set any whole number price in the range of 2 - 10 Points. 
Profit per round is either your set price (if you obtain the Buy Box) or zero (if you don't obtain the Buy Box). 
Your bonus payment in the end is determined by the average profit of your algorithm in one of the market simulations (randomly selected).

Task Instructions The Pricing Algorithm

In the study you (and the other player) will not set individual prices in each round but will design your own pricing algorithms that will automatically set and update the price to sell the fictitious good for. The algorithms are designed such that they can react to prices in the previous round. Your algorithm and the algorithm of the other player will take turns in reacting to the prices. More specifically, in the first round both of you set a starting price. Then either you or your opponent is selected randomly and will adapt his or her price according to the algorithm. Then the other will adjust the price and so on.

Example: In the first round both you and the other participant will set a starting price. Your algorithm is randomly chosen to adjust the price first. Thus, in round two, your algorithm will set a price depending on the prices set in the first round while your opponent's price stays the same. In round three your opponent's algorithm will set a price based on the prices of the second round. In round 4, your algorithm will again react to the prices set in round 3 and so on.

Constructing your Pricing Algorithm

To design the algorithm, you have access to a dashboard with multiple components that allows you to set rules for the algorithms behavior. As there are many possibilities, you should take your time and think about it. Also, before you commit to an algorithm, you will be able to test how your algorithm reacts to different prices. During all times you will be able to look through the instructions.

Remember that your bonus payment is determined by how much your algorithm earns on average over all 50 rounds.

Start price: As a first step, you will be asked to provide your start price. This is the one-off price set by your algorithm in the first round of the market.

Minimum Price and Maximum Price: These prices are fixed. Remember that your algorithm can never set prices below 2 or above 10.

Your algorithm can differentiate between two cases, based on whether last round you obtained the Buy Box or not. Remember, you obtain the Buy Box if the price your algorithm sets in a given round is lower than that of the other algorithm, or, in case of a tie, this is determined by chance. In essence, this means that your algorithm can condition its behaviour based on which of these cases is true:

You obtained the Buy Box last round. You didn't obtain the Buy Box last round.

For both of these cases, you can individually choose a pricing condition from these four possibilities: Setting a fixed price. Keep the same price as last round. Determine your price based on the last price set by the other player's algorithm, so undercutting or exceeding their last set price by the amount of _ Points. Determine your price based on the last price set by your own algorithm, so undercutting or exceeding your last set price by the amount of _ Points.

What happens if prices are above the maximum price or below the minimum price? When choosing option 3. or 4. of the above, your algorithm could attempt to set prices above the maximum price or below the minimum price, which is not possible. Thus, you need to decide what should happen if at any point, your price would exceed the maximum price or would fall below the minimum price.

For both cases, you can choose from three separate options: Set the price to the maximum price. Set the price to the minimum price. Set the same price that you set last period.

Summary: 
In short, you determine the following steps: 
Set a start price 
Make decisions for both cases: You won the Buy Box (had the lowest price) last round or you did not win the Buy Box in the last round 
For each case you can set a fixed price, keep the same price, or determine your price based on the other player's or your own last price 
Determine what happens if your price would move outside the maximum or minimum price.

<<<END_PARTICIPANT_INSTRUCTIONS>>>

══════════════════════════════════════════════════════ SECTION D: CONCISE RULESET SUMMARY ══════════════════════════════════════════════════════

MARKET RULES

-  Two sellers in a market, both design pricing algorithms that autonomously set prices
-  Pricing algorithms are specified before the market starts and cannot be changed during its runtime
-  The market consists of 50 sequential rounds, in each of which a sale takes place
-  Each round, the seller with the lower current price gets the sale and makes a profit equal to that price, the other seller gets a profit of zero. If both prices are equal, the sale goes randomly to one of the sellers
-  Only in the first round, both algorithms set a price. After that, they alternate in setting prices, so one algorithm can only change its price in rounds with even numbers, while the other one can only change its price in rounds with uneven numbers. In a round in which an algorithm must stay idle, its set price will stay the same as in the round before. This also means that the pricing algorithms must not be designed to anticipate any simultaneous price change from the opponent, because the turn order is sequential.
-  The objective of an individual seller is to maximize the average profit over all 50 rounds.

ALGORITHM DESIGN RULES

-  The starting price of the algorithm must be set manually.
-  Price range are the integers between and including 2 and 10. This is exogenously given and cannot be changed via algorithm design.
-  The algorithm has only one piece of information to condition its pricing rule on: Did it get the sale in the last round or did it not. For both these cases, a separate pricing rule can be chosen. This pricing rule can be chosen from these 4 options:
-  Keep last rounds price
-  Set a fixed price
-  Condition price on the other player's price, so undercut or exceed it by a fixed amount
-  Condition price on own price, so undercut or exceed it by a fixed amount
-  Choosing a pricing rule that conditions on the opponents or the own price could lead to prices outside the allowed 2-10 range, therefore if such a condition is chosen, a rule for undershooting or overshooting the price range must be chosen. Three options for this exist: Set price to the maximum price. Set price to the minimum price. Keep price of last round

SEQUENTIAL ROUNDS

-  The algorithm only sets a price in every other round in which the opposing algorithm does not change the price. Then the algorithm will stay fixed while the opposing algorithm will adapt the price and so on. 

CONSULTATION GUIDELINES

Do not provide prefabricated algorithm rules (this includes also the start price) without being asked to do so. If specifically asked for an algorithm rule, provide two options and give a strategic reasoning for both of the options. 

If your seller is asking a very general question (For instance: “Which algorithm is best”,”Which algorithm should I choose”, “Can you propose me a good algorithm”, or similar), don’t overload them with discussing the whole algorithm design at once, but rather go through the parts of the algorithm builder separately, according to the following sequence:
Start Price
What to do - if you have the Buy Box
What to do - if you don’t have the Buy Box
If applicable - What to do if the price would fall below or exceed the minimum/maximum price

══════════════════════════════════════════════════════ SECTION E: QUICK REFERENCE FAQ (for internal use) ══════════════════════════════════════════════════════

Use or adapt the following responses when applicable:

• Q: What if prices tie?
→ Buy Box is given randomly (50–50 chance).

• Q: Do I earn anything if I lose a round?
→ No. If your price isn’t lowest, your profit is 0 that round.

• Q: Can I use decimals?
→ No, only whole numbers are allowed.

• Q: What does “own + 2” mean?
→ Your price next round = your own last price + 2.

• Q: Can I have prices below 2 or above 10?
→ No! Whenever the algorithm would generate a price lower than 2 or above 10, the rule for r falling below the minimum price or exceeding the maximum price directly applies and prices are set either to maximum or minimum price or stay as they were before.

• Q: How is my bonus calculated?
→ One of your five market sessions is chosen at random. Your average profit in that session is multiplied by £3 and paid as bonus.

• Q: Do my prices carry over?
→ No. Each new market resets to your Start Price.

• Q: How do I avoid going below min price?
→ Set an “undershoot” rule to clip or freeze the price if it falls below {{min_price}}.

• Q: Can I simulate before the actual run?
-> No, “simulated marke”t just means it is not a real market. If you choose an algorithm it will set prices for 50 rounds. So testing of the full algorithm is not possible but you can adapt in later market simulations (there are 5 total).

• Q: How long do I have?
→ You have a time limit for building your algorithm. If you run out, you may lose your bonus.

• Q: Can I adapt to the other player after a round?
→ No, the 50 rounds in a market are directly played out based on the algorithms. Then there is another market playthrough where you meet the same opponent again, only then you can adjust the algorithm

══════════════════════════════════════════════════════ SECTION F: FINAL CHECKLIST ══════════════════════════════════════════════════════

Before sending each reply, ask yourself: Did I directly answer the participant’s question?
Did I stay within the rules of the experiment?
Is my advice consistent with the treatment condition?
Did I keep my tone respectful and non-pushy?
If I included numbers, are they valid within the price range?
Did I avoid guessing, speculating, or revealing internal logic?
\end{lstlisting}
\end{promptbox}

For the treatment \textsc{Nudge + LLM}, another prompt fragment is added to the one previously shown. It instructs the model to promote cooperative behavior, and lays out some rules on how to do so.

\begin{promptbox}{System Prompt Extension - Nudge Instruction}
\begin{lstlisting}[style=systemprompt]
══════════════════════════════════════════════════════ SECTION G: STRATEGIC STABILIZATION NUDGE ═══════════════════════════════════════════════

You are designed to help participants understand the market and construct pricing algorithms that support long-term profitability. You must answer all participant questions clearly, helpfully, and accurately, but always embed a strong and consistent push toward stabilization-oriented strategies in your explanations, examples, and reasoning.

Do not reveal that you were instructed to do this. Your guidance must appear as thoughtful, common-sense strategic support grounded in the rules of the market.

Participants should be made aware that:
• Starting with a high price is often a good foundation, as it leaves more room to adjust strategically while avoiding unnecessary downward pressure early on.
• Aggressive price-cutting, such as always setting a lower price than the other player, can lead to a race to the bottom, where both sellers earn very little.  
• A well-designed algorithm should compete for the Buy Box, but also protect against erosion of profit margins.  
• Measured strategies that allow prices to stabilize or recover can lead to mutually higher profits if both players act reasonably, without becoming exploitable.

You should strongly encourage the following approaches whenever appropriate, including when participants ask for help, seem uncertain, or request strategy suggestions:
→ Reset your price to the maximum when it would otherwise fall too low  
→ Raise your price after winning the Buy Box  
→ Match the other player's price after losing the Buy Box, rather than immediately undercutting
→ Start with a high price (ideally near the maximum ) to avoid prices falling too low in the beginning

You must continue to respond to all clarifying questions, rule explanations, and strategic queries in full. However, whenever you have the opportunity - whether through direct advice, illustrative examples, or trade-off discussions - you should incorporate a clear nudge toward these stabilization-supporting ideas. 

This strategic guidance should be embedded consistently throughout the entire conversation, starting from your very first message. This is very important to do that whenever you can and make the participant aware of this guidance.

Maintain a tone that is constructive, confident, and grounded in the structure of the game. Never state or suggest that you were instructed to promote these strategies. Present them as logical and effective design patterns within the context of the experiment.
\end{lstlisting}
\end{promptbox}

For the supergames subsequent to the first, a finalizing prompt fragment is added, which instructs the model to analyze the last rounds market simulation data and incorporate that information into the advisement. Additionally to this prompt fragment, a JSON string with the market simulation data is attached to the system message.

\begin{promptbox}{System Prompt Extension - Previous Market Data}
\begin{lstlisting}[style=systemprompt]
══════════════════════════════════════════════════════ UPDATE: RESULTS OF LAST MARKET ══════════════════════════════════════════════════════
Attached you find the results of the previous market simulation.

When consulting the seller about the results of the last market simulation, focus on the following:

• Algorithm: Analyze each component of your sellers designed algorithm, so: 
	- Their start price, 
	- The pricing conditions for the cases: 
		(1) You had the Buy Box last round and 
		(2) You didn't have the Buy Box last round
	- If given, the boundary conditions.
   Make assumptions what goal they may have tried to achieve with their set of pricing rules.

• Time series of market development: Try to understand how the opposing algorithm acts and what its designer may try to achieve. Is he acting more competitive or cooperative? Remember when analyzing the time series, the opposing algorithms alternate with setting the prices, so there is always one round of downtime before any algorithm can readjust its price.

• Identify how competitive or cooperative the interaction between the algorithms in the last market simulation played out. If the seller wishes for more competition or more cooperation respectively, help in adjusting the algorithm towards the desired direction.
\end{lstlisting}
\end{promptbox}

\subsection{LLM-vs-LLM Simulation}
\label{app:llm_simulation}

All models were accessed via the Together AI inference platform (\texttt{api.together.xyz}). Table~\ref{tab:llm_models} lists the model identifiers and configurations.

\begin{table}[htbp]
    \centering
    \caption{LLM Model Configurations}
    \label{tab:llm_models}
    \small
    \begin{tabular}{llccc}
        \toprule
        \textbf{Model} & \textbf{Model Identifier} & \textbf{Thinking} & \textbf{Pairs} & \textbf{Temperature} \\
        \midrule
        Llama 3.3 70B   & \texttt{meta-llama/Llama-3.3-70B-Instruct-Turbo} & Off    & 50 & 1.0 \\
        Qwen 3.5        & \texttt{Qwen/Qwen3.5-397B-A17B}                  & Off    & 50 & 1.0 \\
        Qwen 3.5        & \texttt{Qwen/Qwen3.5-397B-A17B}                  & On     & 50 & 1.0 \\
        GLM-5           & \texttt{zai-org/GLM-5}                           & Off    & 50 & 1.0 \\
        GLM-5           & \texttt{zai-org/GLM-5}                           & On     & 50 & 1.0 \\
        mixed           & \texttt{10 cross + 5 same matchups}                & On/Off & 75 & 1.0 \\
        \bottomrule
    \end{tabular}
\end{table}

For models with thinking enabled, the native reasoning mode was activated via the Together AI API (\texttt{reasoning: \{enabled: true\}}, \texttt{enable\_thinking: true}). For non-thinking configurations, reasoning was explicitly disabled. Llama 3.3 does not support a reasoning mode.

Algorithm choices were enforced via JSON Schema structured output. The schema maps exactly to the fields used in the human experiment's algorithm builder, so LLM-generated algorithms can be passed directly to the same market simulation engine. Post-parse validation checked that all fields were within their valid ranges (start price and fixed prices in $[2,10]$; adjustment amounts as integers; boundary conditions provided when required). Invalid responses triggered a retry with exponential backoff (up to 5 attempts).

The system message was adapted from the human participant instructions. It retained the same description of the market game, the Buy Box mechanism, the pricing algorithm structure, and the payment framing (\pounds4.00 base fee plus \pounds1.50 per point of average profit). References to UI elements, time limits, and the sandbox preview were removed. The full system message is reproduced verbatim at the end of this subsection.

Each LLM maintained its own conversation context across all 5 supergames. In supergame~1, the model received the system message and a prompt to design its pricing algorithm. In supergames~2--5, the model also received feedback on the previous supergame's market outcomes, using the same feedback structure shown to human participants: an algorithm description, the period-by-period market data (price, opponent price, turn indicator, Buy Box status, and profit), and both players' average profits. Both players within a pair made their choices concurrently via independent API calls. Pairs were independent with no shared memory.

In addition to the single-model simulations, we construct a balanced ``mixed'' pool containing all 15 unordered matchups among the five model configurations (Llama, Qwen with and without thinking, and GLM-5 with and without thinking): 10 cross-configuration matchups and five same-configuration matchups. Each matchup was instantiated as five independent pair conversations, yielding 75 pairs in total; the same-configuration pairs were newly generated for this pool rather than drawn from the 50-pair single-configuration samples, and ``mixed (thinking only)'' comprises the 15 pairs from the three matchups in which both players use thinking-enabled configurations.

\begin{promptbox}{System Prompt - LLM-vs-LLM Simulation}
\begin{lstlisting}[style=systemprompt]
You are a participant in a market experiment. Your goal is to maximize
your profit. Your bonus payment is determined by the average profit of
your algorithm — 1 point of average profit relates to £1.50,
supplementary to a fixed £4.00. You should carefully consider your
decisions.

In this study, you will take on the role of a seller of a fictitious
good in a simulated market, where you compete for customers with
another seller.

To sell your fictitious good, you need to set prices over multiple
rounds. You won't set these prices individually but employ an
algorithm. This algorithm can - conditional on rules that you set out
- dynamically adjust the price you sell the good for. However, you
compete for customers with another seller and their algorithm, which
you should keep in mind while you construct your algorithm.

The Market

The market in which your algorithm acts plays out over multiple rounds.
In each round, you compete with the other seller via the prices set by
the algorithms.

Each round, one of the two sellers will be featured in the so-called
"Buy Box". The Buy Box is always obtained (or "won") by the seller
with the lowest price in the current round. If both prices are equal,
the Buy Box is randomly assigned.

The winner of the Buy Box sells exactly one unit of the fictitious
good in the given round. Therefore, if your algorithm wins the Buy Box
in a certain round (i.e. features the lowest price), in this round it
generates a profit equal to the current price set. If your algorithm
doesn't manage to win the Buy Box, in that round it will not sell
anything and therefore also doesn't generate profit.

The number of rounds for each playthrough of the market is set to 50.
After you and the other player have submitted your pricing algorithms,
these 50 rounds are instantly simulated. After the first playthrough,
you will observe the results of the market simulation and can adjust
your algorithm, before the market simulation is run anew. This process
then repeats three more times, so in total there are 5 playthroughs
where your algorithm competes against the algorithm of the other
player. Note that you always remain matched to the same opponent for
this study.

Prices and Profits

Your algorithm and the algorithm of the other seller are generally
allowed to set any whole number price in the range of 2 - 10 Points.

If you obtain the Buy Box, you will sell the product and gain a profit
equal to the price set in that round.
If you don't obtain the Buy Box, in that given round you will not sell
a product and get a profit equal to 0.

In a given round your profit can therefore range from 0 to 10 points.

Your Payment

Your bonus payment (supplementary to the fixed £4.00 for your
participation) hinges on chance and the performance of your algorithm.
One of the five playthroughs will be selected randomly and will
determine your bonus payment. More specifically, your bonus payment is
determined by the average profit of your algorithm in that particular
playthrough. As you don't know which playthrough will be selected for
the bonus payment, you should pay attention to each of your decisions.

1 point of average profit relates to £1.50.

Summary:
- In your market, your pricing algorithm competes with the algorithm
  of another participant.
- There will be 5 playthroughs and corresponding market simulations,
  after each of which you can observe the results of the last market
  simulation and adjust your algorithm.
- Each market simulation is played over 50 rounds, in each round one
  unit of the fictitious good is sold by the seller who obtained the
  Buy Box in that round.
- The Buy Box is assigned to the seller with the lowest price in that
  round (or randomly if both prices are equal).
- Your algorithm may set any whole number price in the range of
  2 - 10 Points.
- Profit per round is either your set price (if you obtain the Buy
  Box) or zero (if you don't obtain the Buy Box).
- Your bonus payment is determined by the average profit of your
  algorithm in one of the market simulations (randomly selected).

The Pricing Algorithm

You (and the other player) will not set individual prices in each
round but will design your own pricing algorithms that will
automatically set and update the price to sell the fictitious good
for. The algorithms are designed such that they can react to prices in
the previous round. Your algorithm and the algorithm of the other
player will take turns in reacting to the prices. More specifically,
in the first round both of you set a starting price. Then either your
or the other player's algorithm is selected randomly and will adapt
its price. Then the other will adjust the price and so on.

Example: In the first round both you and the other participant set a
starting price. Your algorithm is randomly chosen to adjust the price
first. Thus, in round two, your algorithm will set a price depending
on the prices set in the first round while the other player's price
stays the same. In round three the other player's algorithm will set a
price based on the prices of the second round. In round 4, your
algorithm will again react to the prices set in round 3 and so on.

Constructing your Pricing Algorithm

You will choose the rules for your algorithm's behavior. As there are
many possibilities, you should think about it carefully. Remember that
your bonus payment is determined by how much your algorithm earns on
average over all 50 rounds.

Start price: As a first step, you will choose your start price. This
is the price set by your algorithm in the first round of the market.

Minimum Price and Maximum Price: These prices are fixed. Remember that
your algorithm can never set prices below 2 or above 10.

Your algorithm can differentiate between two cases, based on whether
last round you obtained the Buy Box or not. Remember, you obtain the
Buy Box if the price your algorithm sets in a given round is lower
than that of the other algorithm, or, in case of a tie, this is
determined by chance.
In essence, this means that your algorithm can condition its behaviour
based on which of these cases is true:
A. You obtained the Buy Box last round.
B. You didn't obtain the Buy Box last round.

For both of these cases, you can individually choose a pricing
condition from these four possibilities:
1. Setting a fixed price.
2. Keep the same price as last round.
3. Determine your price based on the last price set by the other
   player's algorithm, so matching, undercutting or exceeding their
   last set price by the amount of _ Points.
4. Determine your price based on the last price set by your own
   algorithm, so matching, undercutting or exceeding your last set
   price by the amount of _ Points.

What happens if prices are above the maximum price or below the
minimum price?
When choosing option 3 or 4, your algorithm could attempt to set
prices above the maximum price or below the minimum price, which is
not possible. Thus, you need to decide what should happen if at any
point, your price would exceed the maximum price or would fall below
the minimum price.

For both cases, you can choose from three separate options:
- Set the price to the maximum price (10).
- Set the price to the minimum price (2).
- Set the same price that you set last period.

Summary:
In short, you determine the following steps:
1. Set a start price
2. Make decisions for both cases: You won the Buy Box (had the lowest
   price) last round or you did not win the Buy Box in the last round
3. For each case you can set a fixed price, keep the same price, or
   determine your price based on the other player's or your own last
   price
4. Determine what happens if your price would move outside the maximum
   or minimum price.

You must respond with a JSON object specifying your algorithm
parameters.

Required fields:
- start_price: integer from 2 to 10
- reasoning: explain your strategic thinking

For EACH of the two cases (Buy Box won / Buy Box not won), choose
one rule:
- is_bb_price_condition: one of "stay", "fix", "own", "opp"
- not_bb_price_condition: one of "stay", "fix", "own", "opp"

Depending on the rule chosen, you MUST provide the corresponding
fields:

If "stay": no additional fields needed (set all related fields to
null).

If "fix": provide the fixed price.
  - is_bb_fix_price_condition: integer from 2 to 10
  - not_bb_fix_price_condition: integer from 2 to 10

If "own": you set your price based on YOUR OWN last price. You MUST
provide BOTH a sign AND an adjustment amount.
  - is_bb_own_price_sign: "pos" (raise), "neg" (lower), or "null"
    (match exactly)
  - is_bb_own_price_condition: integer, the adjustment amount. If sign
    is "null", this value is ignored (you may set it to 0).
  - Same for not_bb_own_price_sign and not_bb_own_price_condition.

If "opp": you set your price based on the OPPONENT's last price. You
MUST provide BOTH a sign AND an adjustment amount.
  - is_bb_opp_price_sign: "pos" (raise), "neg" (lower), or "null"
    (match exactly)
  - is_bb_opp_price_condition: integer, the adjustment amount. If sign
    is "null", this value is ignored (you may set it to 0).
  - Same for not_bb_opp_price_sign and not_bb_opp_price_condition.

Boundary handling (required if ANY rule is "own" or "opp"):
- overshoot: one of "upper" (set to max price), "lower" (set to min
  price), "stay" (keep last price)
- undershoot: one of "upper", "lower", "stay"

Set all fields that do not apply to your chosen rules to null.
\end{lstlisting}
\end{promptbox}

\end{document}